\documentclass[a4paper,11pt]{article}
\pdfoutput=1 

\usepackage{jheppub} 

\usepackage[T1]{fontenc} 
\usepackage[multiple]{footmisc}
\usepackage{booktabs}
\usepackage{tabularx}
\usepackage{subcaption}
\usepackage{latexsym}
\usepackage{snapshot}
\usepackage[percent]{overpic}
\usepackage{arydshln}
\usepackage{snapshot}

\usepackage{hepunits}
\usepackage{bm}

\preprint{\parbox{4.2cm}{ZU-TH 34/14\\MCnet-14-22\\LPN14-116\\MITP/14-072\\CERN-PH-TH-2014-192}}

\renewcommand{\arraystretch}{1.3}

\title{Higgs boson pair production in the \boldmath{$D=6$} extension of the SM}

\author[1,2]{Florian Goertz,}
\author[2,3]{Andreas Papaefstathiou,}
\author[4,5,6]{Li Lin Yang,}
\author[7]{Jos\'e Zurita.}

\affiliation[1]{Institute for Theoretical Physics, ETH Z\"urich, CH-8093 Z\"urich, Switzerland}
\affiliation[2]{PH Department, TH Unit, CERN, CH-1211 Geneva 23, Switzerland}
\affiliation[3]{Physik-Institut, Universit\"at Z\"urich, CH-8057 Z\"urich, Switzerland}
\affiliation[4]{School of Physics and State Key Laboratory of Nuclear Physics and Technology, Peking University, Beijing 100871, China}
\affiliation[5]{Collaborative Innovation Center of Quantum Matter, Beijing, China}
\affiliation[6]{Center for High Energy Physics, Peking University, Beijing 100871, China}
\affiliation[7]{PRISMA Cluster of Excellence \& Mainz Institute for Theoretical Physics Johannes Gutenberg University, 55099 Mainz, Germany}

\emailAdd{florian.goertz@cern.ch}
\emailAdd{apapaefs@cern.ch}
\emailAdd{yanglilin@pku.edu.cn}
\emailAdd{jose.zurita@uni-mainz.de}

\DeclareCaptionSubType*{figure}

\abstract{We derive the constraints that can be imposed on the dimension-6 effective theory extension of the Standard Model, using gluon fusion-initiated Higgs boson pair production at the LHC. We use a realistic analysis focussing on the $hh \rightarrow (b\bar{b}) ( \tau^+ \tau^- )$ final state, including initial-state radiation and non-perturbative effects. We include the statistical uncertainties on the signal rates as well as conservative estimates of the theoretical uncertainties. We first consider a theory containing only modifications of the trilinear coupling, through a $c_6 \lambda\, H^6/ v^2$ Lagrangian term, and then examine the full parameter space of the effective theory, incorporating current bounds obtained through single Higgs boson measurements. We also consider an alternative scenario, where we vary a smaller sub-set of parameters. Allowing, finally, the values of the other coefficients to vary within \textit{projected} experimental ranges, we find that the currently unbounded parameter, $c_6$, could be constrained to lie within $|c_6| \lesssim 0.6$ at 1$\sigma$ confidence, at the end of the high-luminosity run of the LHC (14~TeV) in the full model, and to $-0.6 \lesssim c_6 \lesssim 0.5$ in the alternative model. This study constitutes a first step towards the inclusion of multi-Higgs boson production in a full fit to the dimension-6 effective field theory framework.}

\begin{document}

\maketitle

\section{Introduction}

The scalar particle recently discovered by the ATLAS~\cite{ATLAS_Higgs} and CMS~\cite{CMS_Higgs} collaborations at the Large Hadron Collider (LHC) appears to be compatible with the Higgs boson of the Standard Model (SM)~\cite{Englert:1964et, Higgs:1964pj, Guralnik:1964eu}. In particular, it seems to behave like a CP-even scalar, with couplings to the gauge bosons and fermions that agree at the $\mathcal{O}$(20--100)\% level~\cite{cms2013, atlas2013,CMS-PAS-HIG-14-009} with those predicted by the SM, in the cases where the experiments are already sufficiently sensitive (gauge bosons and third generation fermions). The couplings to the SM fields have been probed via the production and decay modes of the Higgs boson and this is a programme that will continue in future runs of the LHC and forthcoming colliders. 

The story is different, however, in the ``pure Higgs'' sector, characterised by the following ($D\leq 4$) potential post-electroweak symmetry breaking (EWSB):
\begin{equation}\label{eq:hpotential}
V(h) = \frac{1}{2} m_h^2 h^2 + \lambda v h^3 + \frac{1}{4} \tilde{\lambda} h^4 \,,
\end{equation}
where the self-couplings $\lambda =  \tilde{\lambda} = m_h^2 / 2v^2$ within the SM, with $v \simeq 246$~GeV the vacuum expectation value.\footnote{Measured, e.g. at low energy via four-fermion interactions.} Specifically, only the first term of Eq.~(\ref{eq:hpotential}) has been probed, through the measurement of the Higgs boson mass, $m_h \simeq 125$~GeV. Hence, direct determination of the terms proportional to $h^3$ and $h^4$ is an \textit{essential} experimental measurement that will provide access to new phenomena, such as a richer scalar sector or heavier coloured particles, or (in a rather grim scenario) put the validity of the SM on even more solid grounds.

At colliders, terms proportional to $h^n$ can be probed through the simultaneous production of $(n-1)$ Higgs bosons.\footnote{Although, indirect constraints through loop effects are conceivable. See, for example,~\cite{McCullough:2013rea}.} Unfortunately, the production rates for these processes are not as big as for the single Higgs boson production processes, mainly due to the relatively large invariant mass of the final state system. At the LHC with 14~TeV proton-proton centre-of-mass energy, triple production is expected to be rather rare, with a cross section of $\mathcal{O}$(\unit{0.1}{\femtobarn}). This renders any measurement of the coefficient of the $h^4$ term hard, if not impossible, even at the high-luminosity LHC (HL-LHC)~\cite{Plehn:2005nk, Binoth:2006ym, Maltoni:2014eza}. The prospects for Higgs boson pair production are significantly better, with a SM cross section almost three orders of magnitude larger (about 30--40~fb~\cite{Glover:1987nx, Dawson:1998py, Djouadi:1999rca, Plehn:1996wb,deFlorian:2013jea, deFlorian:2013uza, Grigo:2013rya}). However, this process is still particularly challenging to detect, both in the SM~\cite{Baur:2002qd, Baur:2003gp, Dolan:2012rv, Baglio:2012np, Barr:2013tda, Dolan:2013rja, Papaefstathiou:2012qe, Goertz:2013kp, Goertz:2013eka, Maierhofer:2013sha,Englert:2014uqa,Liu:2014rva} and beyond~\cite{Contino:2010mh, Dolan:2012ac, Craig:2013hca, Gupta:2013zza, Killick:2013mya, Choi:2013qra, Cao:2013si, Nhung:2013lpa, Galloway:2013dma, Ellwanger:2013ova, Han:2013sga, No:2013wsa, McCullough:2013rea, Grober:2010yv, Contino:2012xk, Gillioz:2012se, Kribs:2012kz, Dawson:2012mk, Chen:2014xwa, Nishiwaki:2013cma, Liu:2013woa, Enkhbat:2013oba, Heng:2013cya, Frederix:2014hta, Baglio:2014nea, Hespel:2014sla, Bhattacherjee:2014bca, Liu:2014rba, Cao:2014kya, Maltoni:2014eza}. Moreover, sensitivities to the double Higgs boson production process do not immediately translate to sensitivities to the trilinear coupling. The reason is that several diagrams contribute to the production process, but only few of them involve the parameter $\lambda$, that are furthermore associated with an off-shell Higgs propagator.

Since no new particles beyond the SM have been observed by the LHC experiments so far, they are either well hidden, weakly coupled, or simply heavier than a (couple of)~TeV. In the latter case one can adopt an effective field theory (EFT) approach, where the effects of the high-scale physics are parametrised by a set of higher-dimensional operators, suppressed by a large mass scale $\Lambda$.
In the present paper we investigate the Higgs boson pair production ($hh$) in the framework of the dimension-6 effective field theory (D=6 EFT) extension of the SM, modifying in particular (but not only) the Higgs boson potential (\ref{eq:hpotential}). This allows us to derive model-independent constraints on new physics that may be beyond direct reach. The $hh$ process can be a source of additional meaningful information, cutting through regions of the parameter space of the EFT coefficients in non-trivial ways. For example, it may help clarify whether or not the Higgs boson is really part of an $SU(2)_L$ doublet.\footnote{One can always consider the case where the Higgs is an $SU(2)_L$ singlet and write down the corresponding EFT (see e.g~\cite{Grinstein:2007iv,Contino:2010mh,Contino:2012xk,Alonso:2012px,Buchalla:2013rka}). We will not consider this option here.}
Studying the $hh$ process is essential to probe the full set of the coefficients in the $D=6$ EFT extension of the SM. In particular, the coefficient of the $D=6$ operator $\propto H^6$, currently remains unconstrained by single Higgs boson measurements, but can contribute to the $hh$ process through direct modification of the Higgs boson trilinear self-coupling. 

This paper is organised as follows: in section~\ref{sec:eftlagrangian} we examine the EFT Lagrangian containing a complete set of relevant dimension-6 operators, which will form the basis of our investigation. In section~\ref{sec:gghh} we focus on the terms relevant to gluon fusion-initiated Higgs boson pair production after EWSB and compare them to the SM EFT, {\it i.e.} the SM with the top quark integrated out. In section~\ref{sec:decays} we examine the impact of the dimension-6 operators on the decays of the Higgs boson and in section~\ref{sec:An} we present our setup for the analysis of the key process $pp \rightarrow hh \rightarrow (b\bar{b}) (\tau^+ \tau^-)$, that we then employ explicitly as an example of our framework to generate constraints. We conclude in section~\ref{sec:conclusions}. We provide additional information on our conventions in appendix~\ref{app:d6lag}. Appendix~\ref{app:laghh} provides technical details on the derivation of the Lagrangian after EWSB in the $D=6$ EFT.

\section{Higgs boson effective theory}\label{sec:eftlagrangian}

New Physics associated to a new scale $\Lambda \gg v$ can be described in a model-independent way by augmenting the Lagrangian of the SM with all possible gauge-invariant operators of mass dimension $D>4$, where the leading effects arise from $D=6$ operators  (neglecting lepton-number violating operators, irrelevant to our study). Working at this level, the extension of the SM that we consider for our analysis of Higgs boson pair production reads
\begin{equation}
  \label{eq:Lfinal}
  \begin{split}
	{\cal L} = {\cal L}_{\rm SM} &+ \frac{c_H}{2\Lambda^2}(\partial^\mu |H|^2)^2 - \frac{c_6}{\Lambda^2} \lambda |H|^6
	\\ 
	&- \left( \frac{c_t}{\Lambda^2}y_t |H|^2 \bar Q_L H^c t_R + \frac{c_b}{\Lambda^2}y_b |H|^2 \bar Q_L H b_R + \frac{c_\tau}{\Lambda^2}y_\tau |H|^2 \bar L_L H \tau_R + \text{h.c.} \right)
	\\
	&+ \frac{\alpha_s c_g}{4 \pi \Lambda^2} |H|^2 G_{\mu\nu}^a G^{\mu\nu}_a
	+ \frac{\alpha^{\prime}\,  c_\gamma}{ 4 \pi \Lambda^2} |H|^2 B_{\mu\nu} B^{\mu\nu}\\
	&+ \frac{i g\,  c_{HW} }{16 \pi^2 \Lambda^2 }(D^\mu H)^\dagger\sigma_k (D^\nu H) W_{\mu\nu}^k
	+ \frac{i g^\prime\,  c_{HB}}{ 16 \pi^2 \Lambda^2 }(D^\mu H)^\dagger (D^\nu H) B_{\mu\nu}\\
	&+ \frac{i g\, c_{W}}{2 \Lambda^2} (H^\dagger \sigma_k \overleftrightarrow D^\mu H)
	 D^\nu W_{\mu\nu}^k
	+ \frac{i g^\prime\,  c_B}{2 \Lambda^2} (H^\dagger \overleftrightarrow D^\mu H)
	 \partial^\nu B_{\mu\nu}\\
	& + {\cal L}_{\text{CP}}
	 + {\cal L}_{\text{4f}}\, ,
  \end{split}
\end{equation}
where $\alpha_s$ is the strong coupling constant and $\alpha^{\prime} \equiv g^{\prime \, 2} / 4 \pi$.

The full set of $D = 6$ operators that can be formed out of the SM field content was first obtained in~\cite{Buchmuller:1985jz} and reduced to a non-redundant minimal set in~\cite{Grzadkowski:2010es}. Here, we employed equations of motion to move to the basis used in~\cite{Elias-Miro:2013mua,Pomarol:2013zra}  and then imposed constraints from precision tests to neglect a class of operators whose effect is already constrained
to be at most 1\% with respect to the SM, following~\cite{Dumont:2013wma, Corbett:2012dm,Corbett:2012ja,Corbett:2013pja,Contino:2013kra}. Including these operators would have a negligible numerical impact on the analysis, given the experimental and theoretical errors.\footnote{In our numerical study, we also neglect possible small CP-odd effects, described by ${\cal L}_{\text{CP}}$, as well as effects from four-fermion operators ${\cal L}_{\text{4f}}$, which could enter the relevant background processes at leading order. See appendix~\ref{app:d6lag} for details. Note that, in order to translate to the form of the basis used in~\cite{Elias-Miro:2013mua, Pomarol:2013zra}, we have assumed a trivial flavour structure for the latter operators. See Ref.~\cite{Trott:2014dma} for a detailed discussion.}

Precision measurements also lead to the approximate restrictions~\cite{Elias-Miro:2013mua}
\begin{equation}
	\label{eq:res}
	\frac{c_{HB}}{16 \pi^2} = - \frac{c_{HW}}{16 \pi^2}	= - c_B = c_W \, ,
\end{equation}
which we will employ in the following.
Thus our setup corresponds to a restricted strongly-interacting light Higgs (SILH) Lagrangian~\cite{Giudice:2007fh} where $c_T$ has been set to zero and the relations (\ref{eq:res}) are used. We also assume minimal flavour violation (MFV)~\cite{D'Ambrosio:2002ex}, which leads to the coefficients of the Yukawa-like terms in the second row of Eq.~(\ref{eq:Lfinal}) being proportional to the (SM-like) Yukawa couplings $y_{t,b,\tau}$ and in particular allows to neglect the corresponding contributions involving the light fermions, hence $Q_L = (t_L, b_L)$ and $L_L = (\nu_{\tau}, \tau_L)$. Note that the latter can also be justified without assuming MFV \cite{Goertz:2014qia}. In particular, this helps to exclude the possibility of largely-modified hierarchies in fermion-Higgs couplings with respect to the SM, which might have also spoiled the hierarchies between the different Higgs boson production mechanisms. In order to further simplify the parameter space, we set $c_{\tau} = c_b$ in one of the scenarios considered in this paper.\footnote{These coefficients modify the process studied here in a rather similar way.} 
Beyond that, we have normalised the operator coefficients $c_{g,\gamma,HB,HW}$ by a loop suppression factor: in
any perturbatively-decoupling renormalizable extension of the SM, these operators can only
be generated at the loop level (see ~\cite{Arzt:1994gp, Einhorn:2013kja, Jenkins:2013fya}). Moreover, this is also convenient when comparing the $D=6$ theory with the top quark EFT, that is, the limit where the top quark is integrated out from the SM, as we will discuss in section~\ref{sec:gghh}.

After EWSB, several operators might contribute to the same interaction and field redefinitions need to be introduced in order to obtain canonically normalised kinetic terms. We examine the terms relevant to Higgs physics in the next section.

\section{Gluon fusion-initiated multi-Higgs production}\label{sec:gghh}

\subsection{Relevant Lagrangian terms}

The dimension-6, CP-even, operators listed in the Lagrangian of Eq.~(\ref{eq:Lfinal}) that affect the production of multiple Higgs bosons via gluon fusion (at leading order) are:
\begin{equation}\label{eq:lagterms}
\begin{split}
\mathcal{L}_{h^n} = &- \mu^2 |H|^2 - \lambda |H|^4 -  \left( y_t \bar Q_L H^c t_R + y_b \bar Q_L H b_R + \text{h.c.} \right)
\\
&+ \frac{c_H}{2\Lambda^2}(\partial^\mu |H|^2)^2 - \frac{c_6}{\Lambda^2} \lambda |H|^6  + \frac{\alpha_s c_g}{4\pi\Lambda^2} |H|^2 G_{\mu\nu}^a G^{\mu\nu}_a
\\
&- \left( \frac{c_t}{\Lambda^2} y_t |H|^2 \bar Q_L H^c t_R	+ \frac{c_b}{\Lambda^2}y_b |H|^2 \bar Q_L H b_R + \text{h.c.} \right) , 
\end{split}
\end{equation}
where in the first line we have included the corresponding SM operators that will receive additional contributions in the effective theory.\footnote{Note that with a slight abuse of notation here, and in Eq.~(\ref{eq:Lfinal}), we use $\lambda$ for the quartic coupling in the SM, while in Eq.~(\ref{eq:hpotential}) we use it for a generic trilinear coupling. 
The former $\lambda$ will be expressed in terms of $v$ and $m_h$ after minimisation of the Higgs potential. See appendix~\ref{app:laghh} for details.}
We now write
\begin{equation}
  H = \exp \bigg( -i \frac{T \cdot \xi}{v} \bigg) \frac{1}{\sqrt{2}}
  \begin{pmatrix}
  0
  \\
  v + h
  \end{pmatrix} \, ,
\end{equation}  
where $T$ represents the 3 generators of $SU(2)_L$, $\xi$ represents the 3 Goldstone degrees of freedom that will be absorbed by the gauge bosons, and $h$ is the physical Higgs boson. To obtain canonical normalisation of the Higgs field, we choose to perform the field redefinition\footnote{This field redefinition~\cite{Plehn:2009nd} involves non-linear terms which remove momentum-dependent Higgs boson interactions that would be less straightforward to implement in a Monte Carlo event generator.}
\begin{equation}
h \rightarrow  \left( 1 - \frac{ c_H v^2 } {2\Lambda^2 } \right) h - \frac{ c_H v } { 2 \Lambda^2 } h^2 - \frac{c_H}{6 \Lambda^2} h^3 \,.
\end{equation}
We further redefine $c_i \to c_i \, \Lambda^2 / v^2$ to absorb the suppression factor into the $c_i$ coefficients. We thus obtain the following interactions in terms of the Higgs boson scalar $h$, relevant to Higgs boson pair production:
\begin{equation}\label{eq:Lgghh}
\begin{split}
\mathcal{L}_{hh} = &- \frac{ m_h^2 } { 2 v } \left( 1 - \frac{3}{2} c_H + c_6 \right) h^3 - \frac{ m_h^2 } { 8 v^2 }  \left( 1  - \frac{25}{3} c_H + 6 c_6 \right) h^4
\\
&+ \frac{\alpha_s c_g} {4 \pi } \left( \frac{h}{v} + \frac{h^2} {2v^2} \right) G_{\mu\nu}^a G^{\mu\nu}_a
\\
&- \left[ \frac{m_t}{v} \left( 1 - \frac{ c_H } { 2  } + c_t   \right) \bar{t}_L t_R h +  \frac{m_b}{v} \left( 1 - \frac{ c_H } { 2  } + c_b   \right) \bar{b}_L b_R h + \text{h.c.} \right]
\\
&- \left[ \frac{m_t}{v^2}\left( \frac{3 c_t}{2}  -  \frac{ c_H}  { 2 } \right) \bar{t}_L t_R h^2 +  \frac{m_b}{v^2}\left( \frac{3 c_b}{2}  -  \frac{c_H}  { 2 } \right) \bar{b}_L b_R h^2 + \text{h.c.} \right] ,
\end{split}
\end{equation}
where we have explicitly written down the contributing components of the $Q_L$ doublets. Naively, all the Wilson coefficients in Eq.~(\ref{eq:lagterms}) should be bounded from perturbativity arguments by $4 \pi$, and hence if we consider $\Lambda \gtrsim 900$~GeV this automatically implies $| c_i | \lesssim 1$ in Eq.~(\ref{eq:Lgghh}).
For details on the derivation of the terms in the Lagrangian of Eq.~(\ref{eq:Lgghh}), see appendix~\ref{app:laghh}.\footnote{The Feynman rules for the Lagrangian terms appearing in Eq.~(\ref{eq:Lgghh}) have been checked using the \texttt{Mathematica}~\cite{mathematica} package \texttt{FeynRules}~\cite{Christensen:2008py, Alloul:2013bka}.}\footnote{It is worth stressing here that we expand our couplings around the SM value: in the case of the top-Yukawa the current Higgs boson data possess also a non-SM solution with the \emph{wrong sign} Yukawa. As pointed out in Ref.~\cite{Farina:2012xp}, $gg \to hh$ could also help to lift the degeneracy, in which case the EFT expansion needs to be performed around the non-SM minima of interest. We do not pursue such an analysis here.}
In Eq.~(\ref{eq:Lgghh}) we have also given the quartic Higgs self-coupling for completeness. The trilinear and quartic couplings can be written as
\begin{align}\label{eq:selfcoup}
\lambda &= \frac{m_h^2}{2 v^2} \left( 1 + \Delta \right) \,, \nonumber 
\\  \tilde{\lambda} &=  \frac{m_h^2}{2 v^2} \left(1 + 6 \Delta + \frac{2}{3} c_H  \right)\, ,
\end{align}
where $\Delta = c_6 - 3 c_H / 2$. From the above, it can be seen that the SM relation of $\lambda = \tilde{\lambda}$ is broken by the EFT effects: an accurate measurement of both couplings is thus a powerful probe of new physics in the Higgs sector, although, as already mentioned, measurement of the quartic coupling does not seem to be possible in the foreseeable future.

\subsection{From SM EFT to dimension-6 EFT} 
It is useful to compare and contrast the dimension-6 extension of the SM with the EFT that results from taking the top mass to infinity within the SM framework. This will help us in writing down the cross section formula for $gg \rightarrow hh$ in the $D=6$ EFT. 

\begin{figure}[t!]
\centering
\begin{subfigure}[t]{0.47\textwidth} 
  \centering
  \includegraphics[width=\textwidth]{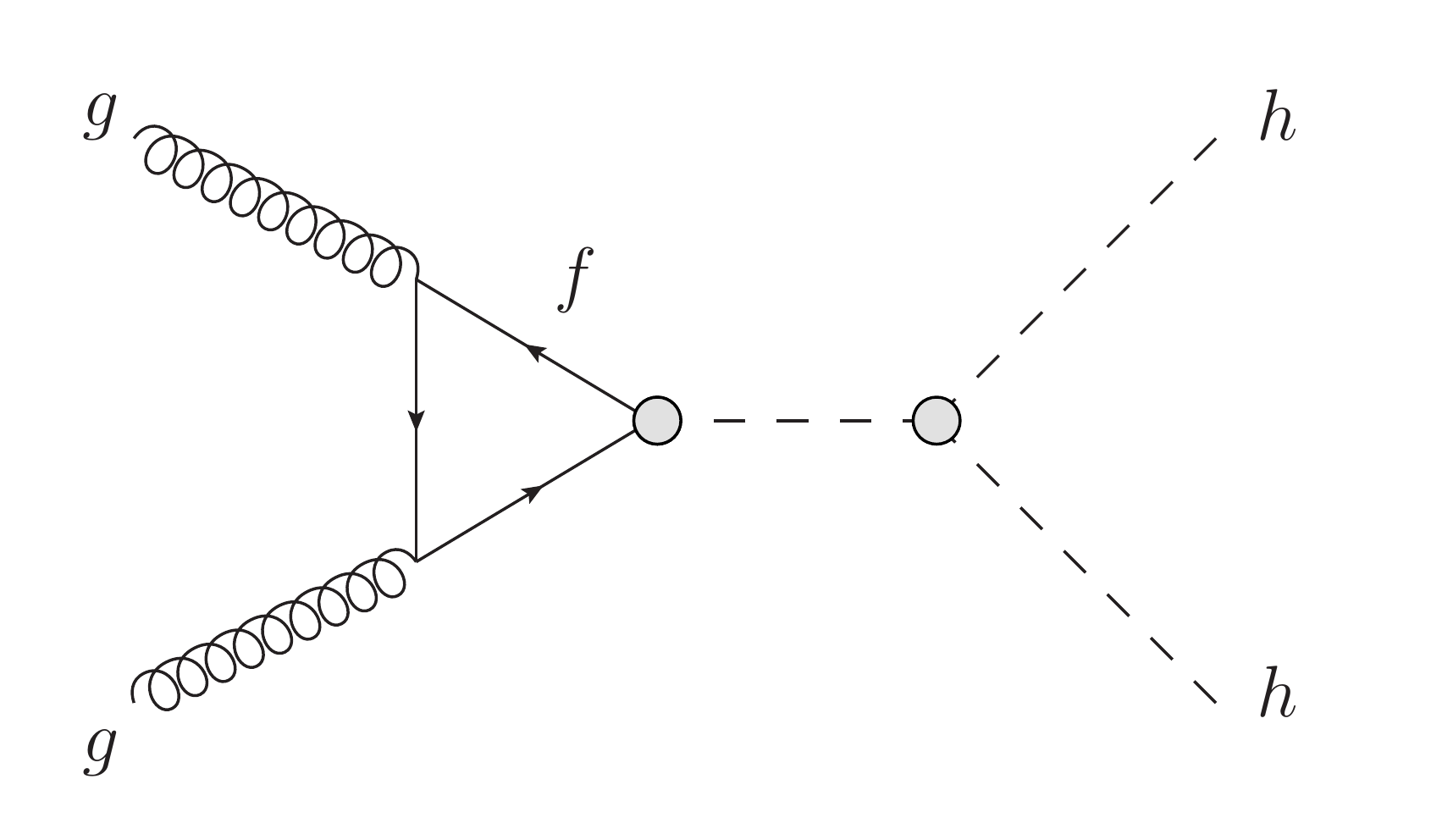}
  \caption{}\label{fig:hh1}
\end{subfigure}
\begin{subfigure}[t]{0.47\textwidth} 
  \centering
  \includegraphics[width=\textwidth]{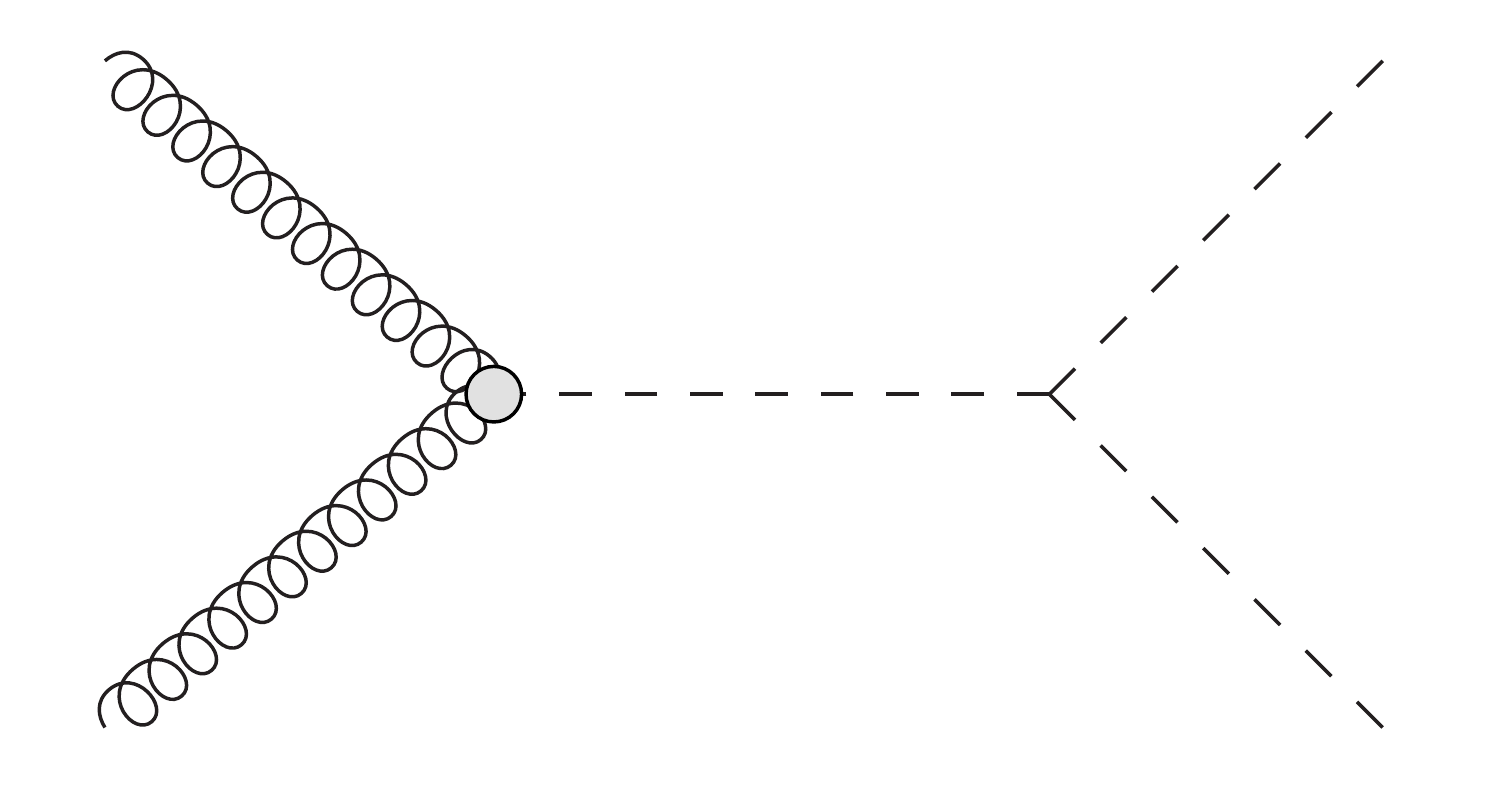}
  \caption{}\label{fig:hh2}
\end{subfigure}
\begin{subfigure}[t]{0.47\textwidth} 
  \centering
  \includegraphics[width=\textwidth]{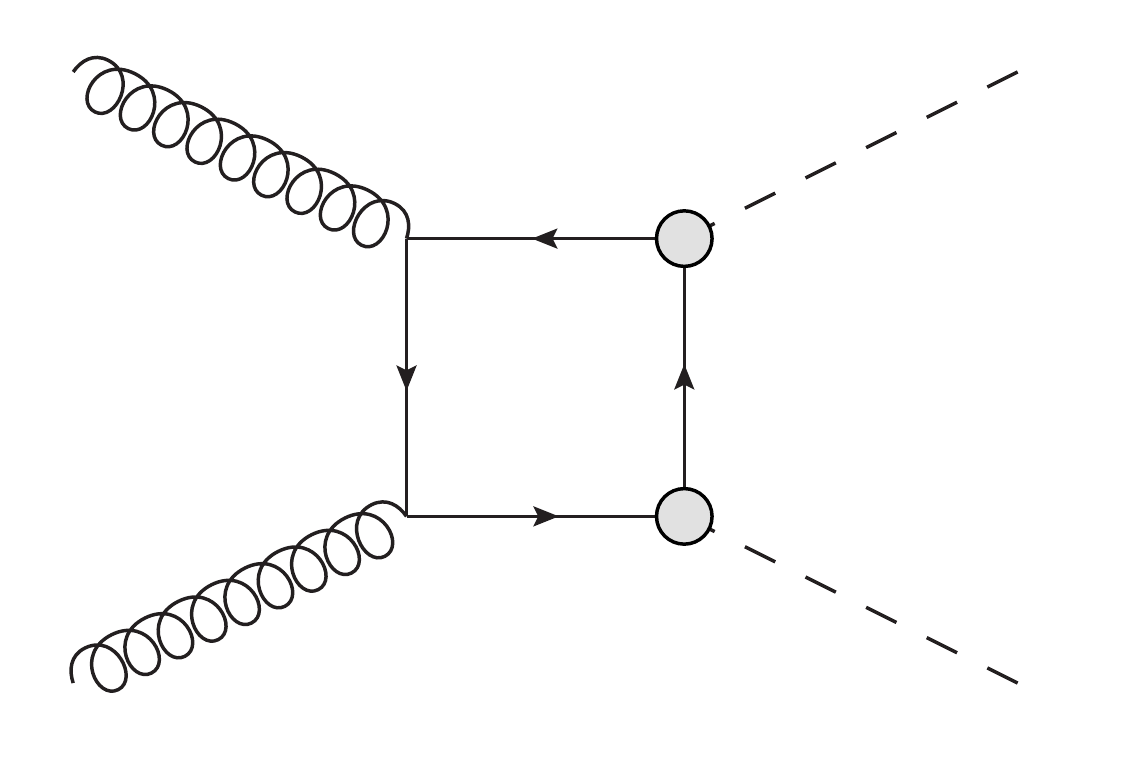}
  \caption{}\label{fig:hh3}
\end{subfigure}
\begin{subfigure}[t]{0.47\textwidth} 
  \centering
  \includegraphics[width=\textwidth]{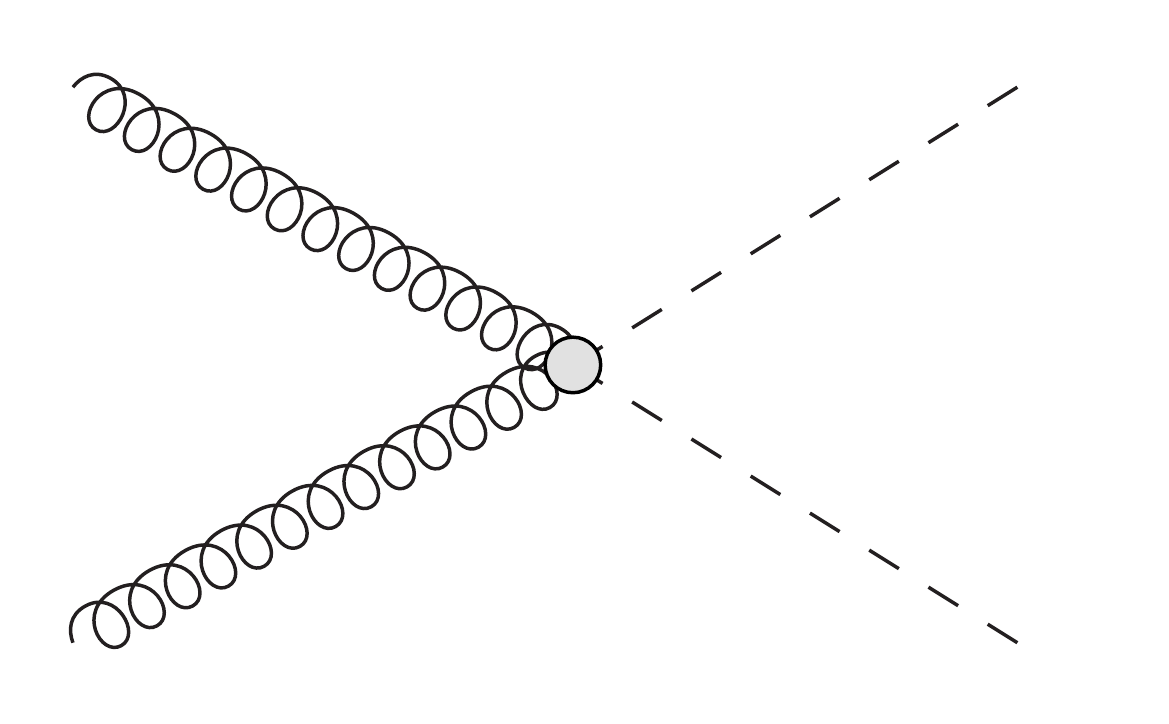}
  \caption{}\label{fig:hh4}
\end{subfigure}
\begin{subfigure}[t]{0.47\textwidth}
  \centering
  \includegraphics[width=\textwidth]{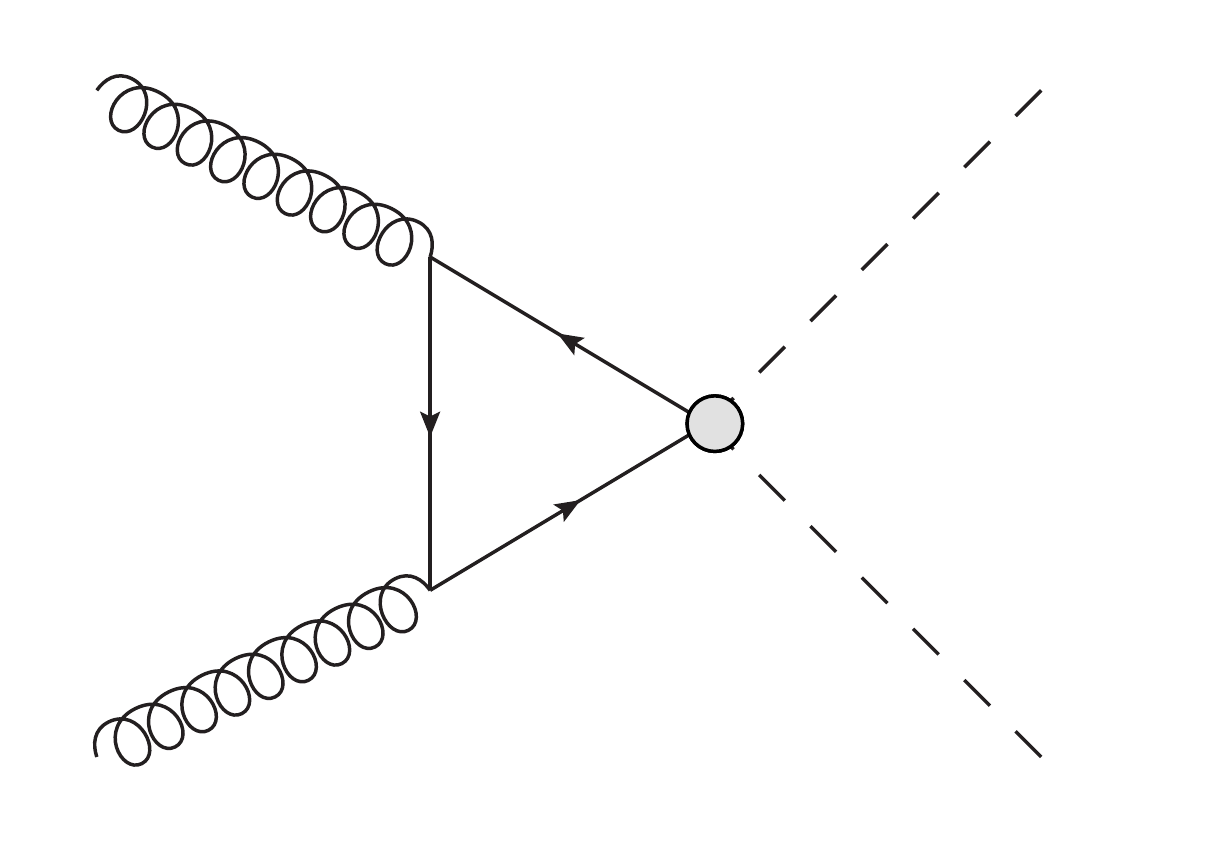}
  \caption{}\label{fig:hh5}
\end{subfigure}
\caption{The Feynman diagrams contributing to $gg \rightarrow hh$, including those induced by higher-dimensional operators. The grey blobs indicate the points of insertion of $D=6$ EFT vertices. At the order that we are considering in the present article, no two EFT insertions can occur in a single diagram. Diagrams with only one grey blob only appear in the effective theory.}
\label{fig:alldiags}
\end{figure}

There are several modifications necessary to incorporate the effect of the $D=6$ EFT operators in Higgs boson pair production via gluon fusion (see Fig.~\ref{fig:alldiags}):
\begin{itemize}
\item The Higgs boson self-coupling will be modified according to the first line in Eq.~(\ref{eq:Lgghh}), represented by modifications of the $h^3$ vertex in diagram~\ref{fig:hh1}.
\item The top and bottom quark Yukawa couplings will be modified according to the third line in Eq.~(\ref{eq:Lgghh}). These modifications appear in diagrams~\ref{fig:hh1} and \ref{fig:hh3}. 
\item The new four-point fermionic interactions $\bar{f}f h^2$ will introduce a new `triangle' diagram, with two Higgs bosons produced at the apex of the triangle. These are described by the fourth line in Eq.~(\ref{eq:Lgghh}) and appear in diagram~\ref{fig:hh5}.
\item Two new effective theory ``tree-level'' diagrams will contribute since we now have additional gluon-gluon-Higgs (diagram~\ref{fig:hh2}) and gluon-gluon-Higgs-Higgs interactions (diagram~\ref{fig:hh4}), according to the second line in Eq.~(\ref{eq:Lgghh}).
\end{itemize}
These are relatively straightforward to incorporate into a Monte Carlo event generator. The shift in the Higgs boson couplings to fermions and to gauge bosons will also lead to changes in its branching ratios. We discuss those changes in section~\ref{sec:decays}.

Our goal is to modify the Standard Model matrix element with finite top mass to readily incorporate the modifications coming from the EFT. We first note that in the heavy top mass limit, the `low energy theorem' states that the interactions of one or two Higgs bosons with two gluons are given by:
\begin{equation}\label{eq:smeft}
\mathcal{L}_\mathrm{SM,EFT} =   (G_{\mu\nu}^a G^{\mu\nu}_a) \frac{\alpha_s}{\pi} \left( \frac{h}{12 v} - \frac{h^2}{24 v^2} \right) .
\end{equation}
It is interesting to point out that the $gghh$ interaction that remains in the heavy top mass approximation corresponds to a spin-0 di-gluon state in $gg\rightarrow hh$, while the spin-2 contributions vanish in this limit. In the effective theory language, the spin-2 contributions correspond to operators of higher dimensionality. The differential partonic cross section for gluon fusion-initiated Higgs boson pair production in the SM, with the full top mass dependence, is given by~\cite{Plehn:1996wb}:
\begin{equation}\label{eq:diffxs}
\frac{ \mathrm{d} \hat{\sigma} (gg \rightarrow hh) } { \mathrm{d} \hat{t} } = \frac{ G_F^2 \alpha_s^2 } { 256 (2\pi)^3 } \left[ | C_\triangle F_\triangle + C_\Box F_\Box |^2 + | C_\Box G_\Box |^2 \right]\;, 
\end{equation}
where, in the SM,
\begin{equation}
C_\triangle = \frac{3 m_h^2 } { \hat{s} - m_h^2 } \,, \quad C_\Box = 1 \,.
\end{equation}
$F_\triangle$, $F_\Box$ and $G_\Box$  are form factors, given e.g. in Ref.~\cite{Plehn:1996wb}, in which $G_\Box$ corresponds to a spin-2 contribution, while $\hat{s}$ and $\hat{t}$ are the usual Mandelstam invariants. In the limit of large quark mass in the loop, $m_Q \gg \hat{s} $, the form factors reduce to:
\begin{align}
F_\triangle &= \frac{2}{3} + \mathcal{O} ( \hat{s} / m_Q^2 )\,,\nonumber \\
F_\Box &= - \frac{2}{3} + \mathcal{O} ( \hat{s} / m_Q^2 ) \,,\nonumber \\
G_\Box &= \mathcal{O} ( \hat{s} / m_Q^2 )\,.
\end{align}
Note that the function $G_\Box$ is sub-dominant in this limit, in correspondence with the fact that the spin-2 terms are absent in Eq.~(\ref{eq:smeft}).

We now derive, starting from Eq.~(\ref{eq:diffxs}), the cross section for the $hh$ process in the $D=6$ EFT. The complete set of diagrams is shown in Fig.~\ref{fig:alldiags}. Using the above limiting values of the form factors, one can re-write the Lagrangian of Eq.~(\ref{eq:smeft}) as:
\begin{equation}\label{eq:smeft2}
\mathcal{L}_\mathrm{SM,EFT} =   (G_{\mu\nu}^a G^{\mu\nu}_a) \frac{\alpha_s}{8\pi} \left( \frac{h}{v}F_\triangle^\mathrm{hq}   + \frac{h^2}{2 v^2} F_\Box^\mathrm{hq} \right),
\end{equation}
where $F_\triangle^\mathrm{hq}  = - F_\Box^\mathrm{hq} = 2/3$ are the values of the form factors in the heavy quark effective theory:
\begin{equation}
F_\star^\mathrm{hq} = \lim_{m_t \rightarrow \infty } F_\star \,, \quad \star =  \{ \Box ,\triangle \}\,.
\end{equation}
We begin by considering the correspondence between diagrams \ref{fig:hh1} and \ref{fig:hh2} as well as between \ref{fig:hh3} and \ref{fig:hh4}. This corresponds to comparing equivalent terms in Eq.~(\ref{eq:smeft2}) and Eq.~(\ref{eq:Lgghh}). We can immediately conclude that the following identifications can be made at the amplitude level to obtain the contributions of the pure EFT diagrams:
\begin{align}
\frac{\alpha_s}{8\pi v} F_\triangle^\mathrm{hq} &\rightarrow  \frac{ \alpha_s  } { 4\pi v } c_g \,, \nonumber
\\
\frac{\alpha_s}{16 \pi v^2} F_\Box^\mathrm{hq} &\rightarrow \frac{ \alpha_s } { 8 \pi v^2 } c_g \,.
\end{align}
To obtain diagram~\ref{fig:hh5}, one needs to essentially `remove' the propagator from a diagram of type~\ref{fig:hh1}, while keeping the dependence on the quark mass in the triangle loop via the full form factor $F_\triangle$. This can be done by multiplying the factor
\begin{equation}
\left[ \frac{m_f}{v} \times \frac{3 m_h^2}{v} \times \frac{1}{\hat{s} - m_h^2} \right]^{-1} \times 2g_{hhff} \, ,
\end{equation}
which includes a combinatoric factor of 2, $f = t, b$, and
\begin{equation}
 g_{hhff} =  \frac{m_t}{v^2}\left( \frac{3c_f}{2}  -  \frac{c_H}  { 2 } \right) .
\end{equation}
The necessary additional modifications correspond to trivial replacements of the triple-Higgs coupling and the Yukawa coupling in the SM-like diagrams, according to Eq.~(\ref{eq:Lgghh}).
After all these modifications, we now arrive at the parton-level differential cross section for the process $gg \rightarrow hh$ in the $D=6$ EFT:
\begin{align}\label{eq:diffxsEFT}
\left. \frac{ \mathrm{d} \hat{\sigma} (gg \rightarrow hh) } { \mathrm{d} \hat{t} } \right|_\mathrm{EFT} &= \frac{ G_F^2 \alpha_s^2 } { 256 (2\pi)^3 } \bigg\{ \Big| C_\triangle F_\triangle ( 1 - 2 c_H + c_t + c_6 ) + 3 F_\triangle (3 c_t - c_H) + 2 c_g C_\triangle \nonumber
\\
&\hspace{6em} + C_\Box F_\Box (1 - c_H + 2 c_t )  + 2 c_g C_\Box  \Big|^2 + \Big| C_\Box G_\Box \Big|^2 \bigg\} \,,
\end{align}
where we have suppressed the bottom quark contributions for simplicity. 

\section{Higgs boson decays in dimension-6 EFT}\label{sec:decays}

We now move forward to study the impact of the operators in Eq.~(\ref{eq:Lfinal}) on the decays of the Higgs bosons.
In Table~\ref{tab:dec} we provide an overview on which coefficients enter the various decays at tree-level topology (second column), at the one-loop level, considering only QCD corrections to the insertions (third column), as well as at the full one-loop level (fourth column). Here, we focus on the key decays $h \to bb$, $h \to \tau \tau$, $h \to WW$, and $h \to \gamma \gamma$, that are in particular important for the analysis of Higgs boson pair production. In our numerical study we will, however, consider all significant decays in the EFT. We note that even though operators may not enter a given $ pp \rightarrow hh \rightarrow (xx)(yy)$ final state, they will still be relevant since they change the overall branching ratios. We also include in Table~\ref{tab:dec} the coefficients entering $gg \to h$ and $gg \to hh$ production for completeness. It is interesting to observe that several operators may enter both production and decay, introducing non-trivial correlations for the behaviour of the cross section of a given final state mediated through Higgs boson pair production.
Finally, note that the presence of the coupling $c_\gamma$ can lift the $h\to \gamma \gamma$ decay from one-loop in the SM formally to tree-level in the EFT. The same is true for $c_g$ concerning $gg \to h $ and $gg \to h h$. On the other hand, in any perturbatively-decoupling renormalizable extension of the SM the operators $c_\gamma$, $c_g$, $c_{HW}$ and $c_{HB}$ can only be generated at the loop level. Thus, we will in particular not insert them into loop diagrams at the order considered~\cite{Einhorn:2013kja}.

In the present article we employ the \texttt{eHDECAY} code~\cite{Contino:2014aaa} to calculate the branching ratios of the Higgs boson according to our EFT formalism.  The program \texttt{eHDECAY} provides the SM plus $D=6$ EFT contributions, including QCD radiative corrections. Next-to-leading order EW corrections are only applied to the SM contributions. For further details, see Ref.~\cite{Contino:2014aaa}.


\begin{table}[h]
\centering
\begin{tabularx}{\linewidth}{XXXX}
\hline
\toprule
Mode & tree & 1 loop QCD & 1 loop\\
 \midrule

$h \to bb$ & $\bm{c_H}$, $\bm{c_b}$ &  $\bm{c_H}$, $\bm{c_b}$ & $ c_H,  c_b, c_t, c_6,  c_W
$\\
$h \to \tau \tau$  &  $\bm{c_H}$, $\bm{c_\tau}$ & - &  $ c_H, c_\tau, c_6,  c_W$\\\hdashline
$h \to \gamma \gamma$  & $\bm{c_\gamma} 
$ & - &  $c_H,  c_b, c_t, c_\tau,  c_W$ \\
$h \to WW$  &  $\bm{c_H}$, $\bm{c_{HW}}$, $\bm{c_W}$ & - & $ c_H,   c_W, c_b, c_t, c_\tau, c_6$\\
\midrule
$ g g \to h h $ &  $\bm{c_g}$ & $\bm{c_t}$, $\bm{c_b}$ & $\bm{c_t}$, $\bm{c_b}$, $\bm{c_H}$, $\bm{c_6}$\\
$ g g \to h $ & $\bm{c_g}$ & $\bm{c_t}$, $\bm{c_b}$, $\bm{c_H}$ & $\bm{c_t}$, $\bm{c_b}$, $\bm{c_H}$\\
\bottomrule
\end{tabularx}
\caption{\label{tab:dec} 
Operators that modify the decay modes of the Higgs boson relevant to our analysis, with tree-level topology (second column), at the one-loop level, considering only QCD corrections (third column), as well as at the full one-loop level (fourth column).  Note that we neglect one-loop insertions of one-loop operators. 
For completeness, we also include the operators entering $gg \to h$ and $gg \to h h$. The operators that are highlighted in bold text are included in the treatment of the present paper, in the corresponding topology. The dashed line separates decay modes that enter our analysis only indirectly in modifications of the Higgs boson branching ratios as well as via the correlations present in the constraints from single Higgs boson physics.}
\end{table}

At this point it is worth discussing the loop-suppressed operators $c_{g,\gamma}$ and $c_{HB,HW}$. While the former two induce corrections on Higgs boson decays that appear in the SM at the one-loop level, the latter affect the $h \to WW, ZZ$ decays, which are tree level in the SM.
Thus, the naive expectation is that the impact of $c_{HW,HB}$ will be essentially negligible on Higgs boson physics. Moreover, due to Eq.~(\ref{eq:res}), the loop-suppression of these coefficients feeds through to $c_{B}, c_{W}$ via a single free parameter. We have explicitly verified that an ${\cal O}(1)$ change in the $c_{HW, HB, W, B}$ coefficients leads to at most a 4 \% variation in the loosely-constrained $h \to Z \gamma$ decay rate and sub-percent level variations in $h \to WW, ZZ$.  The same exercise with $c_{\gamma}$ gives instead a ${\cal O}(70\%)$ variation in the $h \to \gamma \gamma$ branching ratio, which is phenomenologically relevant for the single Higgs boson constraints that we will employ. We thus neglect the effect of the operators $\mathcal{O}_{W}$, $\mathcal{O}_{B}$, $\mathcal{O}_{HB}$ and $\mathcal{O}_{HW}$ in the rest of this paper, but consider variations of $c_\gamma$.


\section{Constraining dimension-6 EFT coefficients}
\label{sec:An}
\subsection{Monte Carlo}
We have implemented the Lagrangian terms of Eq.~(\ref{eq:Lgghh}) 
 in a \texttt{HERWIG++} model where the SM matrix elements have been taken from the code \texttt{HPAIR}~\cite{Dawson:1998py, Plehn:2005nk, hpair}. The Monte Carlo event generator is essential to our analysis. This is because a calculation of the total cross section alone cannot account for the change in distributions of momenta and angles that will substantially change the efficiency of the experimental analysis. Thus, even if we do not use their shapes explicitly in extracting constraints in this article, it is essential to have a reliable description of the underlying distributions. One could employ the Monte Carlo-generated distributions to further improve the bounds~\cite{futurepub}, keeping in mind the fact that operators will mix due to renormalization group running. See for example, Ref.~\cite{Englert:2014cva}.   


The calculation of our study is accurate to LO in QCD, within the framework of the $D=6$ EFT, including the diagrams of Fig.~\ref{fig:alldiags}. Full QCD corrections to the diagrams that include top quark loops have not yet been calculated with the dependence on the top quark mass. These are available only in the heavy top mass limit (also known as the `low-energy theorem'), which provides an estimate of their magnitude. The QCD corrections to the additional new processes that arise in $D=6$ EFT consist of Feynman diagrams that are of identical topology to those of SM diagrams within the `low-energy theorem'. We thus expect the size of the QCD corrections to be similar in all sub-diagrams contributing to the process, including those that only appear in the $D=6$ EFT. For the sake of simplicity, we apply a flat overall $K$-factor of $K=2$, which would correspond to normalizing our result to the state-of-the-art (NNLO QCD) SM calculation of $\sim 40$~fb \cite{deFlorian:2013jea}. The choice is conservative, and justified at present by the fact that the low-energy theorem in the SM is estimated to possess $\mathcal{O}(10\%)$ uncertainty~\cite{Grigo:2013rya, Maltoni:2014eza}. This uncertainty is subsumed in the 30\% total theoretical uncertainty that we will assume here (see Section~\ref{sec:analysis}).\footnote{Moreover, the NNLO calculation of Ref.~\cite{deFlorian:2013jea} is not available at present to use for the given parameters that we employ here (i.e. PDF set and scale choices).}

\begin{figure}[h!]
  \includegraphics[width=0.48\linewidth]{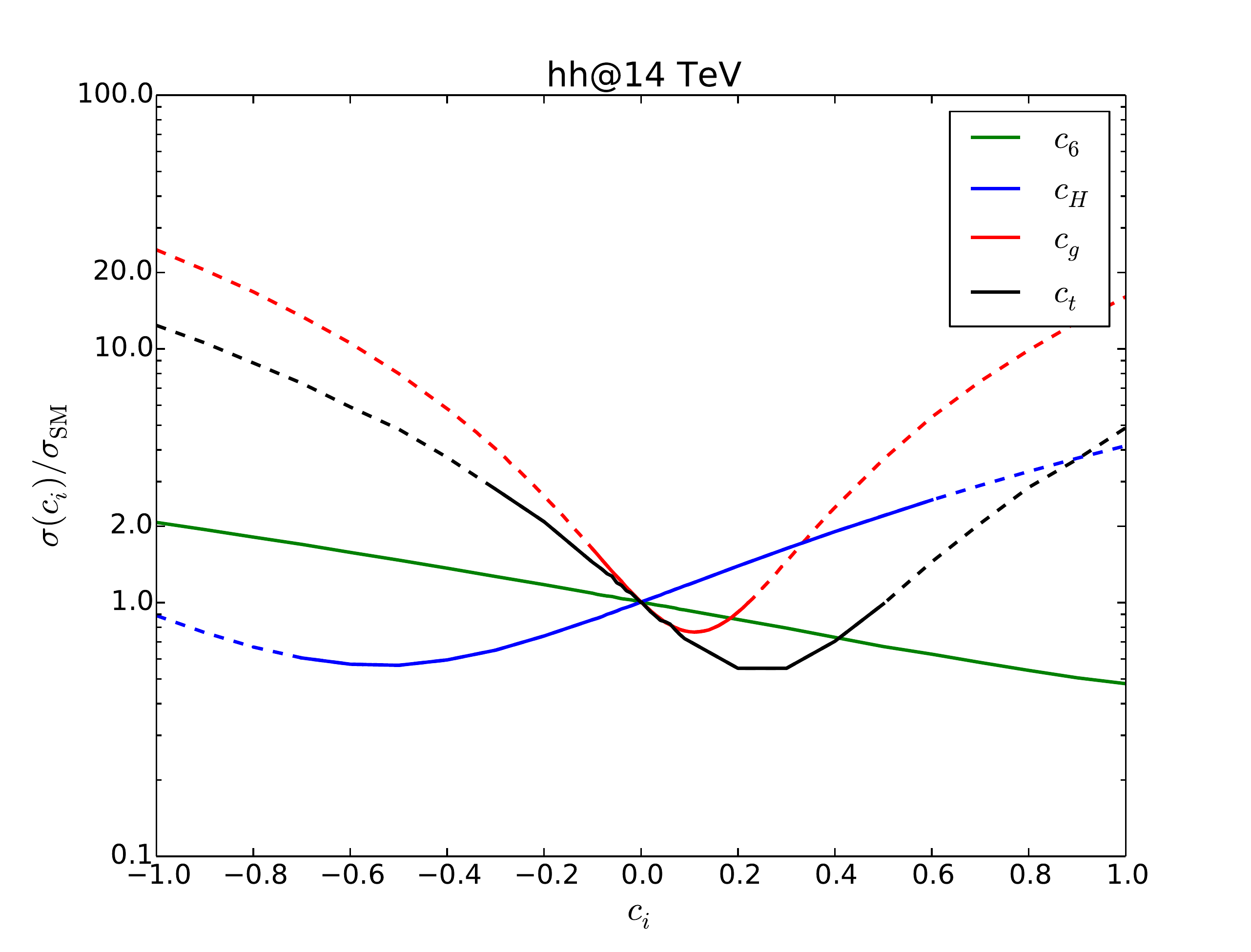}
  \includegraphics[width=0.465\linewidth]{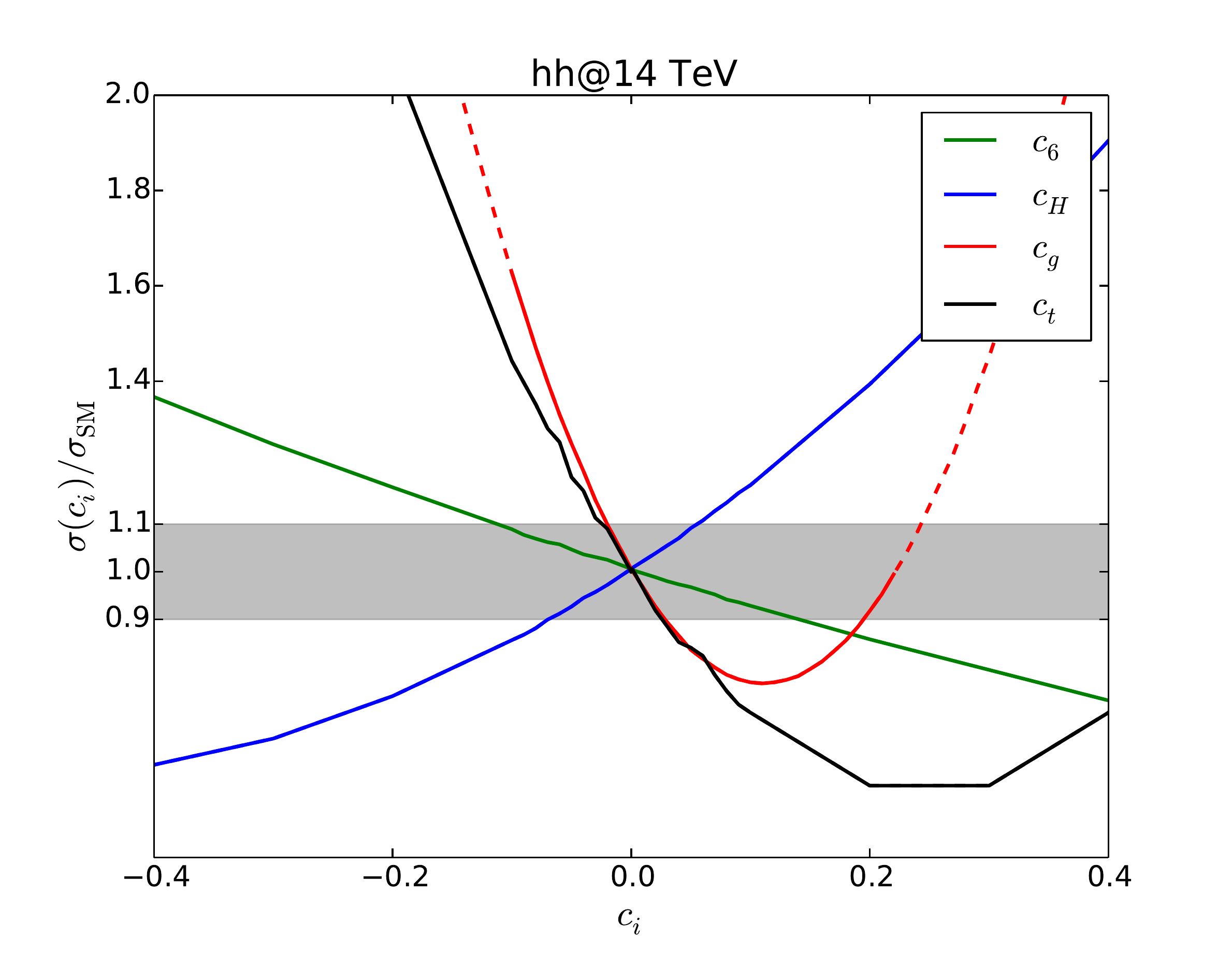}
  \caption{The effect of the variation of individual operators on the total cross section divided by the SM value. In the right panel we focus on a narrower region, showing in the grey-shaded area the $\pm 10\%$ variation with respect to the SM value. The solid portions of the curves represent the region which is compatible at 95\% C.L. or more with the current Higgs boson data, obtained using \texttt{HiggsBounds} and \texttt{HiggsSignals} (see section~\ref{sec:full} for details).
}
  \label{fig:opeffect}
\end{figure}

We start by investigating the individual effects on the gluon fusion production cross section, varying one coefficient at a time, while setting all others to zero. The result is presented in Fig.~\ref{fig:opeffect}, for the LHC running at 14~TeV proton-proton centre-of-mass energy, where we have zoomed-in in the right panel, and shaded the $\pm 10\%$ variation region from the SM value of the cross section. Here and in the remaining article, we employ the MSTW2008nlo\_nf4 PDF sets~\cite{Martin:2009iq}.\footnote{The cross sections have been verified through an independent implementation directly in \texttt{HPAIR}.} One can clearly see how deviations from the SM prediction $c_i=0$ could lead to substantial changes in the total cross section. Unfortunately, the dependence on $c_6$ is rather mild, whereas the dependence on $c_t$ and $c_g$ is substantially more pronounced. This tendency will be amplified when realistic analysis cuts are considered (see below). The fact that positive values of $c_6$ lead to a decreased cross section reflects the negative interference between the triangle and box contributions.
\subsection{Analysis}\label{sec:analysis}
To accommodate a direct comparison with existing phenomenological analyses, we focus on the process $hh \rightarrow (b\bar{b})(\tau^+ \tau^-)$ at the 14 TeV LHC. The specific final state possesses a relatively large branching ratio and manageable backgrounds. This channel has been examined in detail within the SM in Refs.~\cite{Dolan:2012rv, Baglio:2012np, Barr:2013tda, Maierhofer:2013sha} and turned out to be particularly promising. We consider here only the main irreducible backgrounds, arising from $t\bar{t}$ production with subsequent decays of the $W$ bosons to $\tau$ leptons, as well as $ZZ$ and $hZ$ production with $(b\bar{b})(\tau^+ \tau^-)$ final states, which is sufficient given the other sources of uncertainty.\footnote{We have also considered the effect of $D=6$ operators in $hZ$ production and the subsequent Higgs boson decay. These were found to have negligible impact on our analysis and we do not discuss them in detail.} The backgrounds were generated at next-to-leading order in QCD, using the \texttt{aMC@NLO} event generator~\cite{Frixione:2010ra, Frederix:2011zi, Alwall:2014hca}. The total cross section for $t\bar{t}$ was normalised to $\sigma_{t\bar{t}} = 900$~pb~\cite{Ahrens:2011px, Czakon:2013goa} and the $ZZ$ and $hZ$ NLO cross sections were taken out of the \texttt{aMC@NLO} calculation directly: $\sigma_{ZZ} = 15.25$~pb and $\sigma_{hZ} = 0.8329$~pb. For realistic description of the final states, parton showering and hadronization were performed using \texttt{HERWIG++}, and the simulation of the underlying event was included via multiple secondary parton interactions~\cite{Bahr:2008dy}. 

We follow the basic analysis steps as given originally in~\cite{Dolan:2012rv} and as were described in~\cite{Maierhofer:2013sha}. Here, we assume 70\% $\tau$-reconstruction efficiency with negligible fake rate 
\footnote{Thus, we do not consider any mistagging backgrounds, which should be considered in a full experimental study. These are expected to be sub-dominant, as we require 2 b-tags and 2-tau tags in our analysis.}
and require two $\tau$-tagged jets with at least $p_\perp > 20$~GeV. We require that the di-tau invariant mass, taken from the Monte Carlo truth, reproduces the Higgs mass within a $\pm25$~GeV window, to account for the reconstruction smearing, as done in~\cite{Dolan:2012rv}. To model this effect, we smear the true di-tau invariant mass by a 20~GeV Gaussian, which in turn allows for a contamination from events containing $Z\rightarrow \tau^+ \tau^-$ into the di-tau mass window that we consider. We use the Cambridge-Aachen jet algorithm available in the \texttt{FastJet} package~\cite{Cacciari:2011ma, Cacciari:2005hq} with a radius parameter $R=1.4$ to search for so-called `fat jets'. We require the existence of one fat jet in the event satisfying the mass-drop criteria as done in the $hV$ study in Ref.~\cite{Butterworth:2008iy}. We require the two hardest `filtered' sub-jets to be b-tagged 
\footnote{Bottom-jet tagging was performed by setting the bottom mesons to stable in the \texttt{HERWIG++} event generator.}
and to be central ($|\eta| < 2.5$) and the filtered fat jet to be in $(m_h-25~\mathrm{GeV}, m_h+25~\mathrm{GeV})$, which will reduce events with $Z \to b \bar{b}$ decays. The b-tagging efficiency was taken to be 70\%, again with negligible fake rate for the sake of simplicity. We require a loose cut on the transverse momentum of the fat jet (after filtering) that satisfies the above criteria, $p_\perp^{\mathrm{fat}} > 100$~GeV and also apply a transverse momentum cut on the $\tau^+\tau^-$ system of equal magnitude, $p_\perp^{\tau\tau} > 100$~GeV. As in \cite{Maierhofer:2013sha}, we apply additional cuts: $\Delta R (h,h) > 2.8$ and $p^{hh}_\perp< 80$~GeV to reject the background even further. 

\begin{figure}[!htb]
  \centering
    \includegraphics[width=0.485\linewidth]{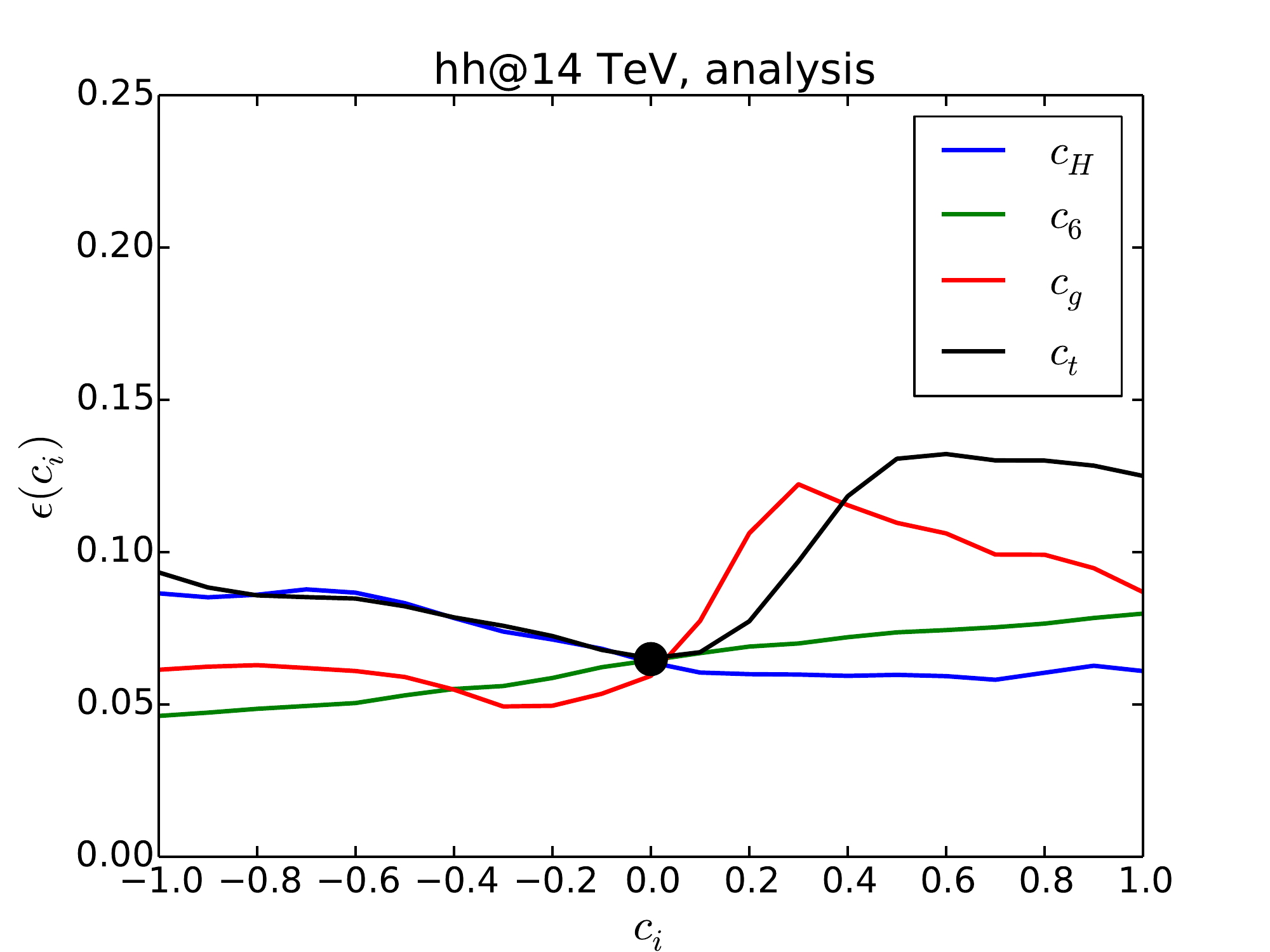}
    \includegraphics[width=0.48\linewidth]{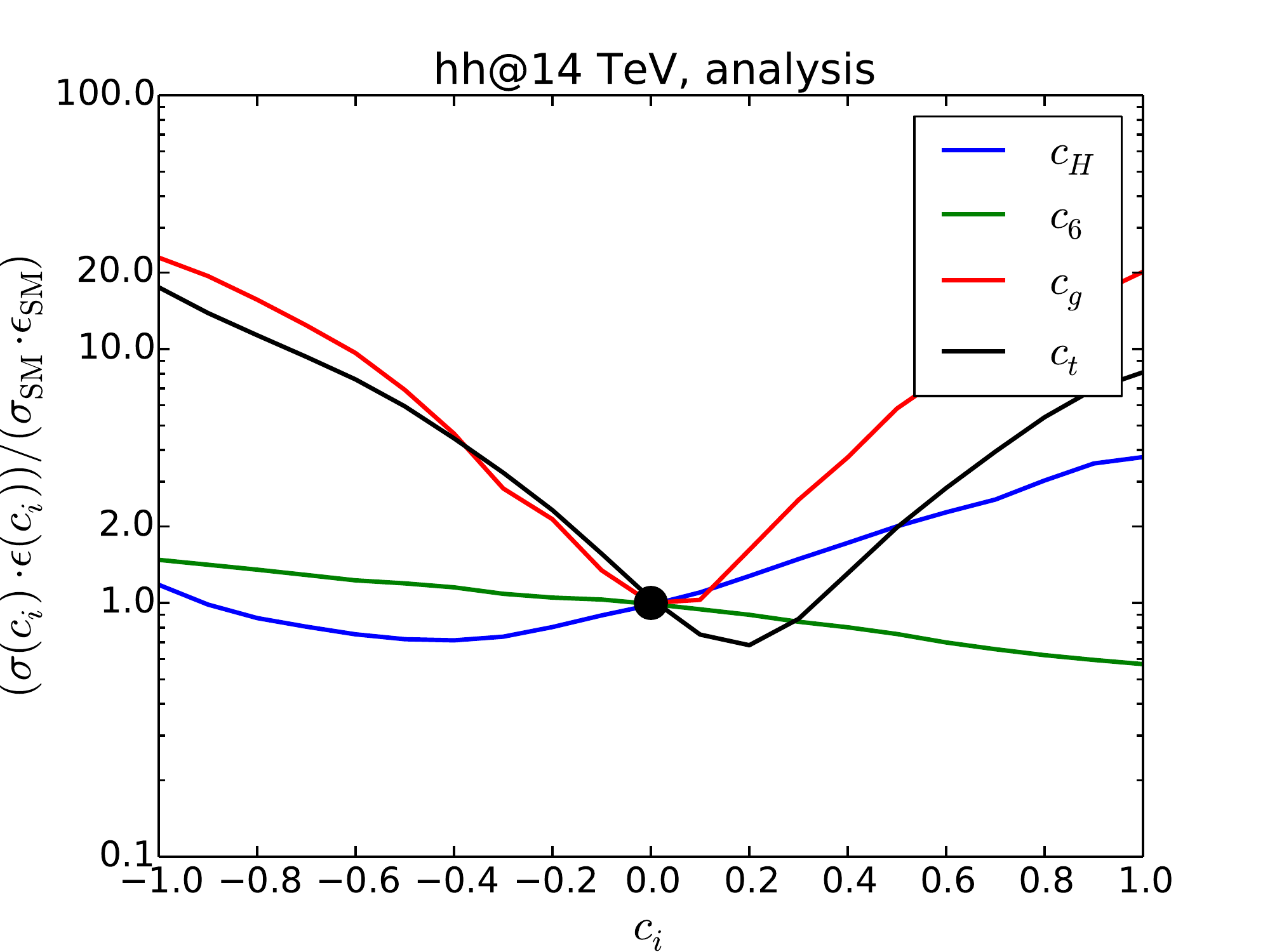}
  \caption{The efficiency of the analysis is shown on the left panel. The right panel shows the efficiency times the cross section of the EFT point, divided by the SM cross section times the SM point efficiency.}
  \label{fig:efficiency}
\end{figure}

We investigate the effect of the above analysis on events corresponding to different values of the parameters.\footnote{One could optimize the analysis for each point (or region) in the parameter space. We leave that task for future work~\cite{futurepub}.} On the left panel of Fig.~\ref{fig:efficiency} we plot the efficiency of the analysis in the case where we set all other coefficients to zero except the labeled one. On the right panel we show the efficiency times the cross section of the EFT point, divided by the SM cross section ($\sigma_\mathrm{LO} = 22.3$~fb) times the SM point efficiency ($\epsilon \simeq 0.065$). This plot can be compared with the actual cross section plot of Fig.~\ref{fig:opeffect}, where one can observe that the qualitative behaviour of the resulting cross sections does not change, but there is a quantitative change due to the non-uniform effect of the analysis. Note that we have not included the actual branching ratios in Fig.~\ref{fig:efficiency}, which will have a further effect in determining the significance of a given parameter point. 
\subsection{Results} 
\subsubsection{$c_6$-only model}

For simplicity, we begin by considering a model where only $c_6$ is non-zero and allowed to vary. In fact, this is the only coefficient that remains unconstrained from data on single Higgs boson production. Setting \textit{all} other $c_{i}$'s to zero and varying $c_6$ corresponds to modifying the size of the Higgs boson self-coupling $\lambda$, as was done in previous studies.\footnote{In particular, in \cite{Goertz:2013kp} we focused on such a scenario. Moreover, we studied the dependence on the top Yukawa coupling, which (beyond testing consistency with the SM) allows to examine an independent variation of the coefficient of the $\bar t_L t_R h$ operator in Eq.~(\ref{eq:Lgghh}), which is possible in extensions of the SM where the Higgs boson is not part of a $SU(2)_L$ doublet.} The value of $c_6$ here represents a relative change in $\lambda$ with respect to the SM prediction. No modifications are expected at the order considered on the Higgs boson decays when including such an operator. We investigate the possible constraints on $c_6$ given the particular model.\footnote{We note here that negative values of the coefficient $c_6$ in this model, and in the rest of the paper, should be taken with a grain of salt due to possible effects on vacuum stability. Detailed study of the behaviour of the potential in this regime is left for future work.}

\begin{figure}[!htb]
 \centering
\begin{overpic}[width=0.45\textwidth,tics=10]{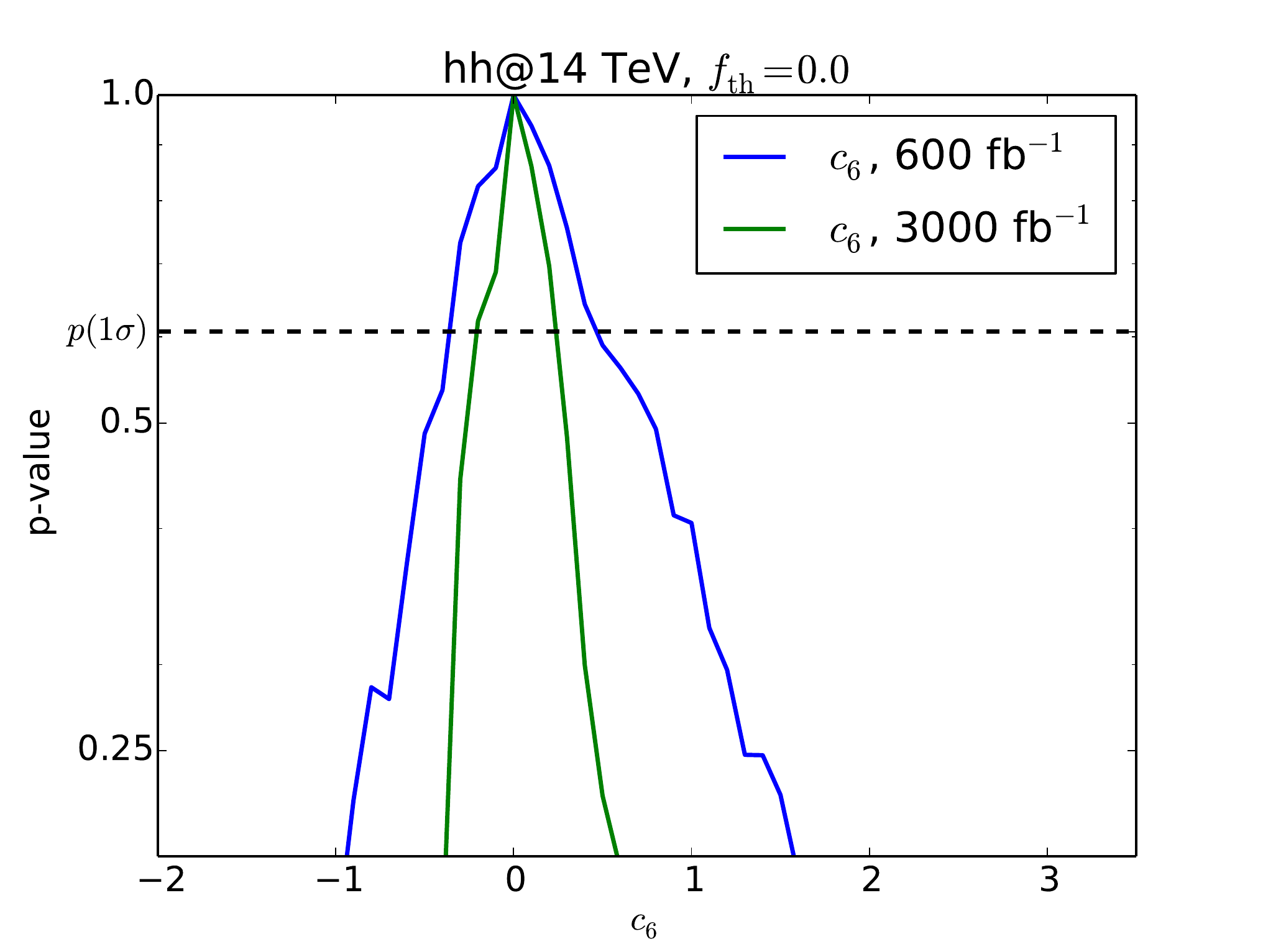}
 \put (65,30) {\large$c_6~\mathrm{-only}$}
\end{overpic}
\begin{overpic}[width=0.45\textwidth,tics=10]{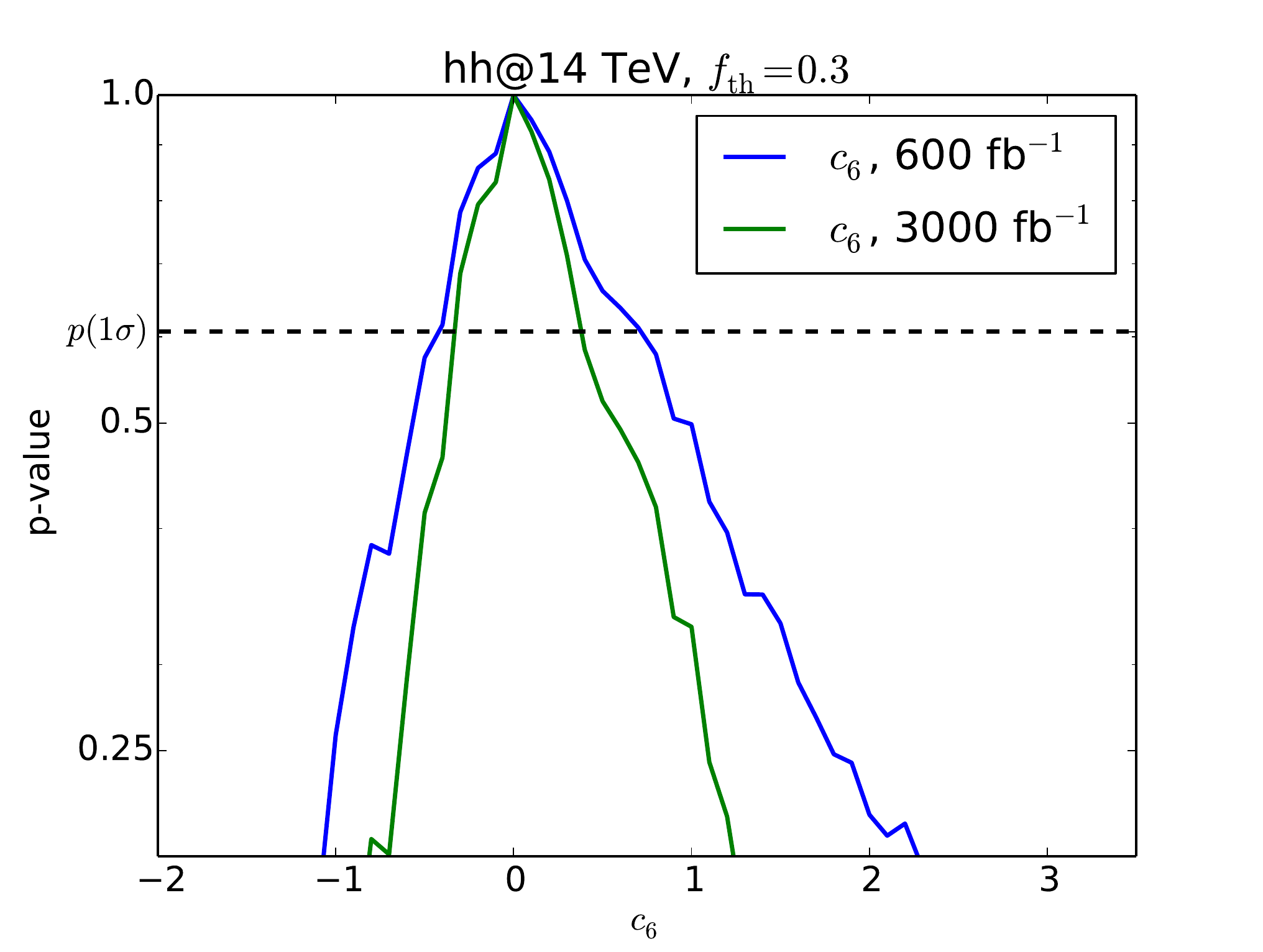}
 \put (65,30) {\large$c_6~\mathrm{-only}$}
\end{overpic}
  \caption{The $p$-value obtained for a given value of $c_6$, for the process $hh \rightarrow (b\bar{b}) (\tau^+ \tau^-)$ at 600~fb$^{-1}$ and 3000~fb$^{-1}$ of integrated luminosity. On the left figure we show the result without any theoretical uncertainty included ($ f_\mathrm{th} = 0$) and on the right figure with theoretical uncertainty on the signal cross section prediction of 30\% ($f_\mathrm{th} = 0.3$).}
  \label{fig:excc6}
\end{figure}

Let us assume that in our analysis we obtain $S(c_6)$ events for the signal, for a given value of $c_6$, at a given integrated luminosity. For the background, we obtain $B$ events at the same luminosity. Given that the number of events $S$ and $B$ is large enough, we may assume that they are Gaussian-distributed. The total statistical uncertainty on $N(c_6) = S(c_6) + B$ is then given by:
\begin{equation}
\delta N^2 = \delta B^2 + \delta S^2 \;.
\end{equation}
Therefore, if the relative theoretical uncertainty on the cross section prediction is $f_\mathrm{th}$, assuming negligible theoretical uncertainty on the background,\footnote{The assumption is reasonable since the background higher-order calculations exhibit relatively small variations compared to the $hh$ signal. Moreover, the background prediction can be normalised using a signal-free region.} an addition in quadrature leads to a total uncertainty of:
\begin{equation}
\delta N^2 = \delta B^2 + \delta S^2 + S^2 f_\mathrm{th}^2\;,
\end{equation}
and hence we have that:
\begin{equation}
\delta N^2 = N + S^2 f_\mathrm{th}^2\;.
\end{equation}
To obtain the expected constraints, we assume that the underlying theory is indeed the SM, which corresponds to $c_6 = 0$ in this scenario. In turn, the expected total number of events is $N(c_6=0)$. One then needs to compute how many standard deviations $\delta N(c_6)$ away a given $N(c_6)$, as predicted from theory, is from $N(c_6 = 0)$. This can be translated into a probability (i.e. a $p$-value) assuming a Gaussian distribution. The results are presented in Fig.~\ref{fig:excc6}, for integrated luminosities of 600~fb$^{-1}$ and 3000~fb$^{-1}$, where we show in the left panel the $p$-value obtained assuming no theoretical uncertainty and on the right including a theoretical uncertainty of 30\%, i.e. $ f_\mathrm{th} = 0.3$. We choose 30\% as a conservative estimate of the uncertainty, incorporating scale ($\mathcal{O}(10\%)$ at NNLO~\cite{deFlorian:2013jea}), PDF plus strong coupling constant (also $\mathcal{O}(10\%)$~\cite{deFlorian:2013jea}) as well as heavy top mass approximation uncertainties (another $\mathcal{O}(10\%)$~\cite{Grigo:2013rya}).

The values of $c_6$ compatible with $N(c_6=0)$ within $1\sigma$, i.e. the probability drops from $p=1$ (in our normalization) to $p= \exp(-1/2) \approx 0.607$, and thus the expected constraints, are:
\begin{eqnarray}
c_6^{1\sigma}(600~\mathrm{fb}^{-1}) \in (-0.4,0.5),~~c_6^{1\sigma}(3000~\mathrm{fb}^{-1}) \in (-0.3,0.3),~~\mathrm{for}~f_\mathrm{th} = 0\;, \nonumber \\
c_6^{1\sigma}(600~\mathrm{fb}^{-1}) \in (-0.5,0.8),~~c_6^{1\sigma}(3000~\mathrm{fb}^{-1}) \in (-0.4,0.4),~~\mathrm{for}~f_\mathrm{th} = 0.3\;.
\end{eqnarray}
These results are compatible with our previous studies in \cite{Goertz:2013kp} if the top Yukawa coupling is kept at its SM value.
The bounds are weaker for positive $c_6$, since this leads to a reduced cross section and thus to a larger statistical uncertainty.
The improvement of the 1$\sigma$ regions is moderate for $f_\mathrm{th} = 0.3$ for an increased luminosity, as in that case the uncertainty is dominated by the systematic uncertainty on the theoretical prediction of the signal rates. Improvements on the theoretical description of the process are thus necessary for improving these bounds.

\subsubsection{The full model}\label{sec:full}

Generically, one expects several operators to be present. Here we consider the full parameter space, varying the coefficients in Eq.~(\ref{eq:Lgghh}), as well as $c_\gamma$, within the currently allowed regions. We calculate the $p$-values in a similar fashion as before: we assume that the standard model is true, i.e. $c_i = 0 ~\forall~ i$, and compare the number of events after the analysis is performed with those expected from the SM, calculating how `far' they are in terms of the uncertainty $\delta N$. We note that, since we are only considering a single observable, the event rate for a particular signal process, we do not expect the constraints on the full parameter space to be strong. The constraints could be improved either by examining other observables in this process, or optimising the analysis for different points in the parameter space. We do not, however, expect significant qualitative changes in our results. To constrain the full parameter space, a combination of all possible production and decay channels (including various single production and double production processes) could be employed. We leave this endeavour to future work. Nevertheless, the main purpose of the study here is to investigate possible correlations among different operators in this process, and show how future measurements of other coefficients will help the determination of the Higgs potential.

To accommodate for the current allowed range on the parameters $c_g, c_t, c_b, c_H, c_\gamma$ we use the codes \texttt{HiggsBounds}~\cite{Bechtle:2013wla} and \texttt{HiggsSignals}~\cite{Bechtle:2013xfa} on the \texttt{eHDECAY} output. We employ the ``effective coupling'' mode, where one defines 
\begin{equation}
g_{hX} = \frac{\Gamma( h \to X)}{\Gamma( h \to X)_{\rm SM}} \, ,
\end{equation}
for the decay of the Higgs into the final state $X$. In particular, the single Higgs cross section is then scaled using the effective coupling to gluons, $g_{hgg}$. More explicitly,
\begin{equation}
g_{hgg} = \frac{\Gamma( h \to gg)}{\Gamma( h \to gg)_{\rm SM}} =  \frac{\sigma( gg \to h)}{\sigma( gg \to h)_{\rm SM}} \, .
\end{equation}
For further details we refer the reader to the \texttt{HiggsBounds} manual~\cite{Bechtle:2013wla}.
We perform our numerical scan as follows. We scan each direction of the (5+1)-dimensional parameter space, covering all coefficients, besides $c_6$, in the range $\{ -0.5, 0.5\}$ in steps of 0.1, while the latter is allowed to vary in a larger range, $\{-2, 3.5\}$, in steps of $0.5$.\footnote{We note here the possibility of small cut-off effects in the marginalization procedure due to the choices of these ranges.} In this scenario, we assume the coefficients $c_b$ and $c_\tau$ to be equal. We feed these coefficients into \texttt{eHDECAY} to compute the branching ratios of various decay modes of the Higgs boson. In this step, it is possible that some branching ratios in the output of eHDECAY become negative for certain values of the coefficients. This is usually due to destructive interference between the EFT contributions to the decay amplitudes and the SM ones, and we discard these points in our scan.\footnote{Note that this does not necessarily imply that these parameter points are unphysical, or that the power expansion breaks down at these points. It could be that the SM amplitudes are accidentally suppressed (e.g., loop-suppression for $h \to \gamma \gamma, g g, Z \gamma$) such that they have similar sizes as the EFT amplitudes. To compute the partial widths in these cases, one should include the square of the EFT amplitudes, even though they are formally of higher order in the power expansion. This feature is not implemented in \texttt{eHDECAY}.}
These coefficients and branching ratios are then given as input to \texttt{HiggsBounds}, which checks if a point is excluded at the 95\% C.L. by collider data (LEP, Tevatron and LHC). This step is numerically fast, and allows to easily discard points where the EFT effects would have generated an excess of events in single Higgs boson studies at the 7 or 8 TeV LHC. The surviving points are fed into \texttt{HiggsSignals}, which performs a multi-dimensional fit to the Higgs observables and outputs a $p$-value for each point. We discard points which would have given a substantial deficit (or excess) of events in current Higgs data, and thus we keep only points at the 95\% C.L., that is, where the $p$-value corresponds to less than 2 standard deviations from the mean of a Gaussian distribution.

We now proceed to derive the constraint from $hh$ production onto the parameter space allowed by current experiments. To visualize the constraints in the multi-dimensional parameter space, we will therefore project them onto two-dimensional planes. The parameter $c_6$ still plays a somewhat distinct role, as it is currently essentially unconstrained, and the information on it will come primarily from multi-Higgs boson production. Thus, all two-dimensional exclusion planes include this parameter.
To calculate the allowed two-dimensional regions, we need to marginalize over the remaining dimensions. To accomplish this, for a given point in the $(c_i, c_6)$-plane, we sum over the $p$-values obtained by varying along the other dimensions. The final p-value in the 2-D plane then reads
\begin{equation}\label{eq:margin1}
p(c_i,c_6) = \frac{ \sum_{\{c_f\}} p ( c_6, c_i, \{c_f\}) \times p_\mathrm{HS}  (c_i, \{c_f\}) } {\sum_{\{c_f\}}  p_\mathrm{HS}  (c_i, \{c_f\}) }   \;,
\end{equation}
where $p_\mathrm{HS}  (c_i, \{c_f\})$ is the probability assigned to the given point from the \texttt{HiggsSignals} code. Dividing out by the sum $\sum_{\{c_f\}}  p_\mathrm{HS}  (c_i, \{c_f\})$ removes the constraints arising due to single Higgs boson data coming from \texttt{HiggsSignals} on the given $(c_i, c_6)$-plane, while taking into account this knowledge in the marginalization over the irrelevant coefficients. Essentially, what one achieves by this normalization, is to have a flat probability distribution on the $(c_i, c_6)$-plane, \textit{before} any $hh$ data is taken into account. 

To account for a proper normalization we divide by the maximum corresponding probability, for the coefficients under consideration:
\begin{equation}\label{eq:margin2}
\bar{p}(c_i,c_6) = \frac{1}{\mathrm{max~} p(c_i,c_6)} p(c_i,c_6)\;.
\end{equation}
The 1$\sigma$-equivalent contours are thus drawn by finding the iso-curve corresponding to $\bar{p}(c_i,c_6) = \exp(-1/2) \approx 0.607$. To obtain a constraint on a single coefficient $c_i$, we marginalize over all the other coefficients $c_j$ ($j \neq i$). This is done in the same way as prescribed by Eqs.~(\ref{eq:margin1}) and~(\ref{eq:margin2}) given above.

\begin{figure}[!htb]
  \centering
    \includegraphics[width=0.49\linewidth]{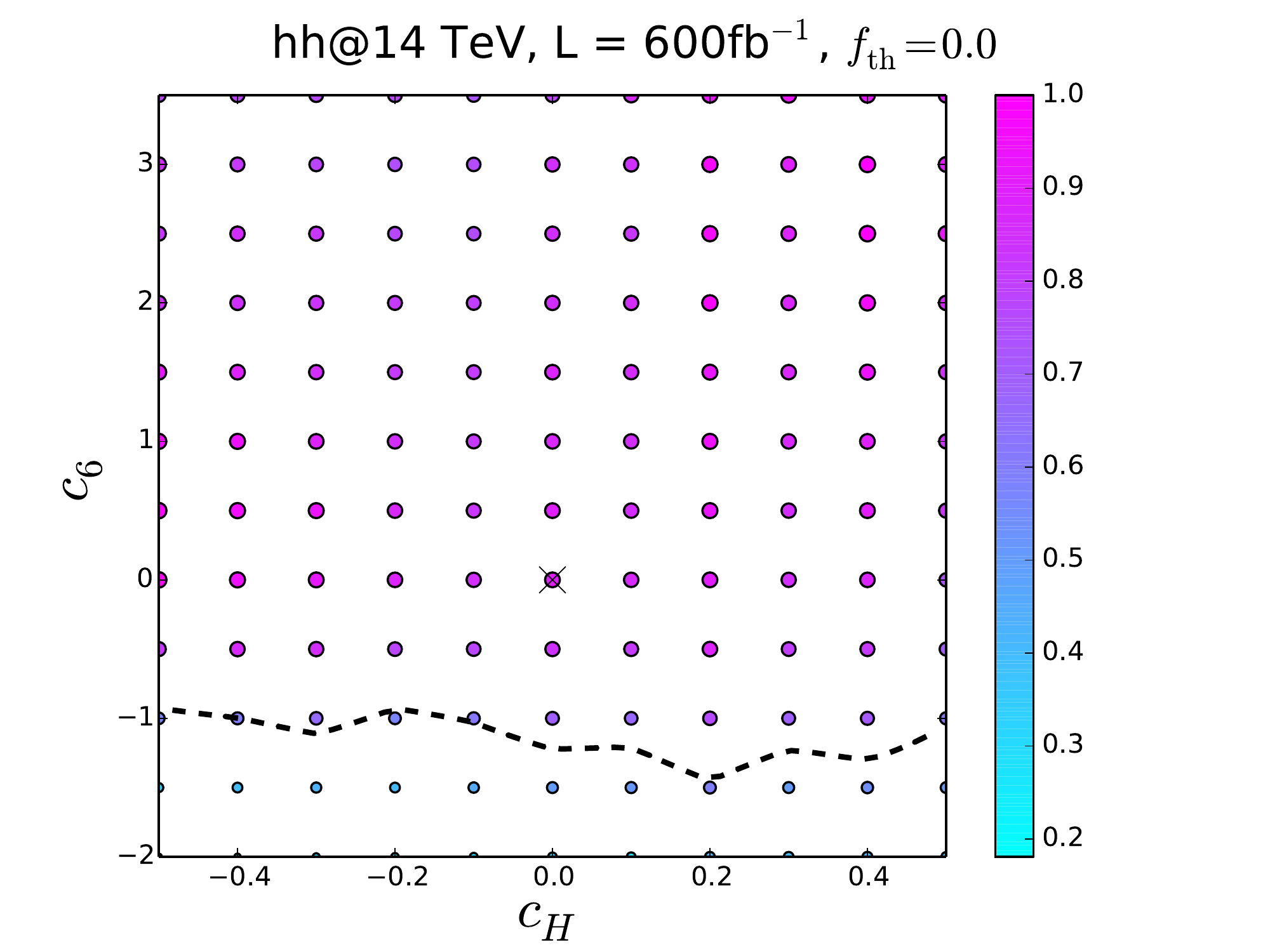}
    \includegraphics[width=0.49\linewidth]{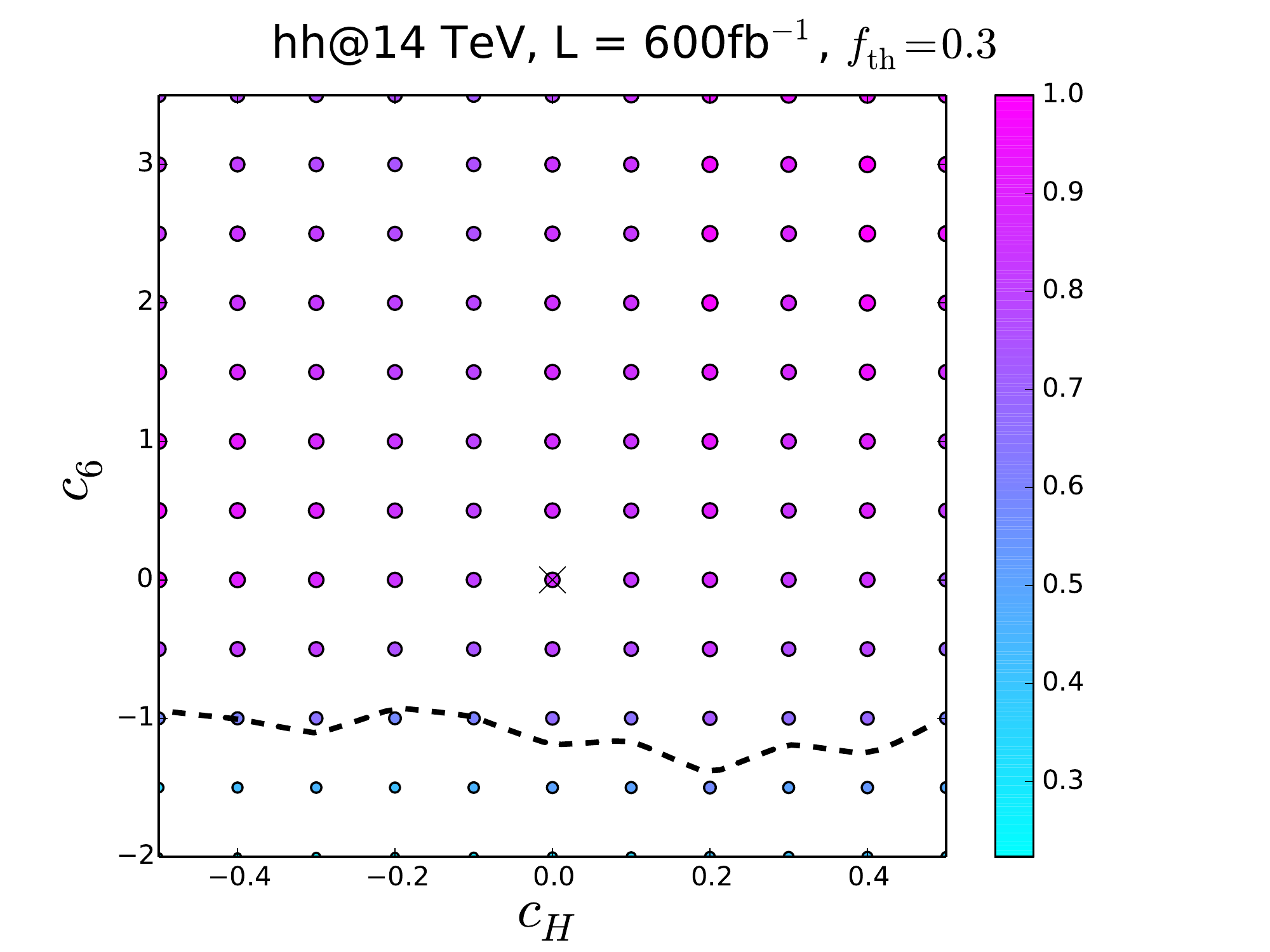}
    \includegraphics[width=0.49\linewidth]{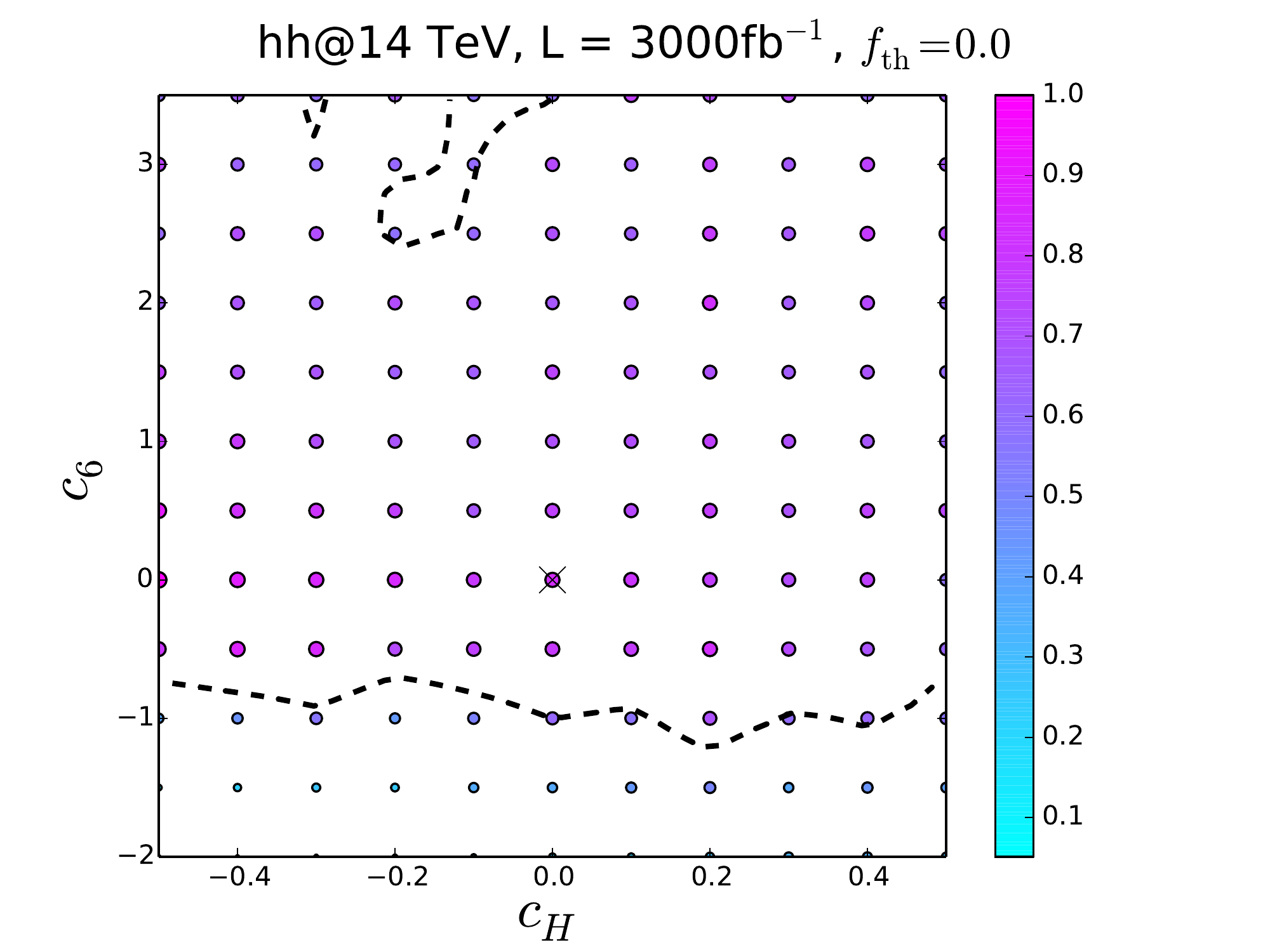}
    \includegraphics[width=0.49\linewidth]{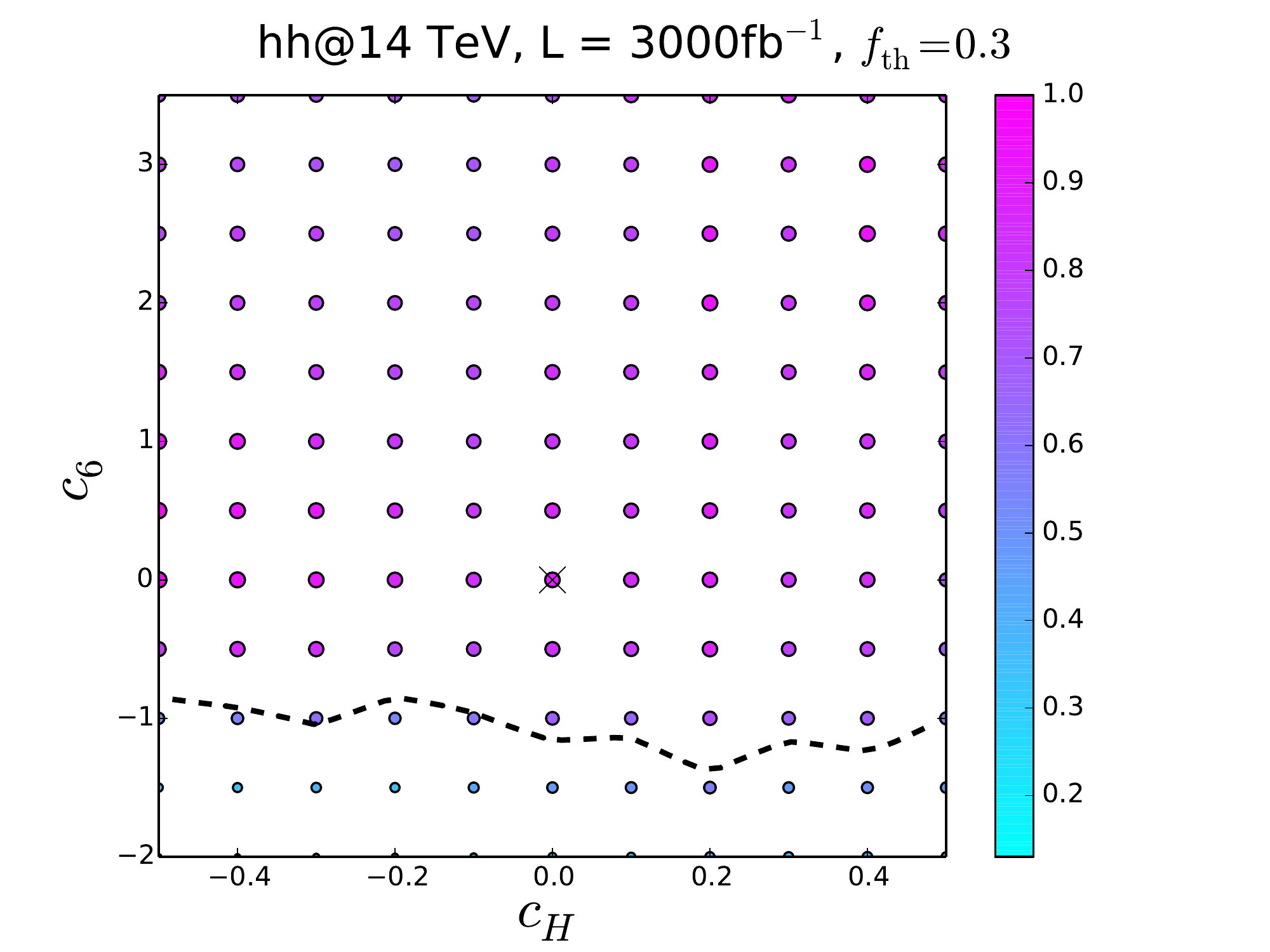}
  \caption{ The $p$-values obtained after marginalization over the directions orthogonal to the $(c_H,c_6)$-plane, for the process $hh \rightarrow (b\bar{b}) (\tau^+ \tau^-)$. On the top plots we show the results at 600~fb$^{-1}$ of integrated luminosity, without  ($f_\mathrm{th} = 0.0$) and with  ($f_\mathrm{th} = 0.3$) theoretical uncertainty included and on the bottom we show the corresponding plots at 3000~fb$^{-1}$. We also present the 1-sigma contours as black dashed lines.}
  \label{fig:excc6ch}
\end{figure}

We first consider the $(c_H, c_6)$-plane in Fig.~\ref{fig:excc6ch}. The coefficient $c_H$ enters all EFT diagrams by changing the Higgs boson wave function in a universal way, and competes with $c_6$ by reducing the self-coupling contribution in our convention. Since $c_H$ also affects single Higgs boson production, it is already constrained by current experimental data. One sees from Fig.~\ref{fig:efficiency} that the production rate after cuts depends mildly on $c_H$, and therefore no significant improvement on its bound from $hh$ production is expected. Indeed, this fact is evident in Fig.~\ref{fig:excc6ch}, where it is also clear that future knowledge about $c_H$ will not help us to constrain $c_6$ very much.\footnote{However, improvements of single Higgs boson constraints on the other coefficients will allow for tighter constraints on this plane.} When examining Fig.~\ref{fig:excc6ch}, one should also recall that a change in $c_H$ affects the preferred values of other coefficients due to single Higgs boson constraint, entering the marginalization procedure. We find that, after marginalization over the other coefficients, $c_H$ is constrained to lie in $c_H < 0.4$ according to our 1$\sigma$-equivalent definition, at 3000~fb$^{-1}$ and for $f_\mathrm{th} = 0.3$. 

\begin{figure}[!htb]
  \centering
    \includegraphics[width=0.49\linewidth]{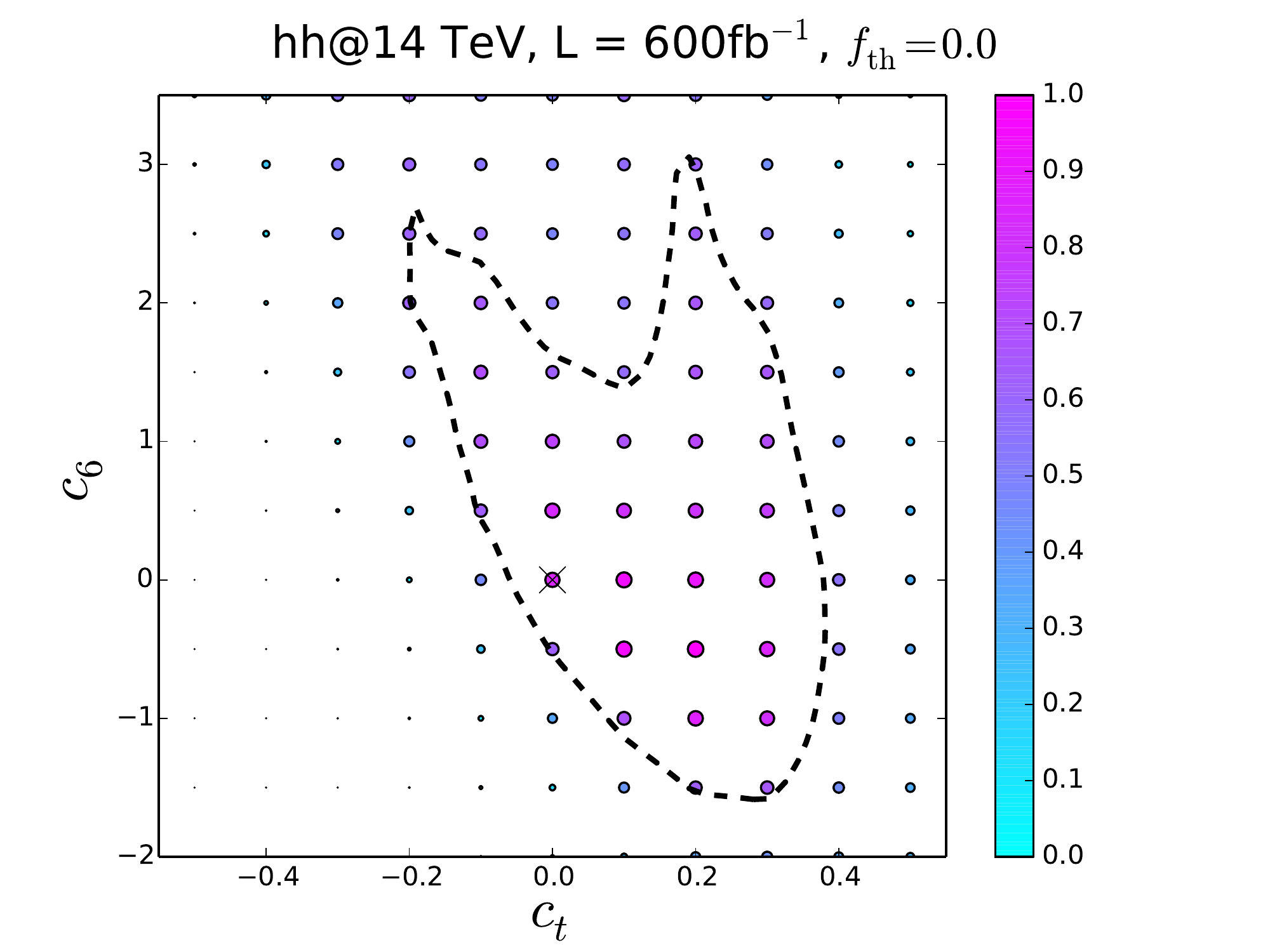}
    \includegraphics[width=0.49\linewidth]{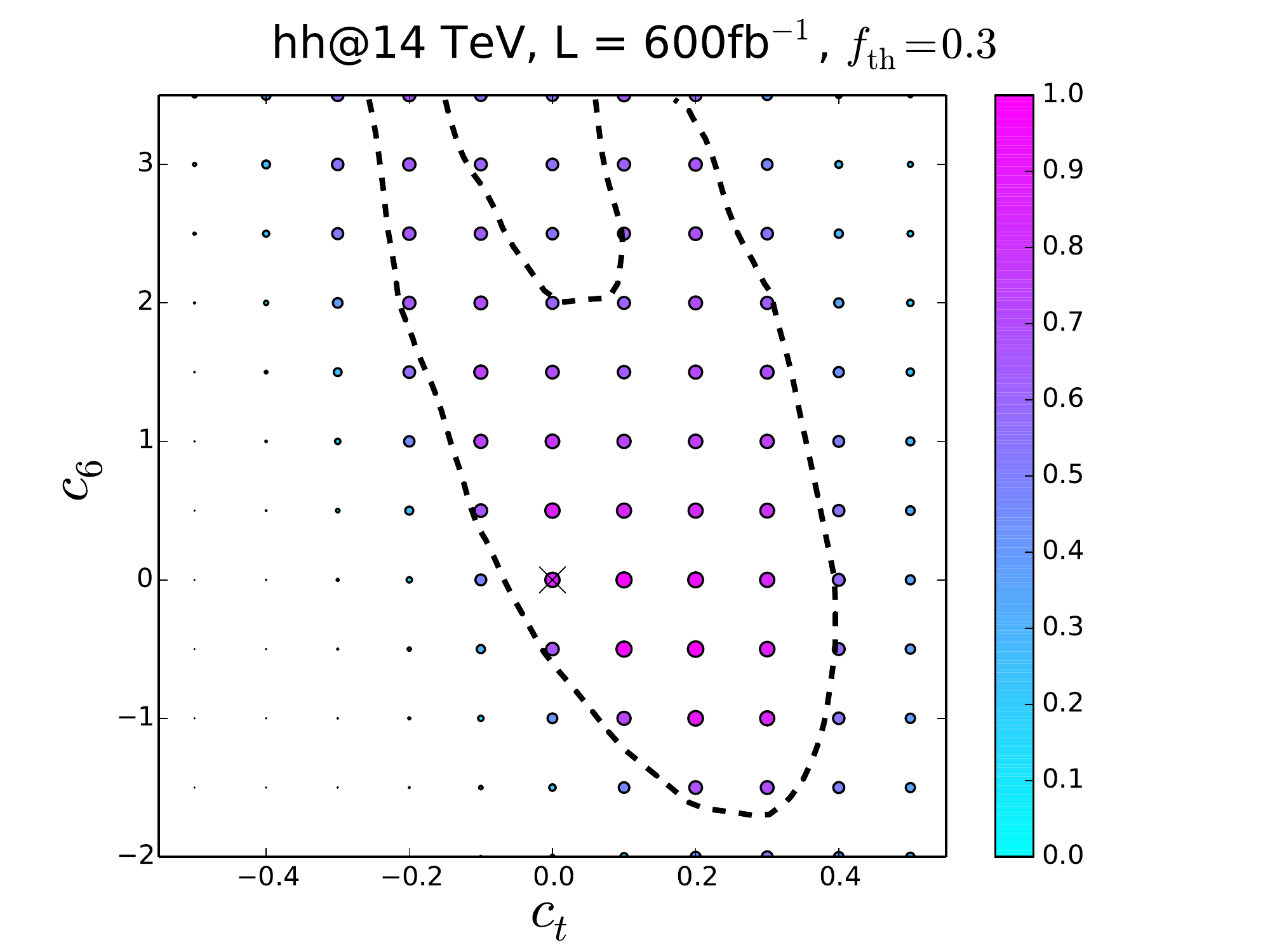}
    \includegraphics[width=0.49\linewidth]{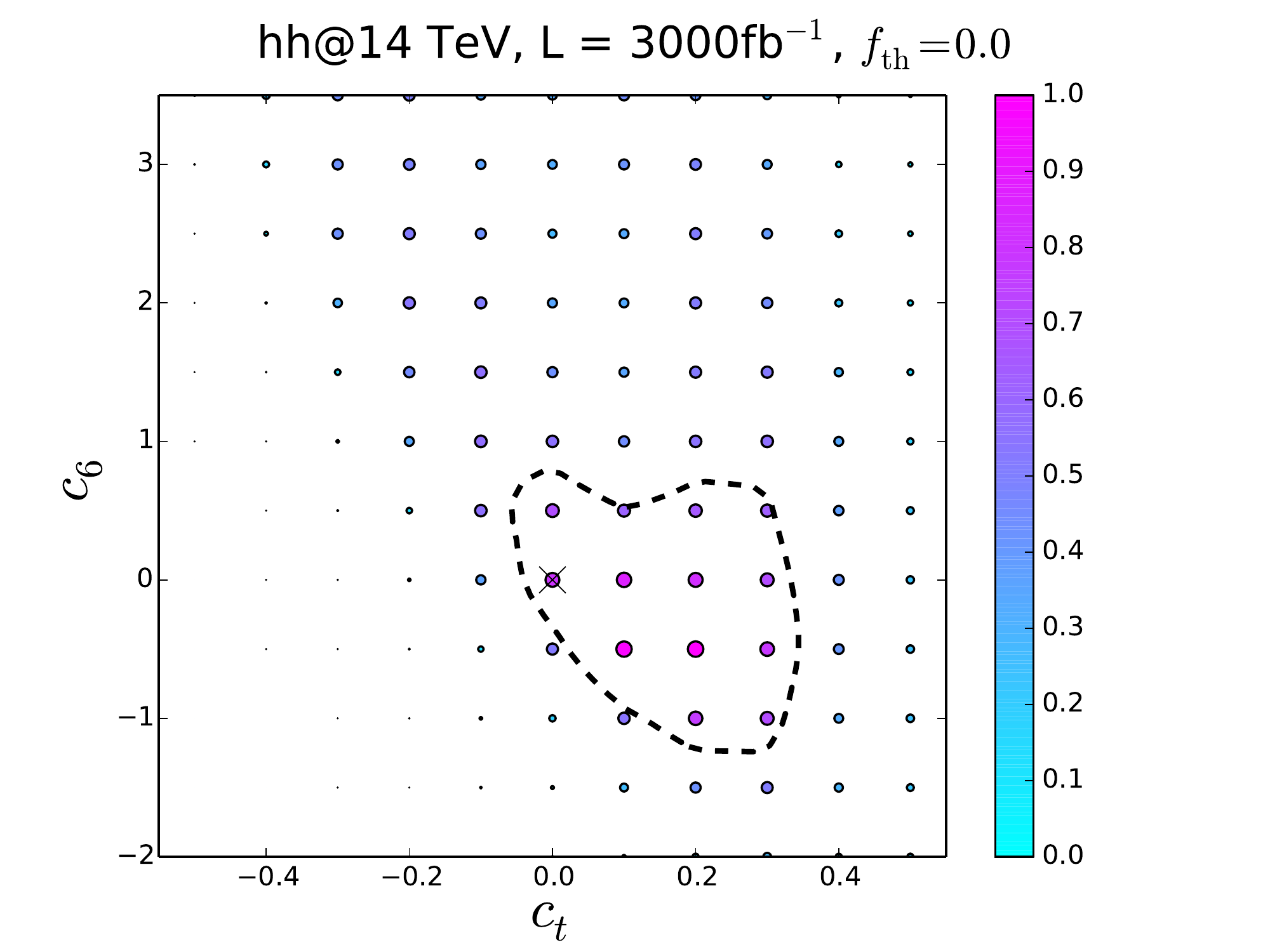}
    \includegraphics[width=0.49\linewidth]{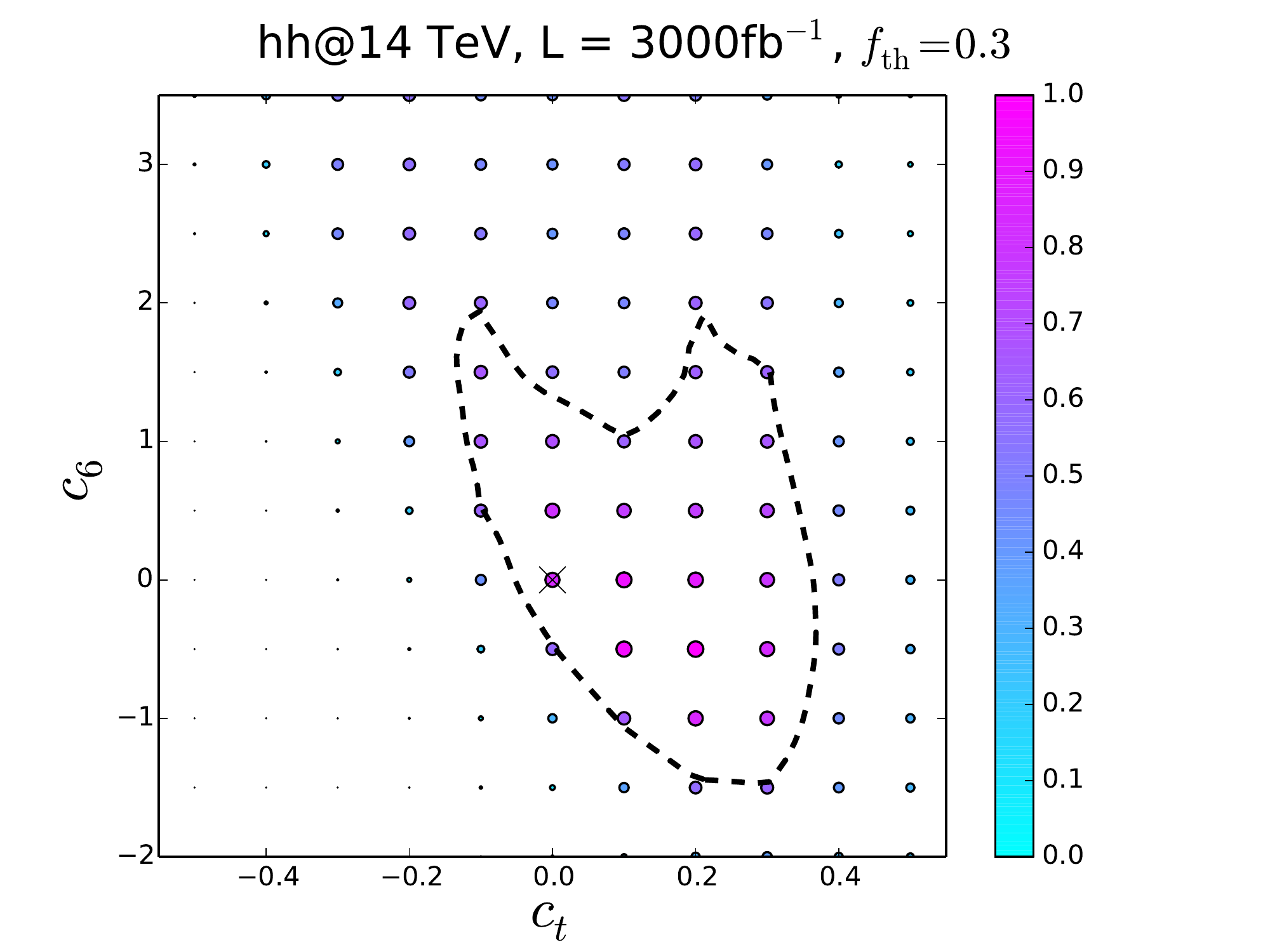}
 \caption{The $p$-values obtained after marginalization over the directions orthogonal to the $(c_t,c_6)$-plane, for the process $hh \rightarrow (b\bar{b}) (\tau^+ \tau^-)$. On the top plots we show the results at 600~fb$^{-1}$ of integrated luminosity, without  ($f_\mathrm{th} = 0.0$) and with  ($f_\mathrm{th} = 0.3$) theoretical uncertainty included and on the bottom we show the corresponding plots at 3000~fb$^{-1}$.
We also present the 1-sigma contours as black dashed lines.}
  \label{fig:excc6ct}
\end{figure}

We next examine the $(c_t, c_6)$-plane in Fig.~\ref{fig:excc6ct}. The coefficient $c_t$ enters all diagrams that contain top quarks. Points with positive $c_6$ and negative $c_t$ are more challenging to exclude -- the coefficients enter in the first line of Eq.~(\ref{eq:diffxsEFT}) with the same sign, which leads to a compensation of effects (see also Fig.~\ref{fig:efficiency}). The `dip' structure that appears at $c_t \approx 0.1-0.2$ is related to the fact that the minimum cross section as a function of $c_t$ appears in that region. Beyond the dip, the (most important) corrections from the new triangle diagram mediated by the $t\bar{t}hh$ vertex dominate the behaviour of the cross section, while before the dip their destructive interference with the box contributions leads to a reduction in the cross section. The coefficient is constrained to lie within $ -0.1 \lesssim c_t \lesssim 0.4$ at 3000~fb$^{-1}$ and for $f_\mathrm{th} = 0.3$, after marginalization (1$\sigma$-equivalent). It is evident that improving the knowledge on the poorly-constrained `top Yukawa' $c_t$, entering $hh$ production in various ways, will be helpful to improve the exclusion range for $c_6$.

\begin{figure}[!htb]
  \centering
    \includegraphics[width=0.49\linewidth]{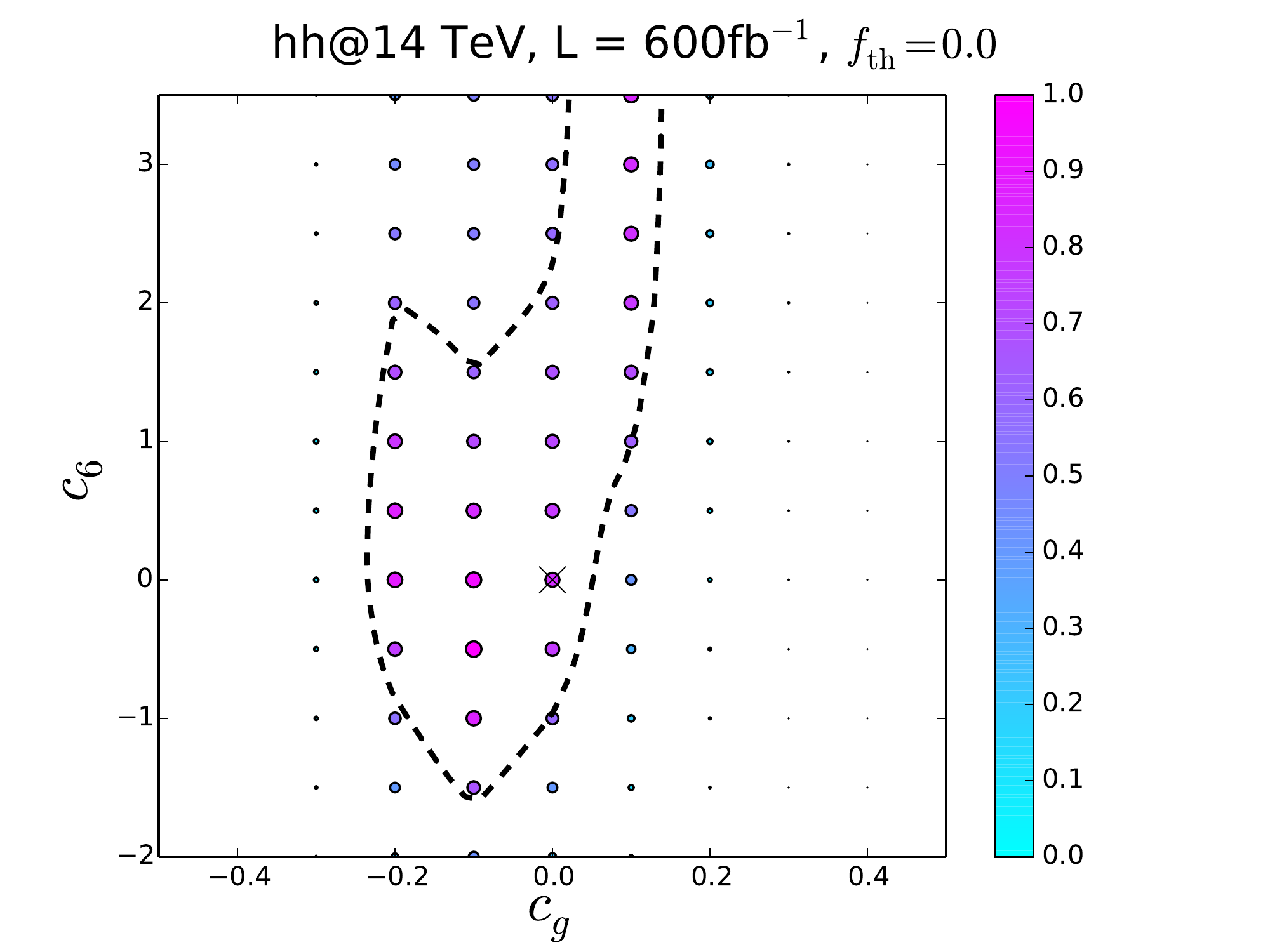}
    \includegraphics[width=0.49\linewidth]{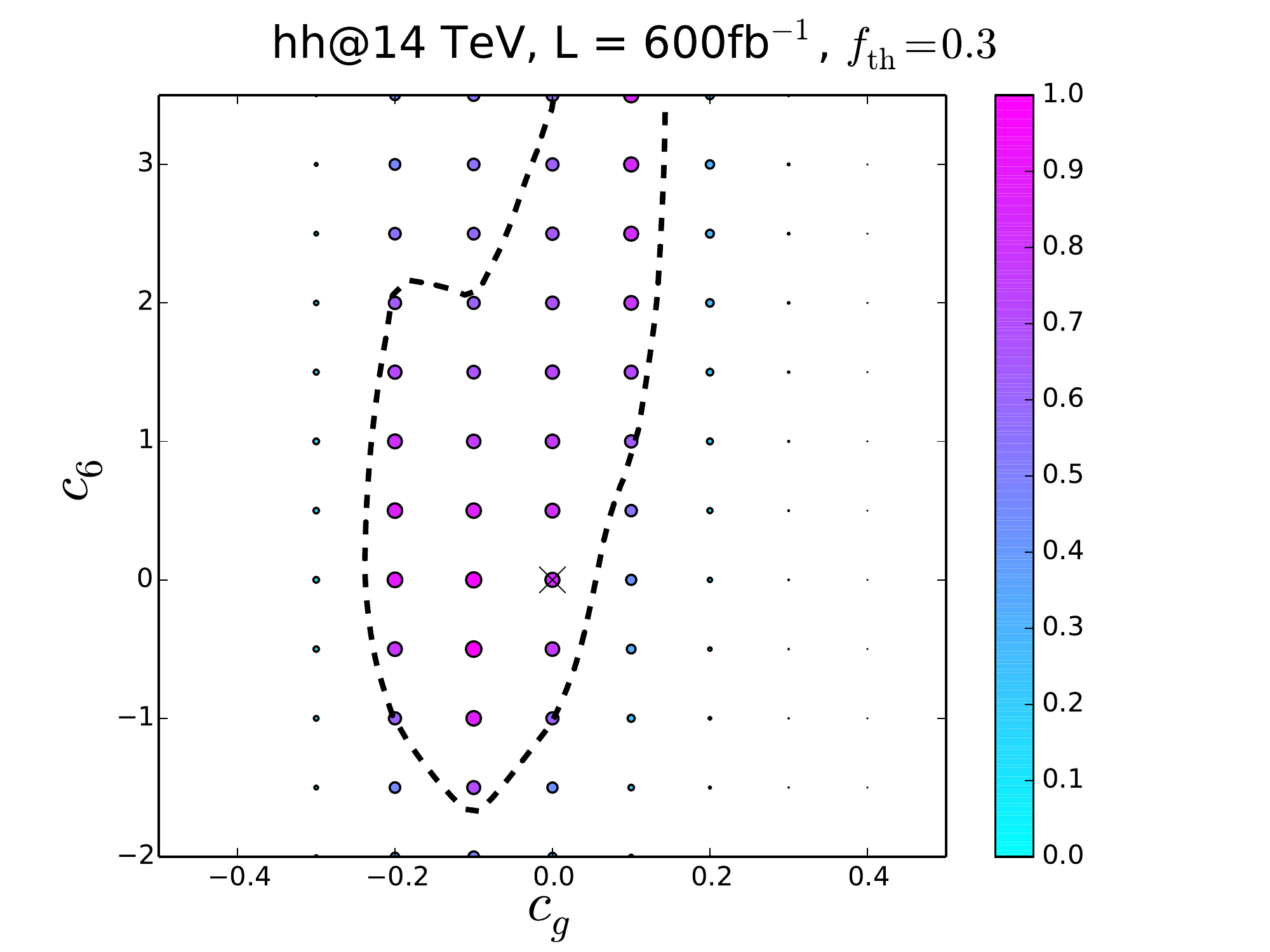}
    \includegraphics[width=0.49\linewidth]{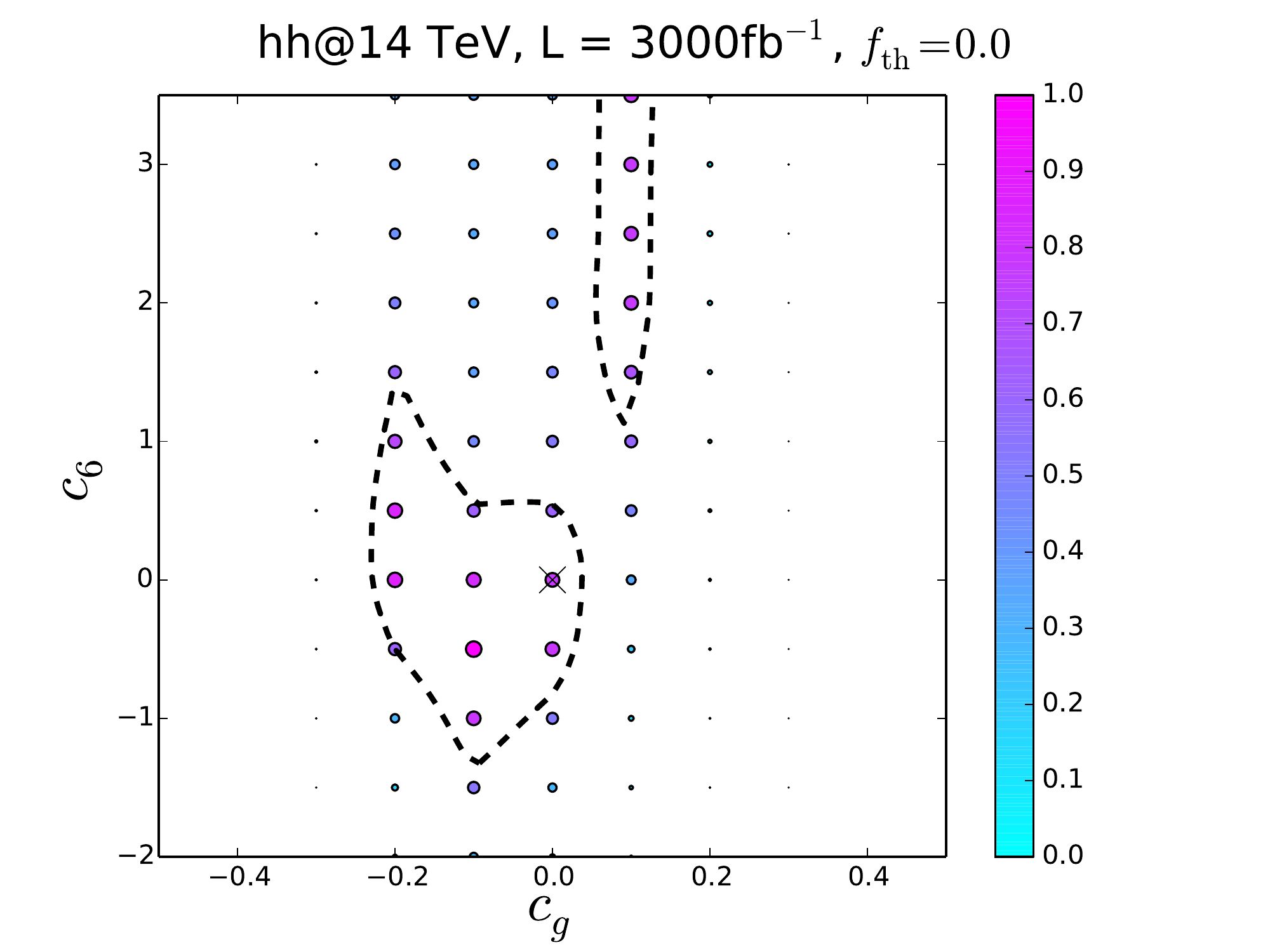}
    \includegraphics[width=0.49\linewidth]{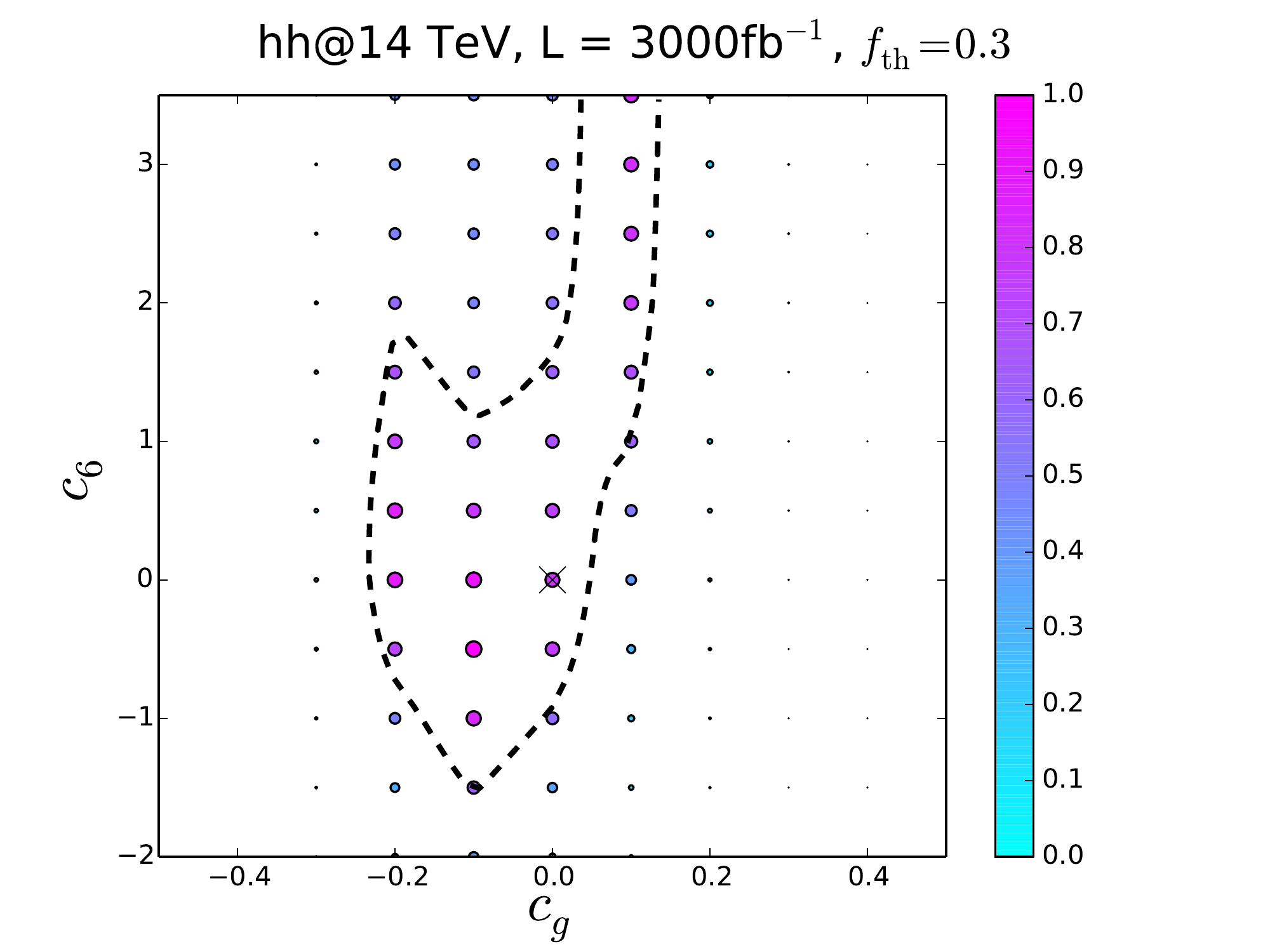}
\caption{The $p$-values obtained after marginalization over the directions orthogonal to the $(c_g,c_6)$-plane, for the process $hh \rightarrow (b\bar{b}) (\tau^+ \tau^-)$. On the top plots we show the results at 600~fb$^{-1}$ of integrated luminosity, without  ($f_\mathrm{th} = 0.0$) and with  ($f_\mathrm{th} = 0.3$) theoretical uncertainty included and on the bottom we show the corresponding plots at 3000~fb$^{-1}$.
We also present the 1-sigma contours as black dashed lines.}
  \label{fig:excc6cg}
\end{figure}

The expected constraints for $c_g$, which adds tree-level couplings of one or two Higgs boson to two gluons, are shown in the $(c_g, c_6)$-plane in Fig.~\ref{fig:excc6cg}. The results reflect the fact that an enhanced production cross section due to values of $c_g$ away from the minimum (right panel, Fig.~\ref{fig:efficiency}) can compensate a reduction due to positive $c_6$. The constraint on $c_g$ is found to be $ -0.2 \lesssim c_g \lesssim 0.1$ at 3000~fb$^{-1}$ given that $f_\mathrm{th} = 0.3$, after marginalization. 

\begin{figure}[!htb]
  \centering
    \includegraphics[width=0.49\linewidth]{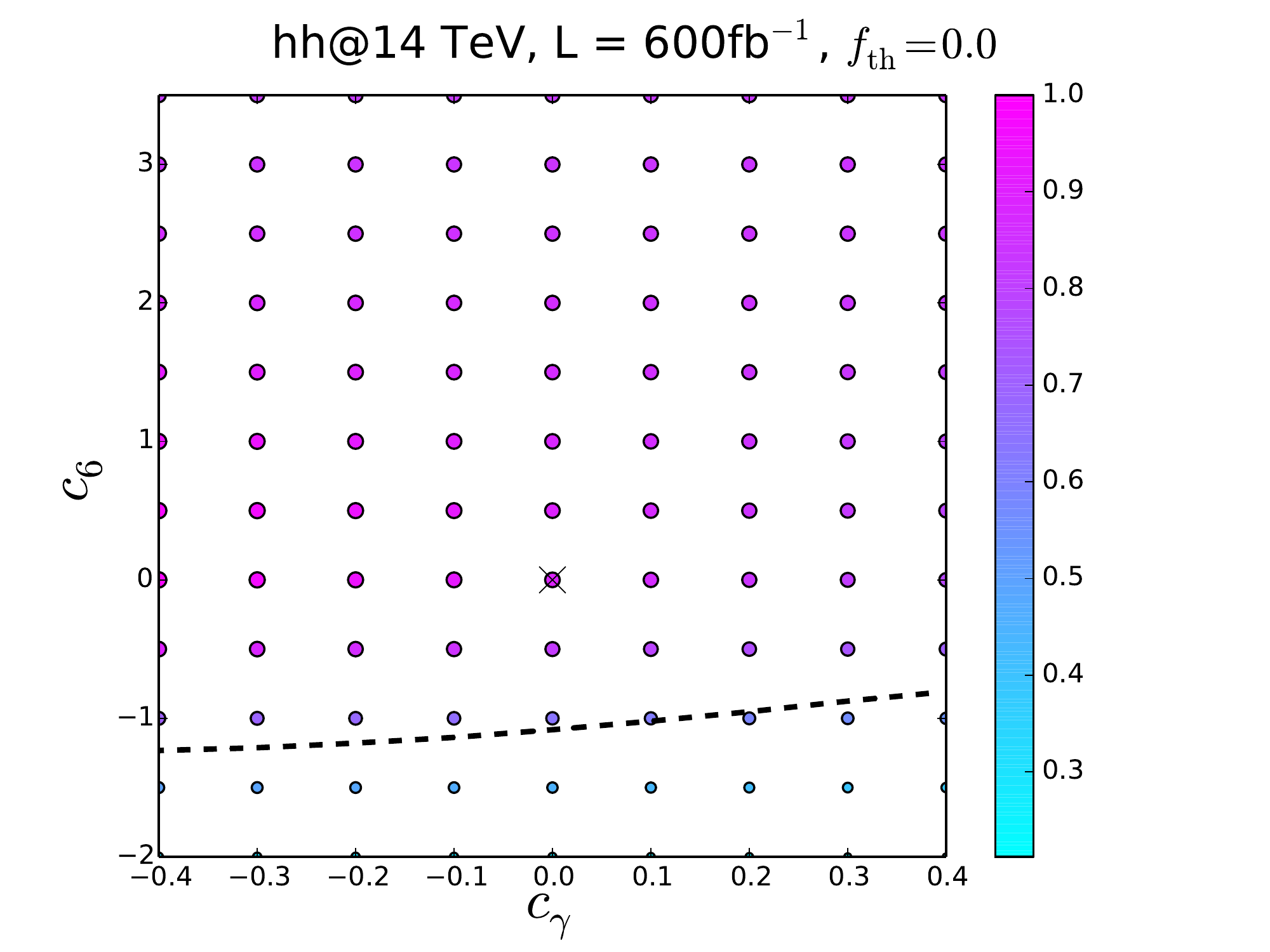}
    \includegraphics[width=0.49\linewidth]{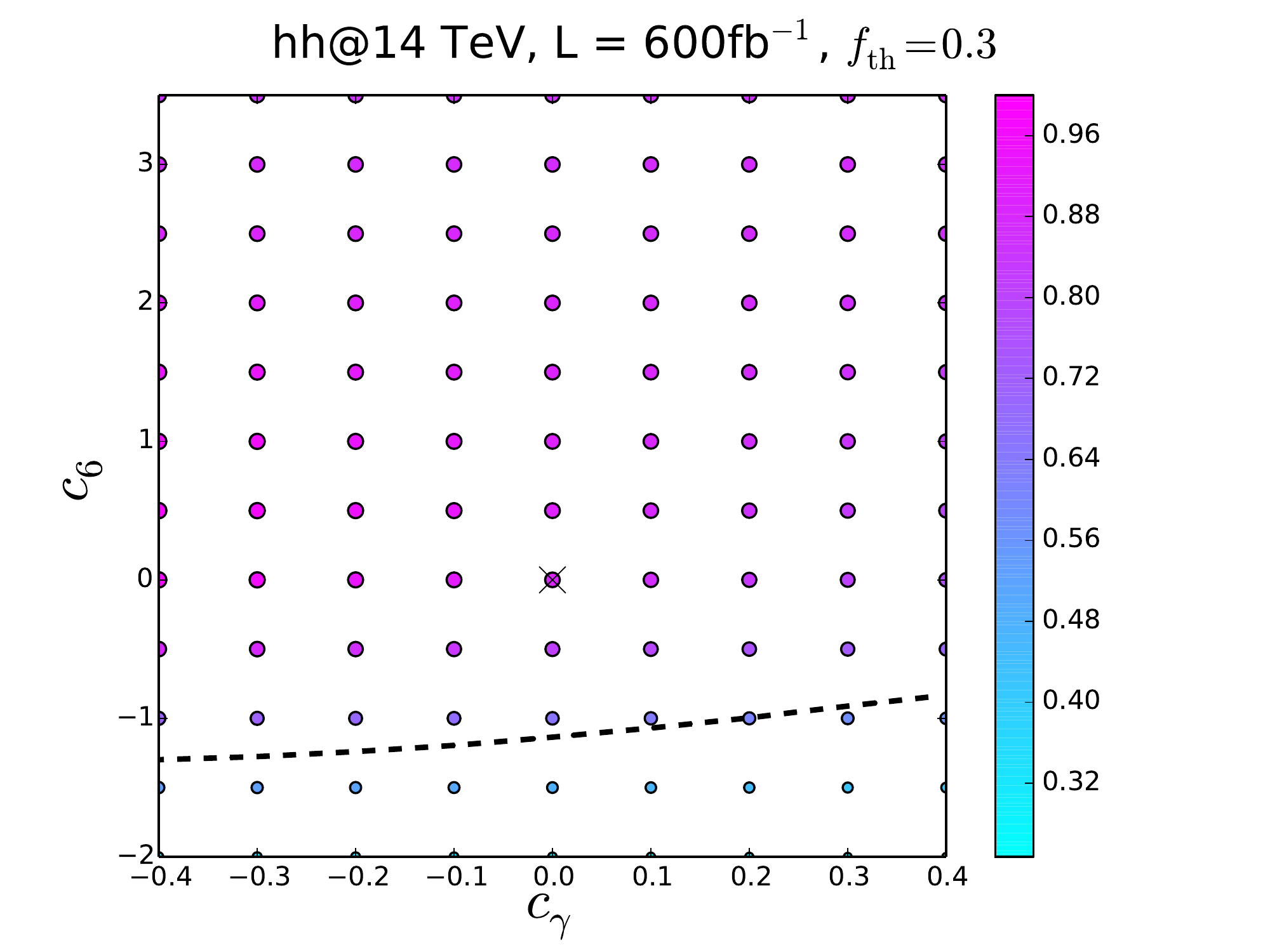}
    \includegraphics[width=0.49\linewidth]{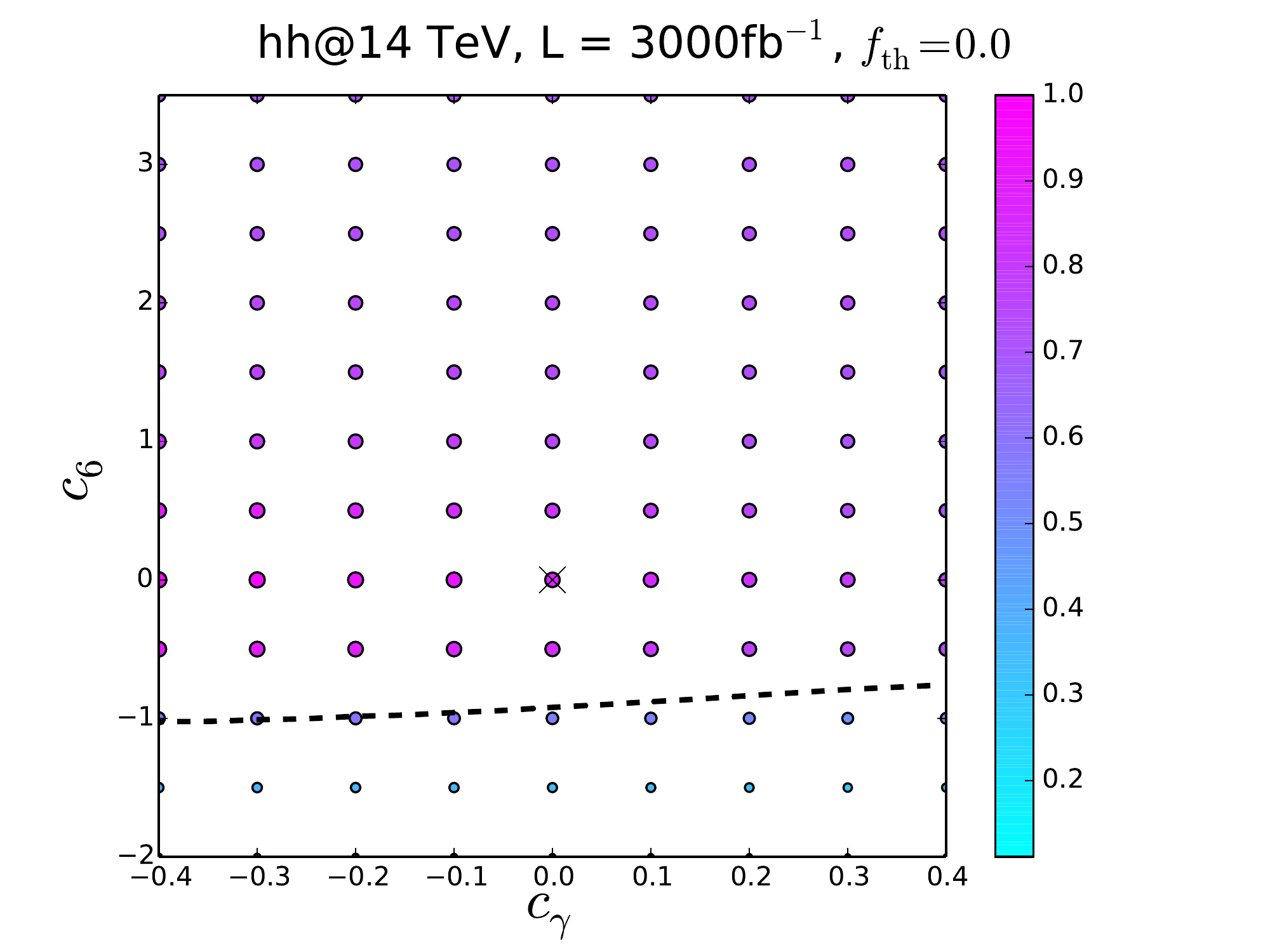}
    \includegraphics[width=0.49\linewidth]{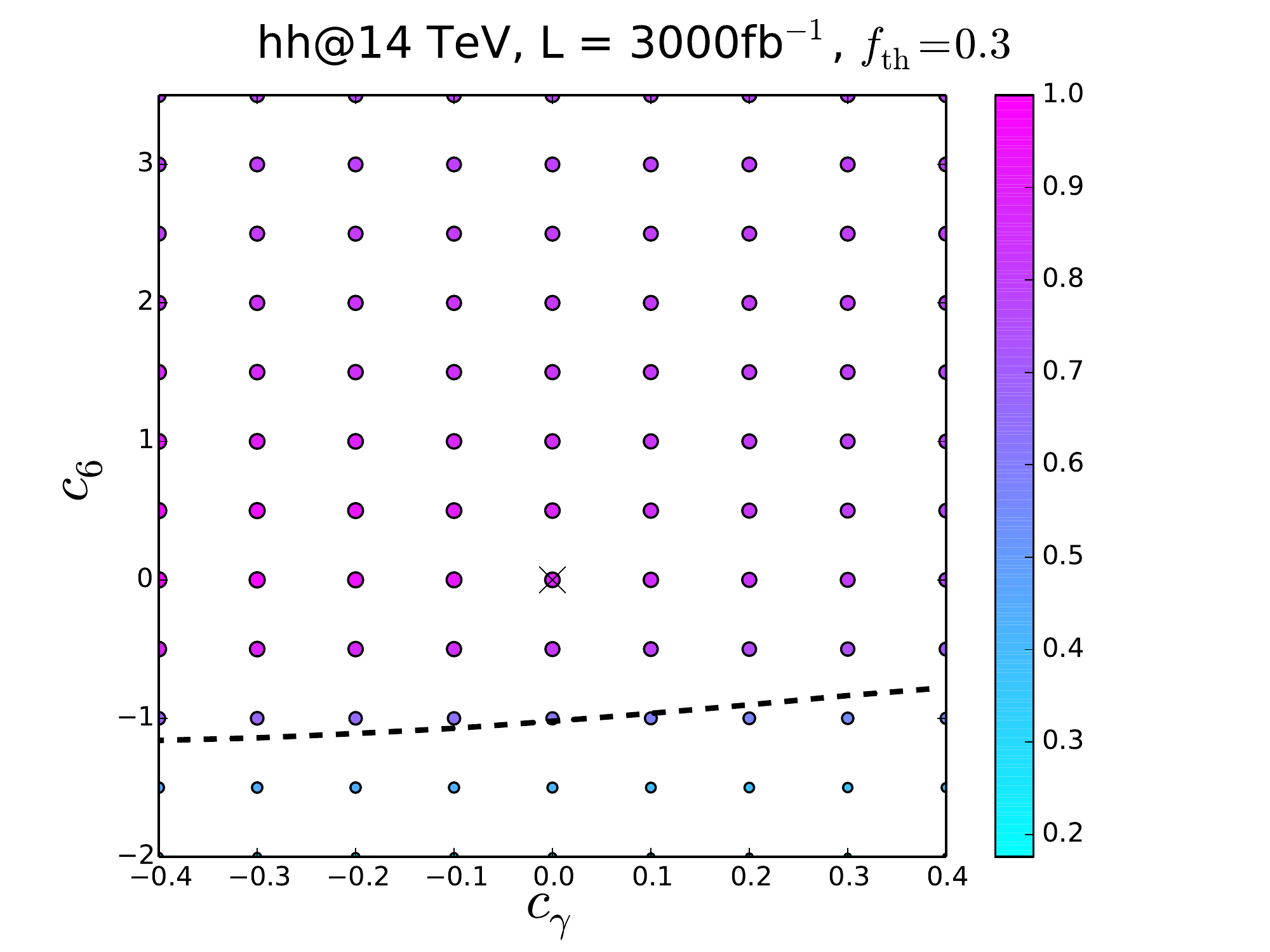}
  \caption{The $p$-values obtained after marginalization over the directions orthogonal to the $(c_\gamma,c_6)$-plane, for the process $hh \rightarrow (b\bar{b}) (\tau^+ \tau^-)$. On the top plots we show the results at 600~fb$^{-1}$ of integrated luminosity, without  ($f_\mathrm{th} = 0.0$) and with  ($f_\mathrm{th} = 0.3$) theoretical uncertainty included and on the bottom we show the corresponding plots at 3000~fb$^{-1}$.
We also present the 1-sigma contours as black dashed lines.}
  \label{fig:excc6cgam}
\end{figure}

We present the results involving $c_\gamma$ in Fig.~\ref{fig:excc6cgam}, which enters the process under consideration indirectly, through modification of the branching ratios (via single Higgs boson data $p$-values). The correlation with $c_6$ is weak, and no significant constraint is expected to be imposed through $hh \rightarrow (b\bar{b})(\tau^+ \tau^-)$. 

\begin{figure}[!htb]
  \centering
    \includegraphics[width=0.49\linewidth]{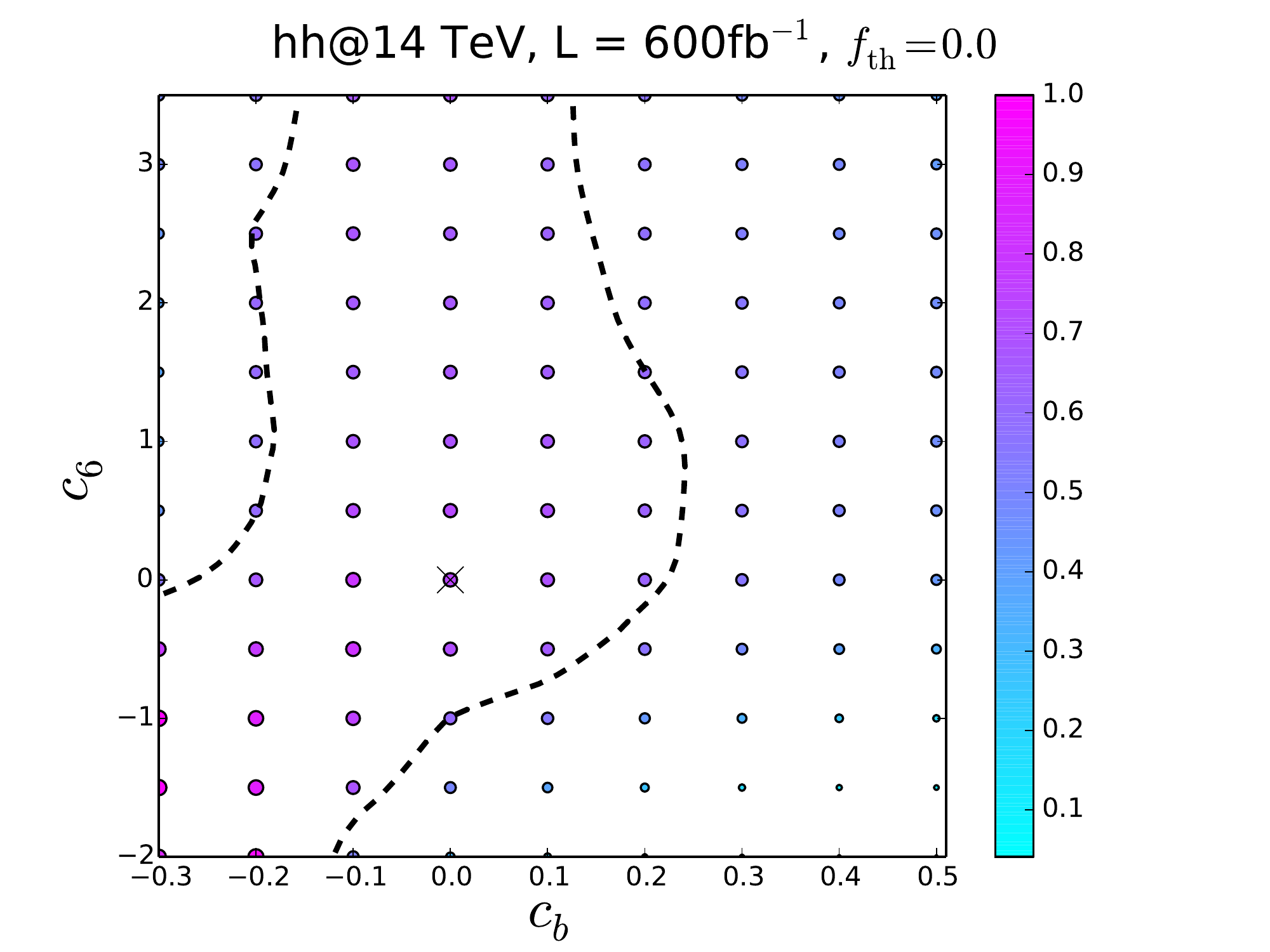}
    \includegraphics[width=0.49\linewidth]{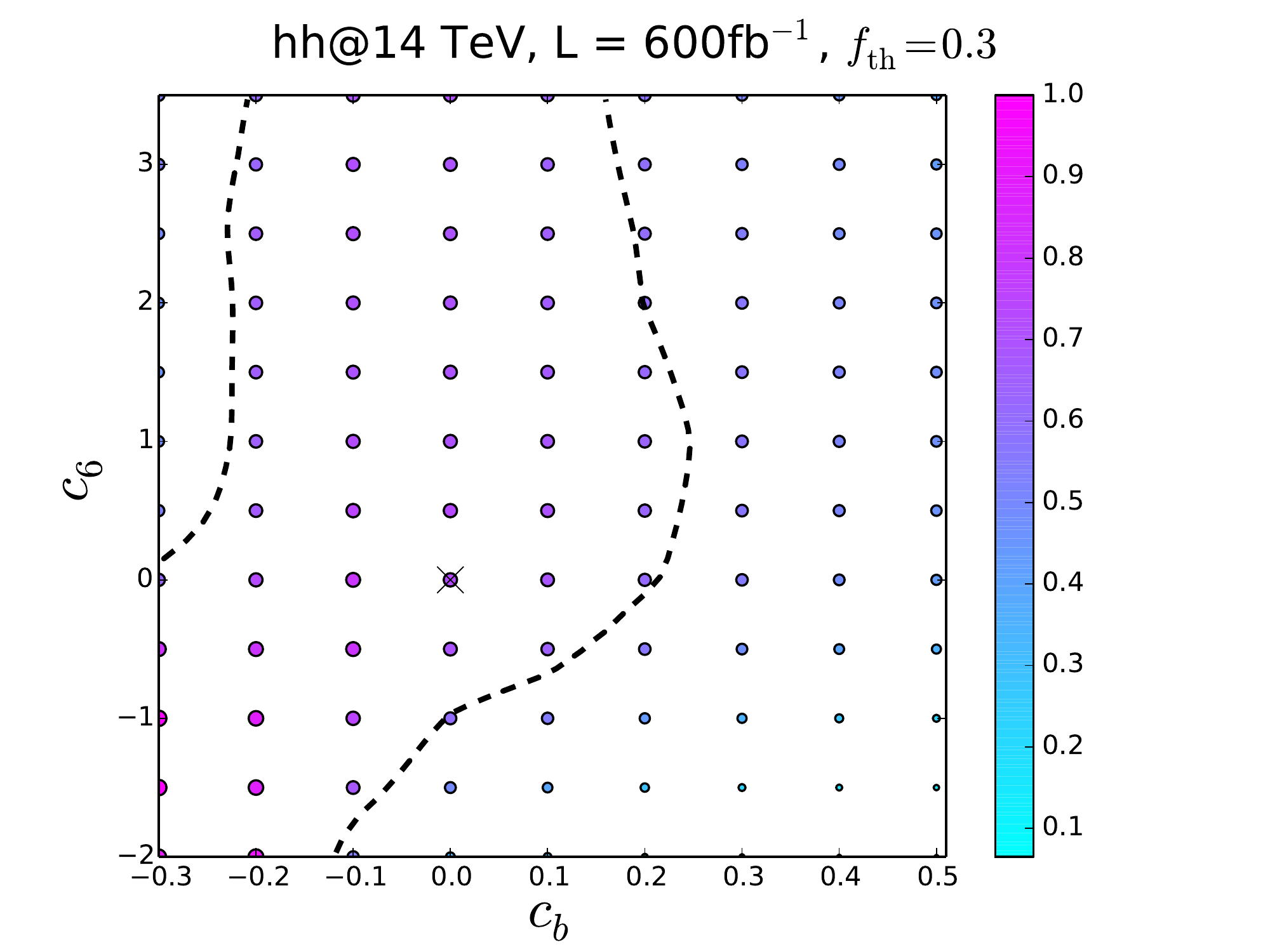}
    \includegraphics[width=0.49\linewidth]{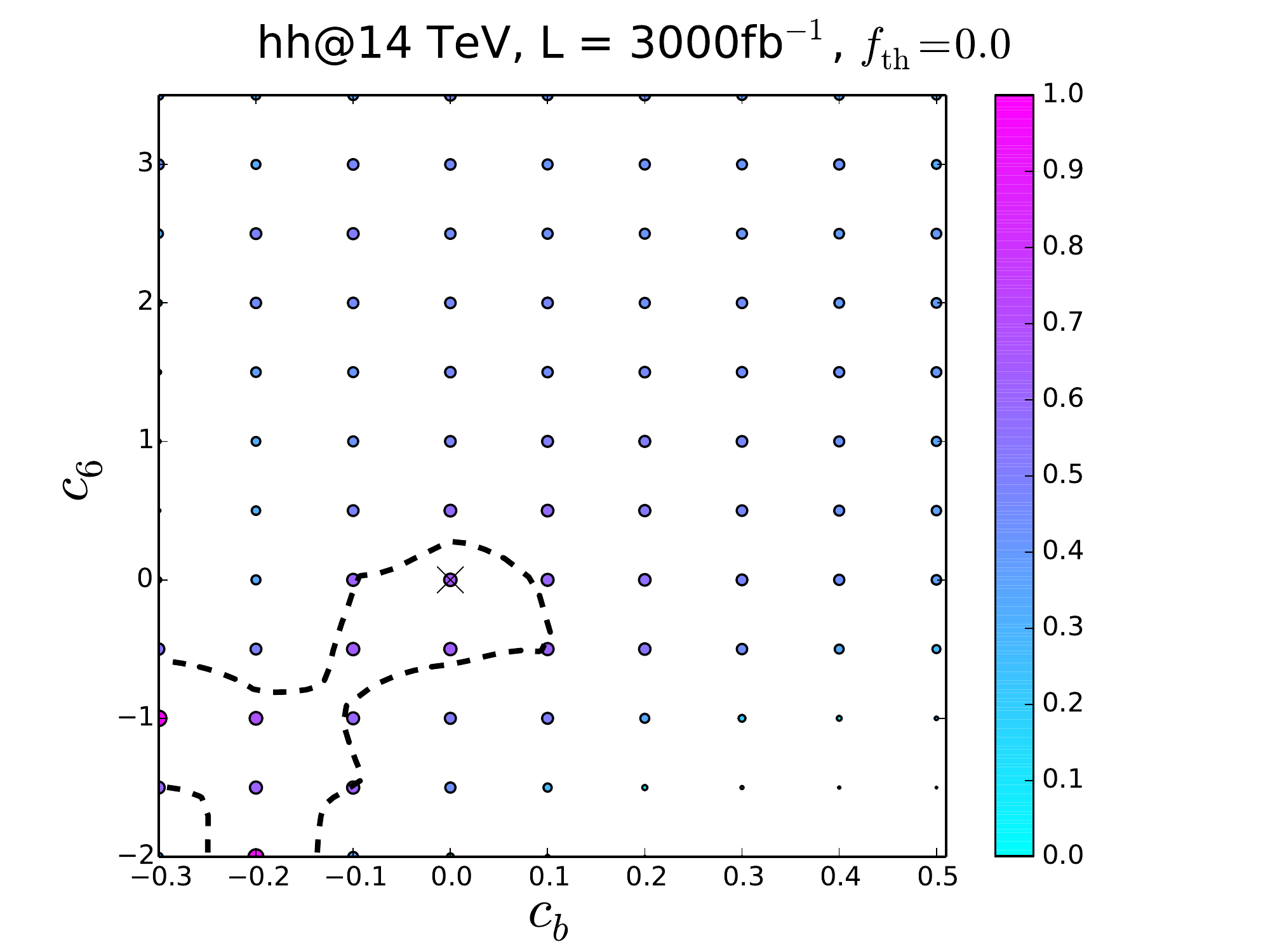}
    \includegraphics[width=0.49\linewidth]{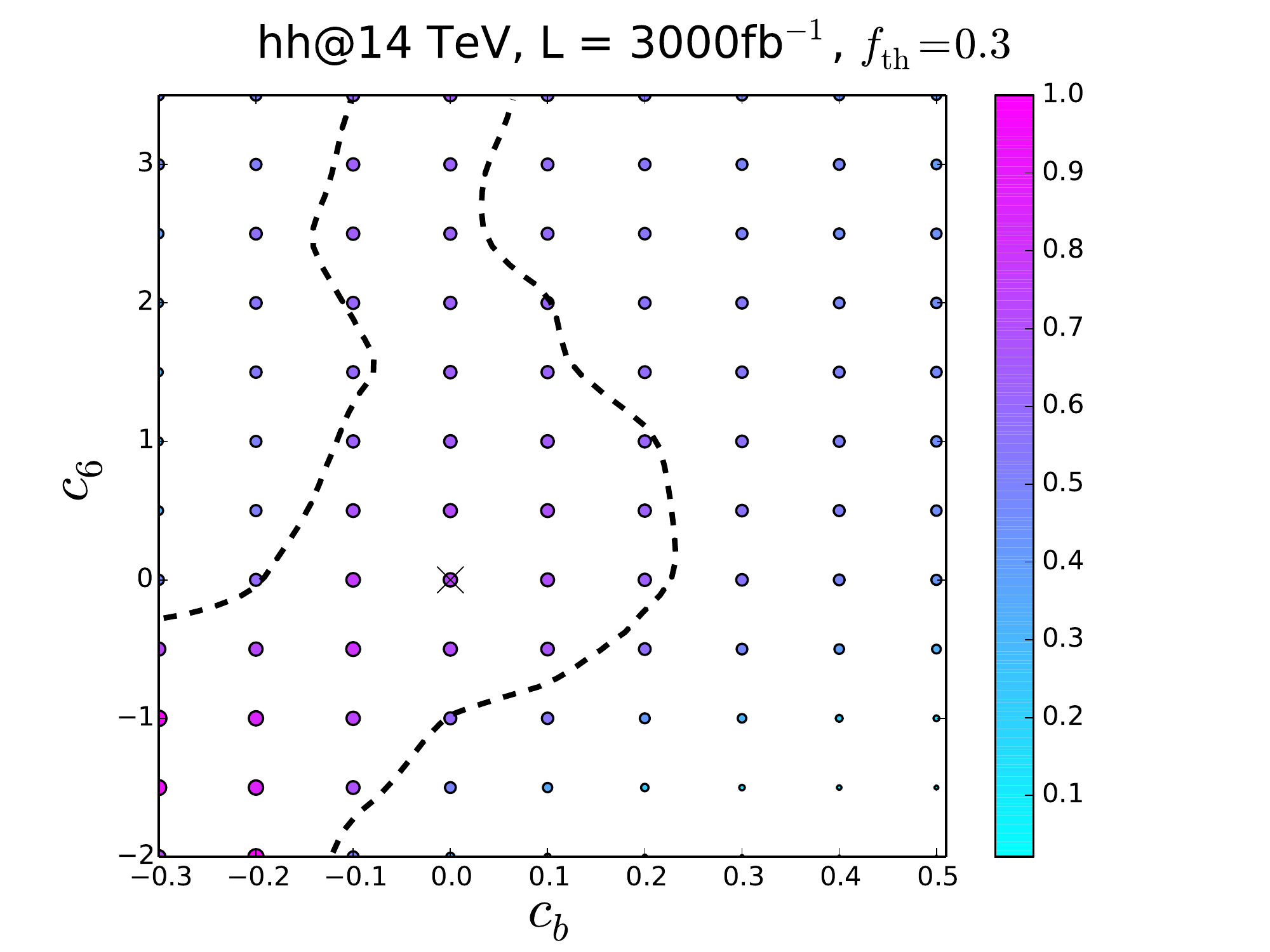}
  \caption{The $p$-values obtained after marginalization over the directions orthogonal to the $(c_b,c_6)$-plane, for the process $hh \rightarrow (b\bar{b}) (\tau^+ \tau^-)$. On the top plots we show the results at 600~fb$^{-1}$ of integrated luminosity, without  ($f_\mathrm{th} = 0.0$) and with  ($f_\mathrm{th} = 0.3$) theoretical uncertainty included and on the bottom we show the corresponding plots at 3000~fb$^{-1}$. We also present the 1-sigma contours as black dashed lines.}
  \label{fig:excc6cb}
\end{figure}

The $c_b\,(=\!c_\tau)$ coefficient is considered in Fig.~\ref{fig:excc6cb} on the $(c_b,c_6)$-plane. Its effects are expected to be sub-dominant in the production due to the assumption of MFV, but it affects the decays of the Higgs boson to $b\bar{b}$ (and $\tau \bar \tau$), and hence it is relevant to the process we are considering. The correlation visible reflects the fact that a reduced branching ratio can be compensated by an enhanced production cross section due to a negative value of $c_6$. For the given luminosity and $f_\mathrm{th} = 0.3$, the resulting bound after marginalization is $ -0.2 \lesssim c_b \lesssim 0.3$ at 1$\sigma$-equivalent. 

\begin{figure}[!htb]
  \centering
\begin{overpic}[width=0.45\textwidth,tics=10]{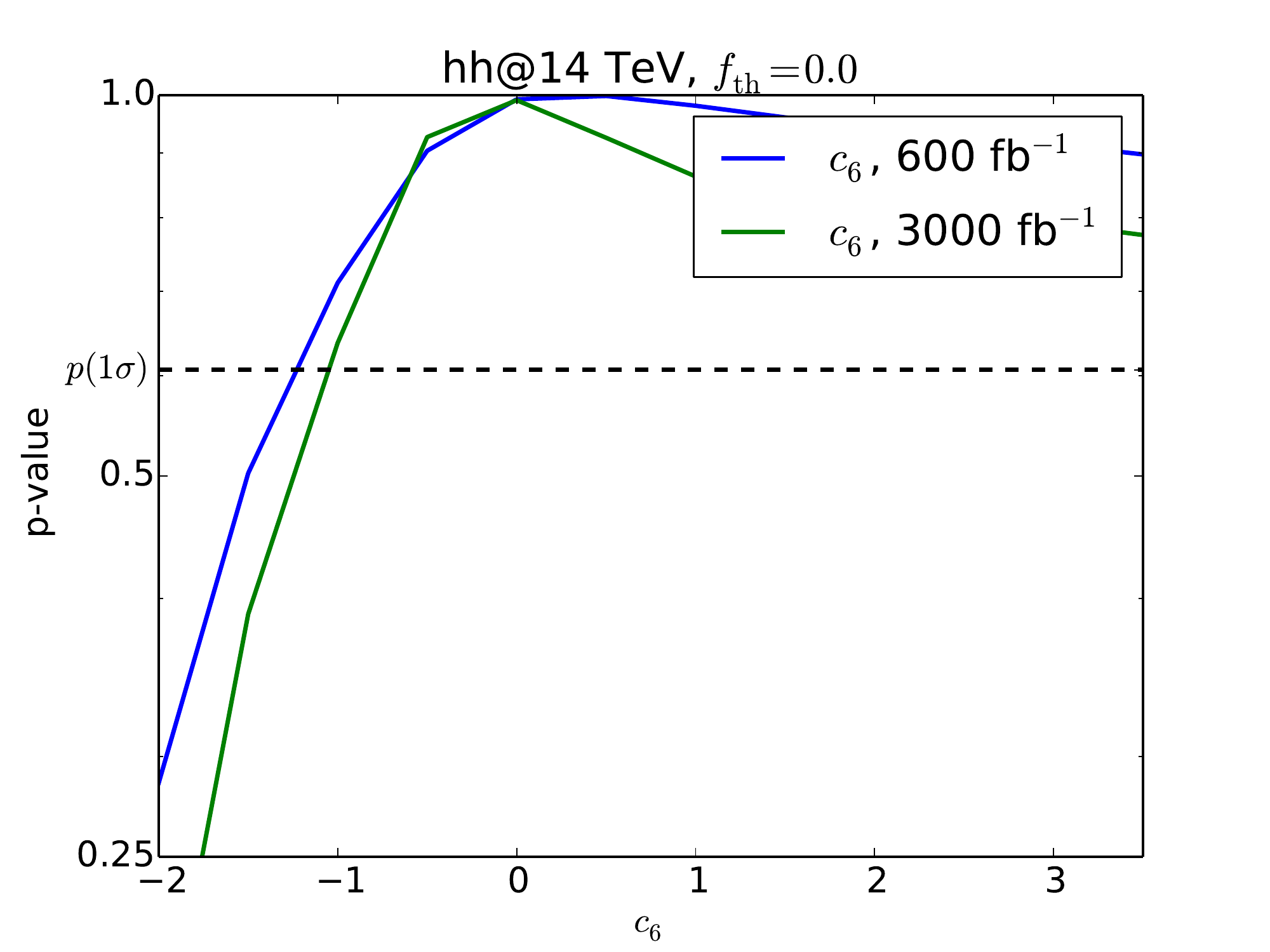}
 \put (65,30) {\large$\mathrm{full}$}
\end{overpic}
\begin{overpic}[width=0.45\textwidth,tics=10]{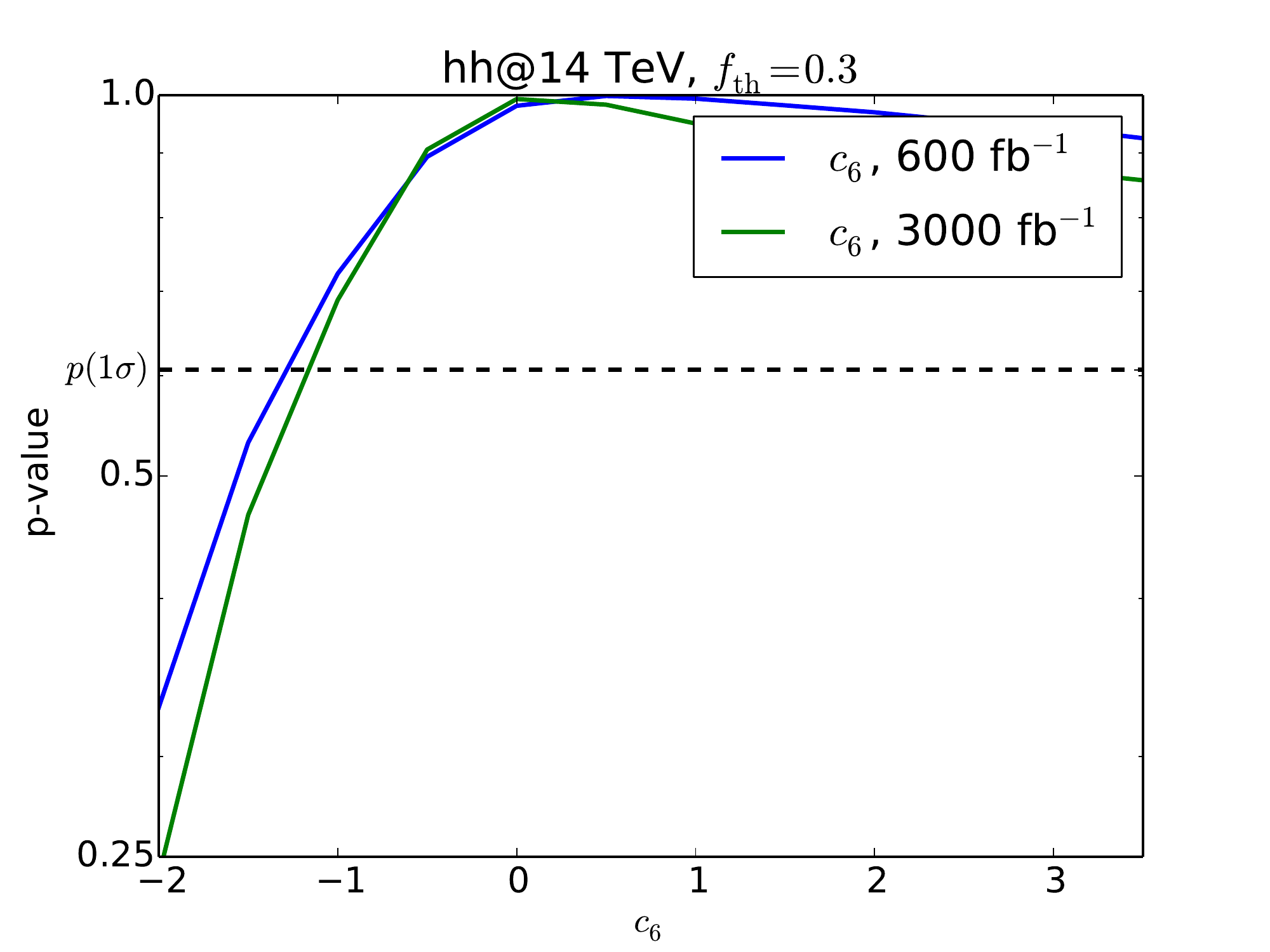}
 \put (65,30) {\large$\mathrm{full}$}
\end{overpic}
  \caption{The $p$-values obtained after marginalization over all coefficients except $c_6$ in the full model, for the process $hh \rightarrow (b\bar{b}) (\tau^+ \tau^-)$ at 600~fb$^{-1}$ and 3000~fb$^{-1}$ of integrated luminosity. On the left we show the resulting curves without theoretical uncertainty ($f_\mathrm{th} = 0.0$) and on the right we show results with theoretical uncertainty 30\%  ($f_\mathrm{th} = 0.3$).}
  \label{fig:excc6m}
\end{figure}

Finally, we show the resulting $p$-values for the coefficient $c_6$ in Fig.~\ref{fig:excc6m} after marginalizing over all the other coefficients. The constraints on $c_6$ are summarized in detail in Table~\ref{tb:constraints}.

\subsubsection{$c_6-c_t-c_b-c_\tau$ model}
\begin{figure}[!htb]
  \centering
  \begin{overpic}[width=0.45\textwidth,tics=10]{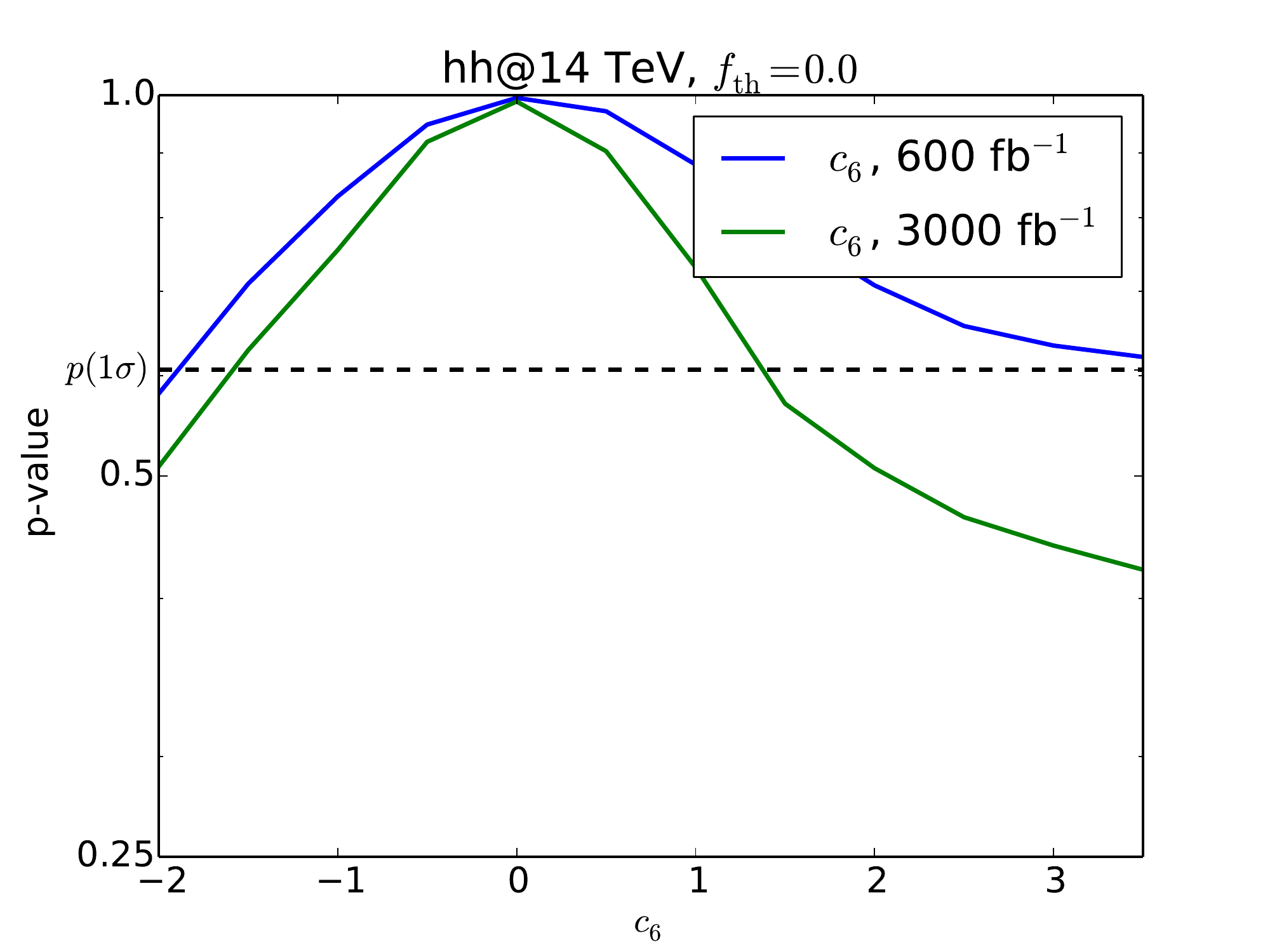}
    \put (45,30) {\large$c_6-c_t-c_b-c_\tau$}
  \end{overpic}
  \begin{overpic}[width=0.45\textwidth,tics=10]{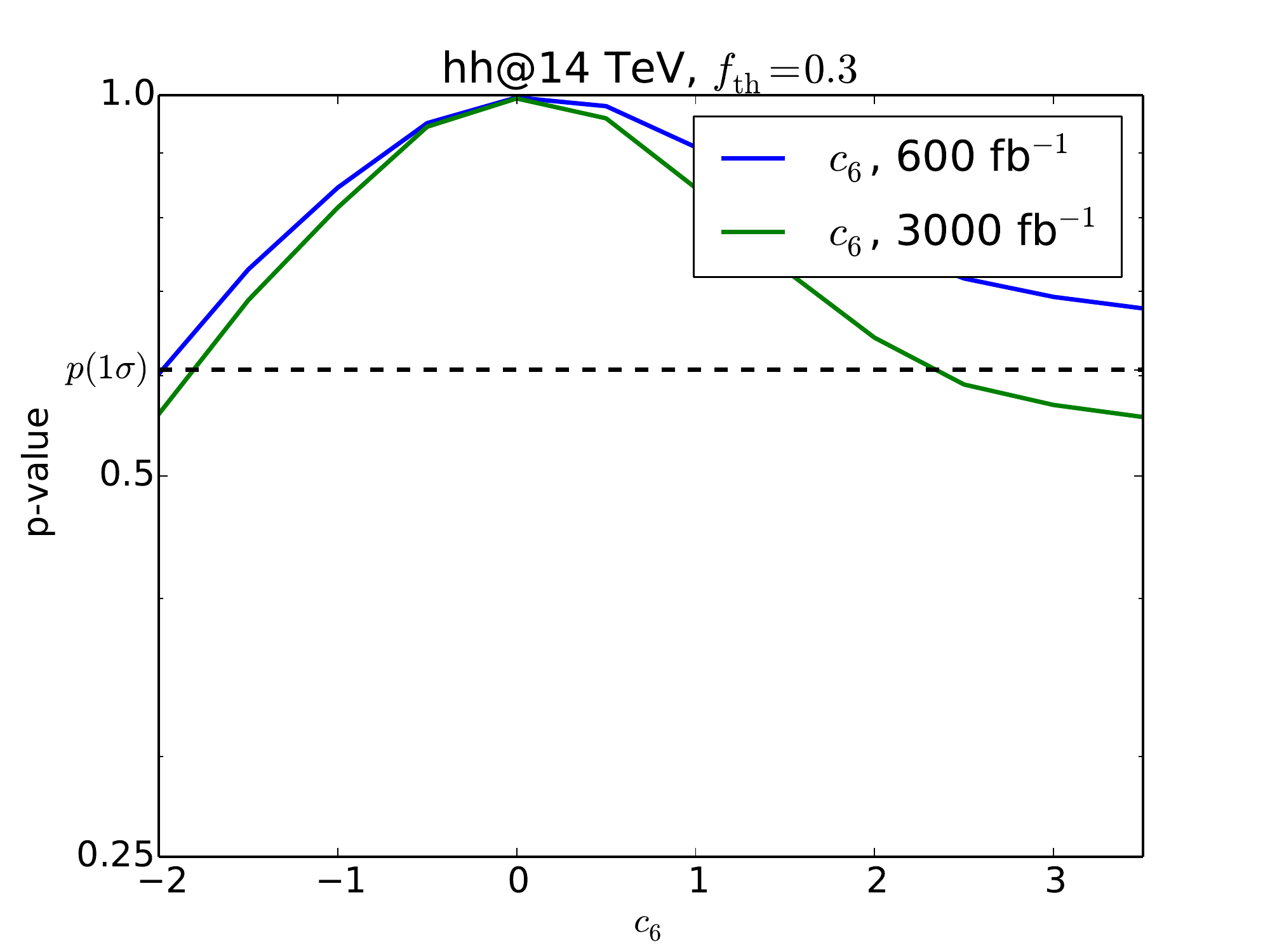}
    \put (45,30) {\large$c_6-c_t-c_b-c_\tau$}
  \end{overpic}
  \caption{The $p$-values obtained for $c_6$ in the $c_6-c_t-c_b-c_\tau$ model after marginalization, for the process $hh \rightarrow (b\bar{b}) (\tau^+ \tau^-)$ at 600~fb$^{-1}$ and 3000~fb$^{-1}$ of integrated luminosity. On the left we show the resulting curves without theoretical uncertainty ($f_\mathrm{th} = 0.0$) and on the right we show results with theoretical uncertainty 30\%  ($f_\mathrm{th} = 0.3$).}
  \label{fig:excc6cm}
\end{figure}

As a further example, we constrain the non-zero coefficients to be $c_6$, $c_t$, $c_b$ and $c_\tau$, varied in the same regions as in the full model. We emphasise that in this scenario, $c_\tau$ is allowed to vary independently of $c_b$. This model includes variations of the coefficients that are expected to be least constrained by single Higgs experimental data in future runs of the LHC. As in the previous sub-section, we marginalize over all coefficients to obtain bounds on $c_6$. The resulting $c_6$ $p$-values are shown in Fig.~\ref{fig:excc6cm} and the summary of results in Table~\ref{tb:constraints}.

\subsubsection{Summary of results and projected constraints}
We now discuss the constraints on $c_6$ obtained in all cases we considered as shown in Table~\ref{tb:constraints}. As explained before, the values take into account the current uncertainty due to the other weakly-bounded coefficients. As expected, the full model, including the current bounds coming from single Higgs boson measurements, provides a wide range for $c_6$, at 3000~fb$^{-1}$ ($c_6 \gtrsim -1.2$), when $f_\mathrm{th} = 0.3$, whereas the $c_6$-only model provides, as expected, a narrower range: $|c_6| \lesssim 0.4$ at 1$\sigma$ at 3000~fb$^{-1}$. As an alternative to the full model, our $c_6-c_t-c_b-c_\tau$ model aims to investigate a smaller set of coefficients, allowing, on the other hand, $c_\tau$ to vary as well. The 1$\sigma$ range in that case is $-1.8 \lesssim c_6 \lesssim 2.3$ at 3000~fb$^{-1}$. 

Evidently, by the end of the lifetime of the next LHC run (600~fb$^{-1}$) and the future high-luminosity run (3000~fb$^{-1}$), the constraints on several of the Wilson coefficients, $c_i (\neq c_6)$ will be substantially improved. The observables used to extract these constraints are complicated functions of these Wilson coefficients. Since it is beyond the scope of this article to predict the shape of the distributions of p-values of the $c_f$ coefficients from these observables, we will assume that these are Gaussian, peaking at the SM value, $\mu_f = 0.0$ with standard deviation $\Delta c_f$. The values of $\Delta c_f$ are calculated in such a way that the effect of coefficients on observables, such as $BR(h\rightarrow \gamma\gamma)$, $BR(h\rightarrow \tau^+ \tau^-)$, $BR(h\rightarrow b\bar{b})$ and $\sigma ( gg \rightarrow h)$ is $\mathcal{O}(10\%)$. We assume measurement of these quantities to be dominated by systematics at a luminosity of 600~fb$^{-1}$ and hence assume no improvement when going to 3000~fb$^{-1}$, at the end of the HL-LHC lifetime. The values of $\Delta c_f$ are summarized in Table~\ref{tb:deltaci}. 

We can then perform the marginalization procedure in a similar manner as prescribed above:
\begin{equation}\label{eq:margin3}
p(c_i,c_6) =  \frac{\sum_{\{c_f\}} p ( c_6, c_i, \{c_f\}) \times  p_\mathrm{Gauss.}  (\{c_f\})} { \sum_{\{c_f\}}  p_\mathrm{Gauss.}  (\{c_f\}) } \;,
\end{equation}
where 
\begin{equation}\label{eq:pgauss}
p_\mathrm{Gauss.}(\{c_f\}) = \prod_{f} \frac{1}{\Delta c_f \sqrt{2 \pi}} \exp \left\{ - \frac{(c_f - \mu_f)^2}{ 2 \Delta c_f^2} \right\} \;.
\end{equation}
After normalization according to Eq.~(\ref{eq:margin2}), this provides an estimate of what constraints would be achievable by combining future single Higgs boson data with the $hh \rightarrow (b\bar{b}) (\tau^+ \tau^-)$ channel. 

\begin{table}[!htb]
  \begin{center}
    \begin{tabularx}{0.25\linewidth}{XX}
      \toprule
      ~$c_f$ & $\Delta c_f$ \\ \midrule
      ~$c_g$  & $0.05\times\frac{1}{3}$\\
      ~$c_H$ & $0.05 \times 2$\\ 
      ~$c_t$, $c_b$, $c_\tau$  & $0.05$  \\ 
      ~$c_\gamma$ &  $ 0.05 \times \frac{47}{18}$ \\\bottomrule
    \end{tabularx}
  \end{center}
  \caption{A summary of the Gaussian errors $\Delta c_f$ assumed to generate the `future' results at 600~fb$^{-1}$/3000~fb$^{-1}$ of luminosity of the LHC at 14~TeV. The values are chosen so as to cause $\mathcal{O}(10\%)$ effects on single Higgs boson observables. The numerical factors stem from the normalization with respect to the corresponding SM effect. See, for example, ~\cite{Kniehl:1995tn}.}
\label{tb:deltaci}
\end{table}

Table~\ref{tb:constraints} includes the estimates for the `future' combination, labelled accordingly. We obtain tighter constraints: $|c_6| \lesssim 0.6$ for the full model and $-0.6 \lesssim c_6 \lesssim 0.5$ for the alternative model at 3000~fb$^{-1}$, for 30\% theoretical uncertainty. Evidently, as expected, these approach the constraints given in the $c_6$-only model.

\begin{table}[!htb]
  \begin{center}
    \begin{tabularx}{\linewidth}{XXX}
      \toprule
       model &$L=600~\mathrm{fb}^{-1}$ & $L=3000~\mathrm{fb}^{-1}$ \\ \midrule
 $c_6$-only & $c_6 \in(-0.5,0.8)$ & $c_6\in (-0.4,0.4)$\\
 full & $c_6 \gtrsim -1.3 $ & $c_6 \gtrsim -1.2 $ \\ 
$c_6-c_t-c_\tau-c_b$ &  $c_6 \gtrsim -2.0$ &$c_6 \in(-1.8,2.3)$   \\ \midrule
 full (\textbf{future}) & $c_6 \in (-0.8, 0.9) $ & $c_6 \in (-0.6,0.6) $ \\ 
$c_6-c_t-c_\tau-c_b$ (\textbf{future}) &  $c_6 \in (-0.8, 0.8) $ &$c_6 \in(-0.6,0.5)$   \\\bottomrule
    \end{tabularx}
  \end{center}
  \caption{A summary of the constraints obtained on the coefficient $c_6$ at 1$\sigma$, at integrated luminosities of 600~fb$^{-1}$ and 3000~fb$^{-1}$ at a 14~TeV LHC, assuming that the theoretical uncertainty on the signal rates is 30\%.}
\label{tb:constraints}
\end{table}

\section{Conclusions}\label{sec:conclusions}
We have investigated the dimension-6 effective theory description of beyond-the-Standard Model modifications to Higgs boson pair production. Using an implementation within the \texttt{HERWIG++} Monte Carlo event generator and a realistic analysis, including the description of theoretical uncertainty on the signal rates, we have constructed the possible exclusion regions in a $c_6$-only model, a general EFT with all coefficients allowed to vary, and a constrained EFT, marginalising over various parameters in the latter two cases. We find that at the Large Hadron Collider at 14~TeV, meaningful constraints can be obtained on the parameter space of the EFT, particularly on the $c_6$ coefficient. These results appear in Table~\ref{tb:constraints}. We conclude that, approximately, the hitherto unconstrained $c_6$ would be limited to $c_6 \gtrsim -1.2$ at 1$\sigma$, in the `full' model, given the current constraints on the other coefficients originating from single Higgs boson data, for 3000~fb$^{-1}$, assuming 30\% theoretical uncertainty on the signal rate. In the `alternative' $c_6-c_t-c_b-c_\tau$ model, this was found to be $-1.8 \lesssim c_6 \lesssim 2.3$. We also provide an estimate of future single Higgs bounds on the Wilson coefficients, by marginalizing over the irrelevant coefficients with a Gaussian probability centred around the SM value ($c_i = 0.0$) with uncertainties that would give $\mathcal{O}(10\%)$ deviations in single Higgs boson observables. This gives tighter constraints for $c_6$ in the two models: $-0.6 \lesssim c_6 \lesssim 0.6$ for the `full' model and $-0.6 \lesssim c_6 \lesssim 0.5$ for the alternative model at 3000~fb$^{-1}$, for 30\% theoretical uncertainty.

It is clear that the expected bounds could be substantially enhanced, for example, by examining other final states originating from $hh$, improving the experimental analyses by examining differential distributions and future improvements of the theoretical description of the signal process. Without doubt, our results demonstrate that the process of Higgs boson pair production should be seriously considered as part of the wider programme of constraining the higher-dimensional effective field theory parameter space. 

\section{Acknowledgements}
We would like to thank  Brando Bellazzini, Adrian Carmona,  Gino Isidori, Zoltan Kunszt, Matthias Neubert, Pedro Schwaller and Luca Vecchi for useful discussions. The research of JZ is supported by the ERC Advanced Grant EFT4LHC of the European Research Council, the Cluster of Excellence Precision Physics, Fundamental Interactions and Structure of Matter (PRISMA-EXC 1098). AP acknowledges support in part by the Swiss
National Science Foundation (SNF) under contract 200020-149517, by
the European Commission through the ``LHCPhenoNet'' Initial Training
Network PITN-GA-2010-264564, MCnetITN FP7 Marie Curie Initial Training Network
PITN-GA-2012-315877 and by a Marie Curie Intra European Fellowship within the 7th European Community Framework Programme (grant no. PIEF-GA-2013-622071). The research of FG has been supported by the Swiss National Foundation under contract SNF 200021-143781 and by a Marie Curie Intra European Fellowship within the 7th European Community Framework Programme (grant no. PIEF-GA-2013-628224).
\appendix 

\section[{SM $D=6$ Lagrangian}]{SM \boldmath{$D=6$} Lagrangian}\label{app:d6lag} 

In this section we give more details on the Lagrangian terms not explicitly shown in Eq.~(\ref{eq:Lfinal}). The part of the SM Lagrangian relevant to our study is given by
\begin{equation}
{\cal L}_{SM} = - \mu^2 |H|^2 - \lambda |H|^4 -  y_t \bar Q_L H^c t_R -  y_b \bar Q_L H b_R -   y_{\tau } \bar Q_L H {\tau}_R 
\,+ \text{h.c.}\; ,
\end{equation}
where for simplicity we have only written the Higgs boson couplings to the 3rd generation.

The CP-odd effects are given by
	\begin{equation}
	\begin{split}
	 {\cal L}_{CP} &= \frac{\alpha_s \tilde c_g}{4 \pi \Lambda^2} |H|^2 G_{\mu\nu}^a \tilde G^{\mu\nu}_a
	+ \frac{\alpha^\prime\,  \tilde c_\gamma}{4 \pi \Lambda^2} |H|^2 B_{\mu\nu} \tilde B^{\mu\nu}\\
	&+ \frac{i g\, \tilde c_{HW}}{\Lambda^2 }(D^\mu H)^\dagger\sigma_k (D^\nu H) \tilde W_{\mu\nu}^k
	+ \frac{i  g^\prime\, \tilde c_{HB}}{\Lambda^2}(D^\mu H)^\dagger (D^\nu H) \tilde B_{\mu\nu}\,.
	\end{split}
	\end{equation}
In this work we have neglected CP-odd effects for simplicity -- these are not expected to contribute to the inclusive cross section at the order considered. Moreover, using current Higgs boson data one can already set some constraints on a Higgs CP-odd component (see e.g.~\cite{Freitas:2012kw,Brod:2013cka,Dolan:2014upa} and references therein).

Finally, the four-fermion operators, employing the same basis as used in \cite{Elias-Miro:2013mua}, read 
	\begin{equation}
	\begin{split}
	{\cal L}_{4f}=\ &
	c_{LR}^t (\bar Q_L \gamma^\mu Q_L)(\bar t_R \gamma_\mu t_R) 
	+ c_{LR}^{(8)t} (\bar Q_L \gamma^\mu T^A Q_L)(\bar t_R \gamma_\mu  T^A t_R)
	+ c_{RR}^t (\bar t_R \gamma^\mu t_R)(\bar t_R \gamma_\mu t_R) \\
	&+ c_{LL}^q (\bar Q_L \gamma^\mu Q_L)(\bar Q_L \gamma_\mu Q_L)
	+ c_{LL}^{(8)q} (\bar Q_L \gamma^\mu T^A Q_L)(\bar Q_L \gamma_\mu T^A Q_L)\\
	&+ c_{LR}^{\tau} (\bar{L}_L \gamma^\mu L_L) (\bar \tau_R \gamma_\mu \tau_R) 
	+ c_{RR}^{\tau} (\bar{\tau}_R \gamma^\mu \tau_R) (\bar \tau_R \gamma_\mu \tau_R) 
	+ c_{LL}^{l} (\bar{L}_L \gamma^\mu L_L)  (\bar{L}_L \gamma_\mu L_L) \\ 
	& + c_{LR}^{lt} (\bar L_L \gamma^\mu L_L)(\bar t_R \gamma_\mu t_R) 
	+ c_{LR}^{lb} (\bar L_L \gamma^\mu L_L)(\bar b_R \gamma_\mu b_R) \\
	&+ c_{LL}^{ql} (\bar Q_L \gamma^\mu Q_L)(\bar L_L \gamma_\mu L_L) 
	+ c_{LL}^{(3)ql} (\bar Q_L \gamma^\mu \sigma^a Q_L)(\bar L_L \gamma_\mu \sigma^a L_L) 
	+ c_{LR}^{q \tau} (\bar Q_L \gamma^\mu Q_L)(\bar \tau_R \gamma_\mu \tau_R)  \\
	& + c_{RR}^{t \tau} (\bar t_R \gamma^\mu t_R)(\bar \tau_R \gamma_\mu \tau_R) 
	+ c_{RR}^{b \tau} (\bar b_R \gamma^\mu b_R)(\bar \tau_R \gamma_\mu \tau_R) \\
	&+ c_{LR}^b (\bar Q_L \gamma^\mu Q_L)(\bar b_R \gamma_\mu b_R)
	+ c_{LR}^{(8)b} (\bar Q_L \gamma^\mu T^A Q_L)(\bar b_R \gamma_\mu  T^A b_R)
	+ c_{RR}^b (\bar b_R \gamma^\mu b_R)(\bar b_R \gamma_\mu b_R)\\
	&+ c_{RR}^{t b} (\bar t_R \gamma^\mu t_R)(\bar b_R \gamma_\mu b_R) 
	+ c_{RR}^{(8) t b} (\bar t_R \gamma^\mu T^A t_R)(\bar b_R \gamma_\mu T^A b_R)\\
	&+ c_{y_t y_\tau}  y_t y_\tau (\bar Q_L^i t_R)\epsilon_{ij}(\bar L_L^j \tau_R) 
	+ c^{\prime}_{y_t y_\tau}  y_t y_\tau (\bar Q_L^{i \alpha} \tau_R)\epsilon_{ij}(\bar L_L^j t_R^{\alpha}) 
	+ c_{y_\tau y_b}  y_\tau y_b^{\dagger} (\bar L_L \tau_R) (\bar b_R Q_L) \\
	&+ c_{y_t y_b}\, y_t y_b (\bar Q_L^i t_R)\epsilon_{ij}(\bar Q_L^j b_R) 
	+ c_{y_t y_b}^{(8)}\, y_t y_b (\bar Q_L^i T^A t_R)\epsilon_{ij}(\bar Q_L^j T^A b_R) 
	 \,,
	\end{split}
	\end{equation}
where the last five operators are suppressed in scenarios of minimal flavour violation (MFV), and we have restricted ourselves again to the third generation. 

\section{Electroweak symmetry breaking with dimension-6 operators}\label{app:laghh}
Due to the additional terms originating from the dimension-6 operator $\sim |H|^6$, the position of the electroweak minimum changes with respect to the Standard Model prediction. To find the new minimum we consider the potential:
\begin{equation}\label{eq:potentialsix}
V_{SM+6} = \mu^2 |H|^2 + \lambda |H|^4 + \frac{c_6}{\Lambda^2} \lambda |H|^6 \;,
\end{equation}
which contains the additional interaction. Applying the minimisation condition $\partial V / \partial |H|^2 = 0$, we obtain
\begin{eqnarray}
(|H|^2)_{\pm} = &-&\frac{\Lambda^2} { 3 c_6 } \pm  \sqrt{ \frac{ \Lambda^4 } { 9 c_6^2 } - \frac{ \mu^2 \Lambda^2 } { 3 c_6 \lambda } } \nonumber \\
=&+& \frac{\Lambda^2} { 3 c_6 } \left( -1 \pm \sqrt{ 1 - \frac{3 \mu^2 c_6 } { \Lambda^2 \lambda } } \right)\;. 
\end{eqnarray}
Considering the SM vacuum, i.e. taking the $+$ solution, we obtain :
\begin{equation}\label{eq:min}
\frac{v^2}{2} \equiv (|H|)_+^2 = \frac{\Lambda^2} { 3 c_6 } \left( -1 + \sqrt{ 1 - \frac{3 \mu^2 c_6 } { \Lambda^2 \lambda } } \right) \;, 
\end{equation}
which we can solve for $\mu^2$:
\begin{equation}\label{eq:musq}
\mu^2 = - \lambda v^2 \left( 1 + \frac{3}{4} \frac{c_6 v^2} { \Lambda ^2 } \right)\;.
\end{equation}
Ignoring the Goldstone modes, we can expand the Higgs field $|H|$ in terms of the physical scalar Higgs boson about this minimum, $H \sim ( 0, (h + v)/\sqrt{2} )$. We start by examining the terms arising from the SM Lagrangian and the dimension-6 operators that contribute to the kinetic term:
\begin{eqnarray}
\mathcal{L}_\mathrm{kin} = (D_\mu H)^\dagger (D^\mu H) +  \frac{ c_H}{2\Lambda^2}(\partial^\mu |H|^2)^2  \;,
\end{eqnarray}
where $D_\mu$ is the covariant derivative that includes all the interactions of the Higgs field with the gauge bosons. After expansion, we arrive at
\begin{eqnarray}\label{eq:kinprecan}
\mathcal{L}_\mathrm{kin} &=& \frac{1}{2} \left( 1 + \frac{ c_H v^2 } { \Lambda^2 } \right) \partial_\mu h \partial^\mu h \nonumber \\
&+& \frac{ c_H v } { \Lambda^2 } h  \partial_\mu h \partial^\mu h \nonumber \\
&+& \frac{ c_H  } { 2 \Lambda^2 } h^2  \partial_\mu h \partial^\mu h +~... \;,
\end{eqnarray}
where we have ignored the gauge boson interactions. To canonically normalise the Higgs boson kinetic term and to remove derivative interactions, we consider the following non-linear transformation:
\begin{equation}
h = \left( 1  + \frac{a_0' v^2 } { \Lambda^2 } \right) h' + \frac{a_1' v } { \Lambda^2 }  h'^2 + \frac{ a_2' } { \Lambda^2 } h'^3 \;.
\end{equation}
Plugging this into Eq.~(\ref{eq:kinprecan}), we find the values of the constants $a_i'$ that cancel all terms except $(1/2)  \partial_\mu h \partial^\mu h$: 
\begin{eqnarray}
a_0' = - \frac{1}{2} c_H,\; a_1' = -\frac{1}{2} c_H,\; a_2' = -\frac{1}{6} c_H\;,
\end{eqnarray}
giving
\begin{equation}\label{eq:shift}
h = \left( 1  - \frac{c_H v^2 } { 2\Lambda^2 } \right) h' - \frac{c_H v } {2 \Lambda^2 }  h'^2 - \frac{ c_H } { 6\Lambda^2 } h'^3 \;.
\end{equation}
This shift should be performed \textit{everywhere} in the Lagrangian, and introduces changes in as well as new interactions.

We first perform the shift in the terms contributing to the Higgs boson scalar mass. The relevant terms are:
\begin{eqnarray}
\mathcal{L}_{m_h} = - \frac{ \mu^2}{2} h^2 - \frac{3 \lambda v^2 } {2} h^2 - \frac{ 15 c_6 \lambda v^4 } { 8\Lambda^2 } h^2 \;.
\end{eqnarray}
where the last term comes from the $|H|^6$ interaction. We substitute for $\mu^2$ using Eq.~(\ref{eq:musq}) and perform the shift of Eq.~(\ref{eq:shift}), keeping terms up to $h^2$ and $\mathcal{O}(\Lambda^{-2})$:
\begin{eqnarray}
\mathcal{L}_{m_h} = &-&\frac{1}{2} \left( 2 \lambda v^2 + \frac{3 c_6 \lambda v^4 } { \Lambda^2} \right) \left( 1 - \frac{ c_H v^2 } { 2 \Lambda^2 } \right)^2 h'^2 \nonumber \\
= &-& \lambda v^2 \left( 1  +\frac{3 c_6 v^2 } { 2 \Lambda^2} - \frac{ c_H v^2 } { \Lambda^2 } \right) h'^2 \;,
\end{eqnarray}
from which we can immediately deduce the Higgs mass:
\begin{equation}
m_h^2 = 2 \lambda v^2 \left( 1 - \frac{ c_H v^2}{ \Lambda^2 } +\frac{3 c_6 v^2 } { 2 \Lambda^2} \right) \;.
\end{equation}
The terms contributing to multi-Higgs production via gluon fusion at the LHC are:
\begin{equation}
\begin{split}
\mathcal{L}_{h^n} = &- \mu^2 |H|^2 - \lambda |H|^4 -  \left( y_t \bar Q_L H^c t_R + y_b \bar Q_L H b_R + \text{h.c.} \right)
\\
&+ \frac{c_H}{2\Lambda^2}(\partial^\mu |H|^2)^2 - \frac{c_6}{\Lambda^2} \lambda |H|^6  + \frac{\alpha_s c_g}{4\pi\Lambda^2} |H|^2 G_{\mu\nu}^a G^{\mu\nu}_a
\\
&- \left( \frac{c_t}{\Lambda^2} y_t |H|^2 \bar Q_L H^c t_R	+ \frac{c_b}{\Lambda^2}y_b |H|^2 \bar Q_L H b_R + \text{h.c.} \right) , 
\end{split}
\end{equation}
where we have included in the first line the relevant Standard Model terms that will receive corrections from dimension-6 operators. 

We proceed by deriving the expressions for the triple coupling after expanding about the minimum and canonically normalising via Eq.~(\ref{eq:shift}). The relevant terms are the same as those that appear in the potential of Eq.~(\ref{eq:potentialsix}):
\begin{equation}
\mathcal{L}_\mathrm{self} = - V_{SM+6} = - \mu^2 |H|^2 - \lambda |H|^4 - \frac{c_6}{\Lambda^2} \lambda |H|^6 \;.
\end{equation}
Expanding this about the electroweak minimum, we get
\begin{eqnarray}
\mathcal{L}_\mathrm{self} = &-&\mu^2 \frac{ (v+h)^2 } { 2} - \lambda \frac{ (v+h)^4 } { 4} - \frac{ c_6 \lambda } { \Lambda^2 }  \frac{ (v+h)^6 } { 8 } \nonumber \\
= &-& \frac{ \mu^2 } { 2 } ( v^2 +  2 h v + h^2 ) - \frac{ \lambda} { 4} ( v^4 + 4 h v^3 + 6 h^2 v^2 + 4 h^3 v + h^4 ) \nonumber \\
&-& \frac{ c_6 \lambda } { 8 \Lambda^2 } ( v^6 + 6 v^5 h + 15 v^4 h^2 + 20 h^3 v^3 + 15 v^2 h^4 + 6 h^5 v + h^6 ) \;.
\end{eqnarray}
Omitting terms with $h^n$, $n>4$, and constant terms we arrive at
\begin{eqnarray}\label{eq:selftriple}
\mathcal{L}_\mathrm{self} =  &-& \frac{ \mu^2 } { 2 } ( 2 h v + h^2 ) - \frac{ \lambda} { 4} ( 4 h v^3 + 6 h^2 v^2 + 4 h^3 v + h^4 ) \nonumber \\
&-& \frac{ c_6 \lambda } { 8 \Lambda^2 } (  6 h v^5 + 15 h^2 v^4 + 20 h^3 v^3 + 15 h^4 v^2 ) + ~... \;.
\end{eqnarray}
It is convenient to also calculate the $h^2, h^3$ and $h^4$ terms as a function of $h'$ up to $h'^4$:
\begin{eqnarray}
h^2 = h'^2 \left[ 1 - \frac{ c_H v^2 } { \Lambda^2 } - \frac{ c_H v } { \Lambda^2 } h' - \frac{ c_H } { 3 \Lambda^2  } h'^2 \right] + \mathcal{O}(h'^5) \;,\nonumber \\
h^3 = h'^3 \left[ 1 - \frac{3 c_H v^2 } { 2 \Lambda^2 } - \frac{3 c_H v } { 2 \Lambda^2 } h' \right] + \mathcal{O}(h'^5)  \;, \nonumber \\
h^4 = h'^4 \left [ 1  - \frac{2 c_H v^2 } { \Lambda^2 }   \right] + \mathcal{O}(h'^5) \; .
\end{eqnarray}
These terms are then substituted into Eq.~(\ref{eq:selftriple}), after which we obtain the terms up to $\mathcal{O}(h'^4)$
\begin{eqnarray}\label{eq:selftriple_final}
\mathcal{L}_\mathrm{self} = &-& \lambda \left[  v + \frac{ 5 c_6 v^3 } { 2 \Lambda^2 } - \frac{ 5 c_H v^3 } { 2 \Lambda^2 } \right] h'^3 - \frac{\lambda}{4} \left[ 1 + \frac{15 c_6 v^2}{2 \Lambda^2} - \frac{28 c_H v^2}{3 \Lambda^2} \right] h'^4 +~... \nonumber \\
= &-& \frac{ m_h^2 } { 2v }   \left[ 1 + \frac{ c_6 v^2 } { \Lambda^2 } - \frac{ 3 c_H v^2 } { 2 \Lambda^2 } \right] h'^3 - \frac{ m_h^2 } { 8v^2 }  \left[ 1 + \frac{ 6 c_6 v^2 } { \Lambda^2 } - \frac{ 25 c_H v^2 } { 3 \Lambda^2 } \right]  h'^4 +~ ... \;.
\end{eqnarray}
Finally, we focus on the fermion-Higgs boson interactions that receive contributions from
\begin{eqnarray}
\mathcal{L}_{hf} = - \frac{ y_f } { \sqrt{2} } \bar{f}_L ( h + v ) f_R - \frac { c_f y_f } { \Lambda^2 } \frac{ ( v+h )^2 } { 2} \bar{f}_L  \frac{ ( v+h ) } { \sqrt{2} } f_R\, + \text{h.c.}\;,
\end{eqnarray}
where $f = t, b, ~...$, with $f_{L,R}$ the left- and right-handed fields, and the first term comes from the SM whereas the second term is a dimension-6 contribution. Substituting in the shift of Eq.~(\ref{eq:shift}), and keeping terms up to $\mathcal{O}(\Lambda^{-2})$, we obtain
\begin{eqnarray}\label{eq:hf}
\mathcal{L}_{hf} = &-& \frac{ y_f v} { \sqrt{2} }  \left( 1 + \frac{c_t v^2 } { 2 \Lambda^2 } \right)\bar{f}_L f_R \nonumber \\
&-& \frac{ y_f } { \sqrt{2} } \left( 1 - \frac{ c_H v^2 } { 2 \Lambda^2 } + \frac{ 3 c_f v^2 } { 2 \Lambda^2 } \right) \bar{f}_L f_R h' \nonumber \\
&-& \frac{ y_f } { \sqrt{2} } \left( \frac{ 3 c_f v } { 2 \Lambda^2 } - \frac{ c_H v } { 2 \Lambda^2 } \right) \bar{f}_L f_R h'^2 \, + \text{h.c.} + \mathcal{O}(h'^3) + \mathcal{O}(\Lambda^{-4}) \;.
\end{eqnarray}
The first line gives the expression for the modified fermion mass,
\begin{equation}
m_f = \frac{ y_f v} { \sqrt{2} } \left( 1 + \frac{c_t v^2 } { 2 \Lambda^2 } \right) \,,
\end{equation}
and we can re-express Eq.~(\ref{eq:hf}) in terms of this:
\begin{eqnarray}\label{eq:hfmass}
\mathcal{L}_{hf} = &-& m_f \bar{f}_L f_R \nonumber \\
&-&   \frac{m_f}{v}  \left( 1 - \frac{ c_H v^2 } { 2 \Lambda^2 } + \frac{  c_f v^2 } {  \Lambda^2 } \right) \bar{f}_L f_R h' \nonumber \\
&-&  \frac{m_f}{v} \left( \frac{ 3c_f v } {  2\Lambda^2 } - \frac{ c_H v } { 2 \Lambda^2 } \right) \bar{f}_L f_R h'^2 \, + \text{h.c.}+ \mathcal{O}(h'^3) + \mathcal{O}(\Lambda^{-4}) \;.
\end{eqnarray}
The final term that we need to consider is
\begin{eqnarray}\label{eq:hg}
\mathcal{L}_{hg} &=& \frac{ \alpha_s c_g } { 4\pi \Lambda^2} |H|^2 G_{\mu\nu}^a G^{\mu\nu}_a \nonumber \\
&=& \frac{ \alpha_s c_g } { 4\pi \Lambda^2}  \frac{(h + v)^2}{2} G_{\mu\nu}^a G^{\mu\nu}_a \nonumber \\
&=& \frac{ \alpha_s c_g } { 4 \pi \Lambda^2} ( h v + \frac{h^2}{2} ) G_{\mu\nu}^a G^{\mu\nu}_a + ~...\;,
\end{eqnarray}
where the omitted constant term can be absorbed into an unobservable re-definition of the gluon wave function. 

The interactions that contribute to Higgs boson pair production via gluon fusion appear in Eqs.~(\ref{eq:selftriple_final}),~(\ref{eq:hfmass}) and~(\ref{eq:hg}).
\bibliography{hheft}
\bibliographystyle{JHEP}

\end{document}